\documentclass[cernpreprint, atlasdraft=false, UKenglish, texmf, orcidlogo]{atlasdoc}
\usepackage{atlaspackage}
\usepackage{atlasbiblatex}
\usepackage{tikz}
\usepackage{multirow}
\usepackage{arydshln}
\usetikzlibrary{arrows.meta, positioning,shapes.geometric, fit, backgrounds, calc}
\usepackage{atlasphysics}
\graphicspath{{logos/}{figures/}}

\usepackage{ANA-HIGP-2024-37-PAPER-defs}
\AtlasTitle{Improved analysis of non-resonant Higgs boson pair production in the $b\bar{b}\tau^+\tau^-$ final state with $196$~fb$^{-1}$ of data collected at $\sqrt{s} = 13$~TeV and $13.6$~TeV with the ATLAS detector}

\AtlasAbstract{%
A search for non-resonant Higgs boson pair production ($HH$) in the $b\bar{b}\tau^+\tau^-$ final state is performed using $140$~fb$^{-1}$ and $56$~fb$^{-1}$ of proton--proton collision data at centre-of-mass energies of $\sqrt{s} = 13$~\TeV and $13.6$~\TeV, respectively, recorded by the ATLAS detector during 2015--2023 at the CERN Large Hadron Collider. Relative to the previous ATLAS searches in the same final state, the analysis benefits from the additional dataset collected at 13.6 \TeV and from improvements in both event reconstruction and analysis techniques. The Higgs boson pair production cross-section divided by the Standard Model (SM) prediction is found to be $\mu_{HH}=  2.6^{+1.4}_{-1.0}$, consistent with the SM expectation. This corresponds to an observed (expected) significance of $2.6$ ($1.2$) standard deviations over the background-only hypothesis of no $HH$ production and 1.65 standard deviations with respect to the SM prediction. Under the assumption that all other Higgs boson couplings take their SM values, the observed (expected) 95\% confidence level interval for the Higgs boson self-coupling modifier is found to be $\kl \in [-3.4, 1.6] \cup [5.5, 10.1]$ ($\kl \in[-1.7, 8.5]$). As a validation of the analysis strategy, measurements of the $ZH$ and $ZZ$ processes are performed in the same final state using the same event selection, yielding observed (expected) significances of 3.5 (2.4) and 1.8 (2.7) standard deviations, respectively, relative to the background-only prediction.
}

\AtlasRefCode{HIGP-2024-37}

\PreprintIdNumber{CERN-EP-2026-227}

\AtlasJournal{EPJC}
\hypersetup{pdftitle={ATLAS document},pdfauthor={The ATLAS Collaboration}}

\begin{document}

\maketitle

\tableofcontents

\section{Introduction}

%

The discovery of the Higgs boson ($H$) in 2012 by the ATLAS~\cite{HIGG-2012-27} and CMS~\cite{CMS-HIG-12-028} experiments at the Large Hadron Collider (LHC)~\cite{Evans:2008zzb} was a landmark achievement in particle physics, providing direct evidence for the mechanism of electroweak symmetry breaking and mass generation described by the Standard Model (SM)~\cite{HIGGS1964132,PhysRevLett.13.321,PhysRevLett.13.508,PhysRevLett.13.585,PhysRev.145.1156,PhysRev.155.1554}. Since then, the mass of the Higgs boson has been measured with increasing precision~\cite{HIGG-2014-14,HIGG-2022-20}, and its spin, parity, and couplings to other particles have been studied extensively and found to be consistent with the SM predictions~\cite{HIGG-2013-17-witherratum,HIGG-2019-01,CMS-HIG-19-009,HIGG-2018-32,CMS-HIG-21-013,HIGG-2021-23,CMS-HIG-22-001}. One key property that remains only weakly constrained is the Higgs boson self-coupling, which is directly related to the shape of the Higgs potential and hence to the mechanism of electroweak symmetry breaking. Its measurement is important for understanding the nature of the electroweak phase transition and for testing the stability of the electroweak vacuum~\cite{PhysRevD.101.075023}.

\begin{figure}[!tbh]
\centering
\subfloat[]{
\includegraphics[width=0.4\textwidth]{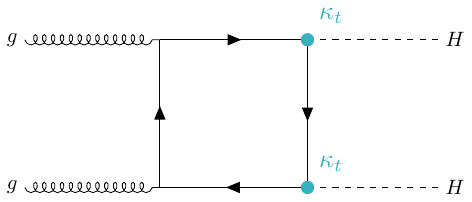}
\label{fig:ggFHH_a}
}
\subfloat[]{
\includegraphics[width=0.4\textwidth]{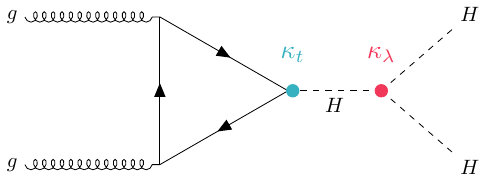}
\label{fig:ggFHH_b}
}
\caption{Leading-order Feynman diagrams of the ggF \HH production process: \protect\subref{fig:ggFHH_a} box and \protect\subref{fig:ggFHH_b} triangle diagrams. The trilinear Higgs boson self-coupling modifier is shown as \kl, while the modifier for the coupling of the Higgs boson to the top quark is shown as \kappat.}
\label{fig:ggFdiagrams}
\end{figure}

\begin{figure}[!tbh]
\centering
\subfloat[]{
\includegraphics[width=0.25\textwidth]{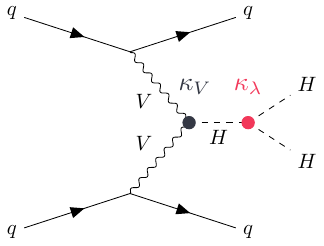}
\label{fig:VBFHH_a}
}
\subfloat[]{
\includegraphics[width=0.25\textwidth]{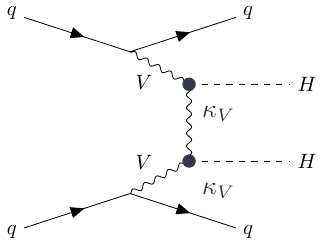}
\label{fig:VBFHH_b}
}
\subfloat[]{
\includegraphics[width=0.25\textwidth]{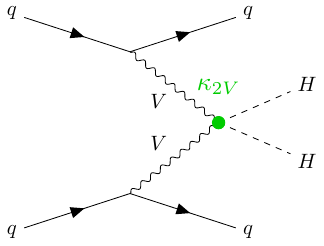}
\label{fig:VBFHH_c}
}
\caption{Leading-order Feynman diagrams of the VBF \HH production process. The trilinear Higgs boson self-coupling modifier is shown as \kl, the modifier for the coupling of the Higgs boson to the SM vector bosons $V$ as \kv, and the quartic $HHVV$ coupling modifier as \kvv.}
\label{fig:VBFdiagrams}
\end{figure}

The most direct way to access the Higgs boson trilinear self-coupling at the LHC is through measurements of Higgs boson pair (\HH) production in proton--proton ($pp$) collisions. In the SM, \HH production at the LHC is dominated by gluon--gluon fusion (ggF), followed by vector-boson fusion (VBF); the corresponding leading-order Feynman diagrams are shown in Figures~\ref{fig:ggFdiagrams} and~\ref{fig:VBFdiagrams}, respectively. The ggF process receives contributions from two interfering amplitudes: one containing two top-quark Yukawa couplings in a box diagram, and one containing the trilinear Higgs boson self-coupling $\lhhh$ via an off-shell Higgs boson in the $s$-channel. The VBF process receives contributions from three amplitudes: one with the trilinear self-coupling $\lhhh$, one with the $HVV$ coupling between a single Higgs boson and two vector bosons, and one with the quartic $HHVV$ coupling between two Higgs bosons and two vector bosons. The predicted SM cross-sections, computed at next-to-next-to-leading order (NNLO) in QCD including finite top-quark mass effects for ggF, and at next-to-next-to-next-to-leading order (N$^3$LO) in QCD and next-to-leading order (NLO) in the electroweak coupling for VBF, are $\sigma_{\textup{ggF}}^{\textup{SM}} = 30.8^{+2.0}_{-7.1}$~fb and $\sigma_{\textup{VBF}}^{\textup{SM}} = 1.69 \pm 0.05$~fb at $\sqrt{s} = 13$~\TeV, and $\sigma_{\textup{ggF}}^{\textup{SM}} = 34.1^{+2.2}_{-7.9}$~fb and $\sigma_{\textup{VBF}}^{\textup{SM}} = 1.87 \pm 0.05$~fb at $\sqrt{s} = 13.6$~\TeV~\cite{Dawson:1998py,Borowka:2016ehy,Baglio:2018lrj,deFlorian:2013jea,Shao:2013bz,deFlorian:2015moa,Grazzini:2018bsd,Baglio:2020wgt,Baglio:2012np,Frederix:2014hta,Ling:2014sne,Dreyer:2018rfu,Dreyer:2018qbw,fixedPow:2023}. Deviations of the Higgs boson couplings from their SM values modify both the inclusive and differential \HH production cross-sections. Such deviations are parameterised by the coupling-strength modifiers $\kl = \lhhh/\lhhh^{\textup{SM}}$ for the trilinear self-coupling, $\kv = g_{HVV}/g_{HVV}^{\textup{SM}}$ for the coupling between a single Higgs boson and two vector bosons, and $\kvv = g_{HHVV}/g_{HHVV}^{\textup{SM}}$ for the quartic $HHVV$ coupling ($V = W, Z$), each defined to be unity in the SM. When probing for values of these couplings beyond the SM (BSM), the corresponding modifications of the \HH production cross-sections and kinematics are taken into account~\cite{PhysRevD.101.075023}.

Searches for Higgs boson pair production have been performed extensively at the LHC by both the ATLAS and CMS Collaborations in various final states, including \bbbb~\cite{CMS-HIG-20-005,HDBS-2019-29,HDBS-2022-02,CMS-HIG-24-010}, \bbyy~\cite{CMS-HIG-19-018,HDBS-2021-10,HIGP-2025-10}, \bbtt~\cite{HDBS-2018-40,CMS-HIG-20-010,HDBS-2019-27}, multilepton~\cite{HDBS-2019-04,CMS-HIG-21-002}, as well as other final states with lower sensitivity~\cite{HDBS-2019-02,CMS-HIG-21-005}. The most stringent constraints are obtained through the statistical combination of the most recent analyses~\cite{HDBS-2021-18,CMS-HIG-20-011} exploiting the full Run~2 datasets at $\sqrt{s} = 13$~\TeV collected by the two experiments~\cite{HIGP-2025-18}. The observed (expected) upper limit at 95\% confidence level (CL) on the \HH production signal strength $\mu_{HH} = (\sigma_{\textup{ggF}} + \sigma_{\textup{VBF}})/(\sigma_{\textup{ggF}}^{\textup{SM}} + \sigma_{\textup{VBF}}^{\textup{SM}})$ is 2.5~(1.7) times the SM prediction. The observed (expected) 95\% CL constraints on \kl and \kvv are $\kl \in [-0.7, 6.1]$~($[-1.3, 6.7]$) and $\kvv \in [0.7, 1.3]$~($[0.7, 1.4]$), respectively, with all other Higgs boson couplings fixed to their SM values.

The \bbtt final state is one of the most sensitive channels for \HH production, along with \bbbb and \bbyy. It accounts for 7.3\% of all $HH$ decays and provides a good compromise between expected signal yield and purity. The latest search for $HH\to\bbtt$ by the ATLAS experiment with the full Run~2 dataset~\cite{HDBS-2019-27}  yields an observed upper limit on the HH production cross-section of 5.9 times the SM prediction at 95\% CL,  while the expected limit at 95\% CL under the background-only hypothesis is 3.3 times the SM prediction. The observed (expected) 95\% CL \kl and \kvv intervals are $\kl \in [-3.1, 9.0]$~($[-2.5, 9.3]$) and $\kvv \in [-0.5, 2.7]$~($[-0.2, 2.4]$). The most recent CMS result in this final state~\cite{CMS-HIG-20-010} sets an observed (expected) upper limit on the $HH$ production cross-section of 3.3~(5.2) times the SM prediction at 95\% CL, with observed (expected) 95\% CL constraints on the coupling-strength modifiers $\kl \in [-1.7, 8.7]$~($[-2.9, 9.8]$) and $\kvv \in [-0.4, 2.6]$~($[-0.6, 2.8]$).

In this paper, an updated search for non-resonant Higgs boson pair production in the \bbtt final state is presented, combining the full Run~2 ATLAS dataset of $140$~fb$^{-1}$ of $\sqrt{s} = 13$~\TeV $pp$ collisions collected during 2015--2018 with the partial Run~3 dataset of $56$~fb$^{-1}$ of $\sqrt{s} = 13.6$~\TeV $pp$ collisions collected during 2022--2023. Relative to the previous ATLAS result~\cite{HDBS-2019-27}, the analysis incorporates several methodological improvements in addition to the extended dataset: a new multivariate analysis (MVA) strategy based on transformer architectures for event categorisation and signal--background discrimination, updated object definitions including the state-of-the-art GN2 transformer-based flavour-tagging algorithm~\cite{FTAG-2023-05}, and an extended trigger strategy in Run~3 to increase the signal acceptance.

Additionally, the full analysis chain is validated through measurements of the $ZZ$ and $ZH$ production cross-sections in the same final state, \ZZbbtt and \ZHbbtt,\footnote{The $ZH$ process contributes through both $Z\to\bbbar,\,H\to\tau^{+}\tau^{-}$ and $H\to\bbbar,\,Z\to\tau^{+}\tau^{-}$ decay configurations.} using the same event selection and analysis strategy as the $HH$ search. These processes share the $\bbtt$ final state and a topology similar to that of $HH$ production, making them a natural benchmark for the analysis. As they are reconstructed from the same dataset, the resulting measurements are not statistically independent of the $HH$ search. Recently, the CMS Collaboration also studied $ZZ$ and $ZH$ production in the \bbtt final state, reporting no evidence for either process and setting upper limits on their production cross-sections~\cite{CMS-B2G-25-004}.

This paper is organised as follows. Section~\ref{sec:detector} provides an overview of the ATLAS detector. The collected dataset and the simulated event samples are presented in Section~\ref{sec:samples}. The object selection and event preselection are detailed in Sections~\ref{sec:object} and~\ref{sec:event}, respectively. The background estimation and MVA strategy are described in Sections~\ref{sec:background} and~\ref{sec:mva}, respectively. Treatment of systematic uncertainties and the statistical interpretation are discussed in Sections~\ref{sec:sys} and~\ref{sec:fit}, respectively. Finally, Section~\ref{sec:result} summarises the results, and Section~\ref{sec:conclusion} concludes the paper.
%


\section{ATLAS detector}

\label{sec:detector}

\newcommand{\AtlasCoordFootnote}{%
ATLAS uses a right-handed coordinate system with its origin at the nominal interaction point (IP) in the centre of the detector and the \(z\)-axis along the beam pipe. The \(x\)-axis points from the IP to the centre of the LHC ring, and the \(y\)-axis points upwards. Polar coordinates \((r,\phi)\) are used in the transverse plane, \(\phi\) being the azimuthal angle around the \(z\)-axis. The pseudorapidity is defined in terms of the polar angle \(\theta\) as \(\eta = -\ln \tan(\theta/2)\) and is equal to the rapidity $ y = \frac{1}{2} \ln \left( \frac{E + p_z}{E - p_z} \right) $ in the relativistic limit. Angular distance is measured in units of \(\Delta R \equiv \sqrt{(\Delta y)^{2} + (\Delta\phi)^{2}}\).}

The ATLAS experiment~\cite{PERF-2007-01,GENR-2019-02} at the LHC is a multipurpose particle detector designed to study a wide range of physics processes.
It has a forward--backward symmetric cylindrical geometry and a nearly \(4\pi\) coverage in solid angle.\footnote{\AtlasCoordFootnote}
The detector subsystems consist of an inner tracking detector surrounded by a superconducting solenoid providing a \qty{2}{\tesla} axial magnetic field, electromagnetic and hadronic calorimeters, and a muon spectrometer with three large superconducting air-core toroidal magnets.

The inner tracking detector covers the pseudorapidity range \(|\eta| < 2.5\) and consists of silicon pixel, silicon microstrip, and transition radiation tracking detectors.
The electromagnetic calorimeters provide coverage up to \(|\eta| = 4.9\) and consist of lead/liquid-argon (LAr) sampling layers.
Hadronic calorimetry in the central region (\(|\eta| < 1.7\)) is provided by a
steel/scintillator-tile calorimeter, while the endcap and forward regions,
extending the coverage up to \(|\eta| = 4.9\), use liquid-argon calorimeters
with copper and tungsten absorbers.
The integrated bending power of the toroidal magnets ranges from \num{2.0} to \qty{6.0}{\tesla\metre} across most of the detector.
The muon spectrometer includes precision tracking chambers and fast detectors for triggering.

Events are selected using a two-level trigger system~\cite{TRIG-2016-01,TRIG-2022-01}, with the first level implemented in hardware and the second level implemented in software.
The hardware-based first-level trigger (L1) uses a subset of the detector information to reduce the event rate to below \qty{100}{\kHz}, while the software-based high-level trigger (HLT) further reduces the accepted event rate to an average of \qty{1.25}{\kHz} in Run~2 and \qty{3}{\kHz} in Run~3. Events accepted by the HLT are recorded in the main stream, which is reconstructed promptly. In Run~3, an additional delayed stream is recorded~\cite{TRIG-2022-01}, in which events are stored and reconstructed later to collect events from lower-threshold triggers that would otherwise exceed the available prompt-reconstruction bandwidth.
An extensive software suite~\cite{SOFT-2022-02} is used for simulation, reconstruction, and analysis of real and simulated data, detector operations, and the trigger and data acquisition systems of the experiment.

%


\section{Data and simulated event samples}

\label{sec:samples}

The analysis uses $pp$ collision data recorded by the ATLAS detector during Run~2 (2015--2018) at $\sqrt{s} = 13$~\TeV and during the early Run~3 period (2022--2023) at $\sqrt{s} = 13.6$~\TeV, corresponding to integrated luminosities of $140$~fb$^{-1}$ and $56$~fb$^{-1}$, respectively. Only data recorded under stable beam conditions and with all relevant detector subsystems fully operational are analysed~\cite{DAPR-2018-01}. Multiple inelastic $pp$ interactions occur per bunch crossing (pile-up), with a mean of approximately 34 interactions per crossing in Run~2~\cite{DAPR-2021-01} and 52 (62) in the 2022 (2023) data-taking period of Run~3~\cite{GENR-2019-02}.

The signal and background processes are modelled using Monte Carlo (MC) simulated event samples. The interaction of the generated particles with the ATLAS detector is modelled using a full simulation~\cite{SOFT-2010-01} based on \GEANT~\cite{Agostinelli:2002hh} except for signal samples for which a faster simulation relying on a parameterisation of the calorimeter response is used~\cite{SIMU-2018-04}. Pile-up is simulated by overlaying minimum-bias events generated with \Pythia[8] and matched to the distribution observed in data~\cite{SIMU-2020-01}. The simulated events are reconstructed and analysed using the same software as data. Corrections are applied to the simulated events to account for differences between data and simulation in the trigger, reconstruction, identification, and isolation efficiencies, as well as in the energy scale and resolution of jets, electrons, muons, and hadronically decaying $\tau$-leptons (\tauhad). Table~\ref{tab:samples} summarises the generators used to simulate the signal and background processes, along with the order of the perturbative cross-section calculations, parton distribution function (PDF) sets, parton shower (PS) and hadronisation models, and underlying event (UE) tunes.

The signal samples for non-resonant \HH production are generated separately for the ggF and VBF production modes, with one Higgs boson decaying into \bb and the other into $\tau^+\tau^-$. The ggF samples are generated at NLO in QCD using the \POWHEGBOX[v2]~\cite{Alioli:2010xd} generator, with a correction applied to correct a known bug in the generator for non-SM values of \kl~\cite{fixedPow:2023}, with the \PDFforLHC[21]~\cite{Ball:2022oua} PDF set, and interfaced to \Pythia[8]~\cite{Bierlich:2022pfr} for parton showering and hadronisation using the \NNPDF[2.3lo] PDF set~\cite{Ball:2012cx} and the A14 set of UE tune~\cite{ATL-PHYS-PUB-2014-021} for both Run~2 and Run~3. The ggF signal samples are normalised to the
NNLO QCD cross-section calculated in the full top-quark mass ($m_t$) scheme using the \textsc{FTapprox} method~\cite{Grazzini:2018bsd}, which includes finite $m_t$ effects up to NLO and soft-gluon resummation at next-to-next-to-leading-logarithmic (NNLL) accuracy. The VBF signal samples are generated at leading order (LO) in QCD using \MGNLO[3.3.1]~\cite{Alwall:2014hca} in the five-flavour scheme (5FS), with the \NNPDF[3.0nlo]~\cite{Ball:2014uwa} PDF set, and interfaced to \Pythia[8.308] using the \NNPDF[2.3lo] PDF set and A14 tune. For both production modes, samples with modified coupling values are also simulated. Three ggF signal samples with $\kl = 0$, $1$, and $5$ are generated. For the VBF mode, five additional \HH samples are generated with the same setup as the nominal sample but with $(\kl, \kvv, \kv) = (1, 1, 0.5)$, $(-5, 1, 0.5)$, $(10, 1, 1)$, $(2, 1, 1)$, and $(1, 0.75, 0.5)$. Together with the SM sample, these form a basis for a linear combination, following the procedure of Ref.~\cite{HDBS-2019-27}, which is used to determine the expected ggF and VBF \HH yields and their kinematic distributions for alternative coupling-modifier values.

All background processes are estimated from simulation, except the background from misidentified hadronic $\tau$-lepton candidates, which is sometimes estimated using data-driven methods as described in Section~\ref{sec:background}. Top-quark pair production (\ttbar) and single top-quark production~\cite{Frixione:2007nw} in the $t$-channel~\cite{Alioli:2009je}, $s$-channel~\cite{Alioli:2009je}, and $tW$~\cite{Re:2010bp} production modes are simulated separately using the \POWHEGBOX[v2] generator with the \NNPDF[3.0nlo] PDF set, and interfaced to \Pythia[8]~\cite{Bierlich:2022pfr,Sjostrand:2014zea} for parton showering and hadronisation using the A14 tune and the \NNPDF[2.3lo] PDF set. The interference between $tW$ and \ttbar production is modelled using the nominal diagram-removal scheme~\cite{Frixione:2008yi}.

The production of a $W$ boson in association with jets ($W$+jets) is simulated using \Sherpa[2.2.11]~\cite{Bothmann:2019yzt} (\Sherpa[2.2.14]) for Run~2 (Run~3), while $Z$+jets production is simulated using \Sherpa[2.2.14] for both Run~2 and Run~3. Both samples are generated using the MEPS@NLO prescription~\cite{Hoeche:2012yf} in the 5FS with the \NNPDF[3.0nnlo] PDF set, using NLO matrix elements for up to two partons and LO matrix elements for up to five partons, calculated using the OpenLoops~\cite{Buccioni:2019sur,Cascioli:2011va,Buccioni:2017yxi,Denner:2016kdg} and Comix~\cite{Gleisberg:2008fv} matrix-element generators. Diboson events ($WW$, $WZ$, and $ZZ$) are simulated with \Sherpa[2.2.16] for $VV\rightarrow\ell\ell\ell\ell$ decay modes and \Sherpa[2.2.14] for other decay modes, for both Run~2 and Run~3. The $\ttbar W$ process is simulated with \Sherpa[2.2.11] (\Sherpa[2.2.14]) for Run~2 (Run~3). All \Sherpa samples described above use the \NNPDF[3.0nnlo] PDF set with a dedicated parton shower tuning developed by the \Sherpa authors. The $\ttbar Z$ process is simulated with the \MGNLO generator, version 2.8.1 for Run~2 and version 3.5.1 for Run~3, interfaced with \Pythia[8] using the \NNPDF[3.0nnlo] PDF set and A14 tune.

Single-Higgs-boson production is simulated with the \POWHEGBOX[v2] generator interfaced to \Pythia[8] following the same approach as in Ref~\cite{HDBS-2019-27}. The ggF~\cite{Hamilton:2013fea,Hamilton:2015nsa} and VBF~\cite{Nason:2009ai} processes use the \PDFforLHC[15nnlo] and \PDFforLHC[15nlo]~\cite{Butterworth:2015oua} PDF sets, respectively, while the $VH$~\cite{Luisoni:2013cuh} processes use the \NNPDF[3.0nlo] PDF set. For these processes, the AZNLO tune~\cite{STDM-2012-23} is used in Run~2 and the A14 tune for Run~3. The $\ttbar H$~\cite{Hartanto:2015uka} process uses the \NNPDF[3.0nlo] (\PDFforLHC[21] in Run~3) PDF set and the A14 tune for both run periods.

\begin{table}[tbhp!]
\captionsetup{skip=2pt}
\caption{Generators used to simulate the signal and background processes, with Run~2 settings listed as the baseline. Differences between Run~2 and Run~3 samples are given in brackets where applicable. If not otherwise specified, the order of the cross-section calculation refers to the expansion in the strong coupling constant ($\alpha_S$). The acronyms ME, PS, and UE refer to matrix element, parton shower, and underlying event, respectively. More details on the simulation of signal and background samples are given in the text and in Ref.~\cite{HDBS-2018-40}. ($\dagger$) The NNLO(QCD)+NLO(EW) cross-section calculation for the $pp\rightarrow ZH$ process already includes the $gg\rightarrow ZH$ contribution. The $qq\rightarrow ZH$ process is normalised to the NNLO(QCD)+NLO(EW) cross-section for the $pp\rightarrow ZH$ process after subtracting the $gg\rightarrow ZH$ contribution.}
\protect\label{tab:samples}
\centering{\small
\resizebox{\textwidth}{!}{\begin{tabular}{lllllll}
\toprule
\hspace*{0.3cm} Process & ME generator & \shortstack{ME QCD\\order} & PDF set & \shortstack{PS and\\hadronisation} & \shortstack{UE model\\tune} & \shortstack{Cross-section\\order} \\
\midrule
\multicolumn{5}{l}{\BF{Signal}} \\
\hspace*{0.3cm} $gg\rightarrow HH$ (ggF)   &\POWHEGBOX[v2]~\cite{Alioli:2010xd} & NLO & \PDFforLHC[21]~\cite{Ball:2022oua} & \Pythia[8.309]~\cite{Bierlich:2022pfr} & A14~\cite{ATL-PHYS-PUB-2014-021} \NNPDF[2.3lo]~\cite{Ball:2012cx} & NNLO FTApprox \\
\hspace*{0.3cm} $qq\rightarrow qqHH$ (VBF)   & \MGNLO[3.3.1]~\cite{Alwall:2014hca} & LO & \NNPDF[3.0nlo]~\cite{Ball:2014uwa} & \Pythia[8.308] & A14 \NNPDF[2.3lo] & N$^3$LO(QCD)+NLO(EW)\\ %
\midrule
\multicolumn{5}{l}{\BF{Top-quark}} \\
\hspace*{0.3cm} \ttbar      & \POWHEGBOX[v2] & NLO & \NNPDF[3.0nlo] & \Pythia[8.230]~\cite{Sjostrand:2014zea} [8.308] & A14 \NNPDF[2.3lo] & NNLO+NNLL\\
\hspace*{0.3cm} Single top ($t$-chan.) & \POWHEGBOX[v2] & NLO & \NNPDF[3.0nlo] & \Pythia[8.230][8.308] & A14 \NNPDF[2.3lo] & NLO \\
\hspace*{0.3cm} Single top ($s$-chan.) & \POWHEGBOX[v2] & NLO & \NNPDF[3.0nlo] & \Pythia[8.230][8.308] & A14 \NNPDF[2.3lo] & NLO \\
\hspace*{0.3cm} Single top ($tW$)        & \POWHEGBOX[v2] & NLO & \NNPDF[3.0nlo] & \Pythia[8.230][8.308] & A14 \NNPDF[2.3lo] & NLO \\
\hspace*{0.3cm} $\ttbar Z$  & \MGNLO[2.8.1][3.5.1] & NLO & \NNPDF[3.0nnlo] & \Pythia[8.244][8.309] & A14 \NNPDF[2.3lo] & NLO \\ %
\hspace*{0.3cm} $\ttbar W$  & \Sherpa[2.2.11][\Sherpa[2.2.14]]~\cite{Bothmann:2019yzt} & NLO & \NNPDF[3.0nnlo] & \Sherpa[2.2.14] & - & NLO \\
\midrule
\multicolumn{5}{l}{\BF{Vector boson + jets}} \\
\hspace*{0.3cm} $W+$jets & \Sherpa[2.2.11][\Sherpa[2.2.14]] & \shortstack{NLO($\leq 2$ jets)\\ LO($3,4,5$ jets)} & \NNPDF[3.0nnlo] &  \Sherpa[2.2.11][\Sherpa[2.2.14]] & - & NNLO \\
\hspace*{0.3cm} $Z+$jets & \Sherpa[2.2.14] & \shortstack{NLO($\leq 2$ jets)\\ LO($3,4,5$ jets)} & \NNPDF[3.0nnlo] & \Sherpa[2.2.14] & - & NNLO \\
\midrule
\multicolumn{5}{l}{\BF{Diboson}} \\
\hspace*{0.3cm}$WW,WZ,ZZ$ & \Sherpa[2.2.14], \Sherpa[2.2.16] & \shortstack{NLO($\leq 1$ jet) \\ LO($2,3$ jets)} & \NNPDF[3.0nnlo] &  \Sherpa[2.2.14], \Sherpa[2.2.16] & - & NLO \\
\midrule
\multicolumn{5}{l}{\BF{Single Higgs boson}} \\
\hspace*{0.3cm} ggF         &\POWHEGBOX[v2] & NNLO & \PDFforLHC[15nnlo]~\cite{Butterworth:2015oua} &  \Pythia[8.212][8.310] & AZNLO~\cite{STDM-2012-23} [A14 \NNPDF[2.3lo]] & N$^3$LO(QCD)+NLO(EW) \\
\hspace*{0.3cm} VBF         &\POWHEGBOX[v2] & NLO & \PDFforLHC[15nlo] &  \Pythia[8.230][8.310] & AZNLO [A14 \NNPDF[2.3lo]] & NNLO(QCD)+NLO(EW) \\
\hspace*{0.3cm} $qq\rightarrow WH$  &\POWHEGBOX[v2] & NLO & \NNPDF[3.0nlo] &  \Pythia[8.212],8.245 [8.309] & AZNLO [A14 \NNPDF[2.3lo]] & NNLO(QCD)+NLO(EW) \\
\hspace*{0.3cm} $qq\rightarrow ZH$  &\POWHEGBOX[v2] & NLO & \NNPDF[3.0nlo] &  \Pythia[8.244],8.245 [8.309] & AZNLO [A14 \NNPDF[2.3lo]] & NNLO(QCD)+NLO(EW)$^{\dagger}$ \\
\hspace*{0.3cm} $gg\rightarrow ZH$  &\POWHEGBOX[v2] & NLO & \NNPDF[3.0nlo] &  \Pythia[8.307][8.309] & AZNLO [A14 \NNPDF[2.3lo]] & NLO+NLL \\
\hspace*{0.3cm} $\ttbar H$ &\POWHEGBOX[v2] & NLO & \NNPDF[3.0nlo][\PDFforLHC[21]] & \Pythia[8.230][8.308] & A14 \NNPDF[2.3lo] & NLO(QCD)+NLO(EW) \\
\bottomrule
\end{tabular}}}
\end{table}

%
%
%
%
%

%


\section{Object selection}

\label{sec:object}
The reconstruction of the \HH$\rightarrow$\bbtt signature relies on multiple physics objects. Higgs bosons decaying into $b$-quarks are reconstructed from pairs of $b$-tagged jets. The $\tau$-lepton decays are classified as either hadronic (\tauhad), reconstructed from the visible decay products (\tauhadvis), or leptonic (\taulep), reconstructed as electrons or muons. Two analysis channels are defined based on the decay modes of the two $\tau$-leptons used to reconstruct the $H\rightarrow \tau^+\tau^-$ candidate: the \lephad channel, where one $\tau$-lepton decays leptonically and the other hadronically, and the \hadhad channel, where both $\tau$-leptons decay hadronically. To reconstruct these final states, electrons, muons, jets, \tauhad candidates, and missing transverse momentum are identified as described in this section. Tracks measured in the inner detector are used to reconstruct interaction vertices~\cite{IDTR-2021-01}, of which the one with the highest sum of squared transverse momenta of associated tracks is selected as the primary vertex of the hard interaction.

Electrons are reconstructed from energy deposits in the electromagnetic calorimeter matched to tracks in the inner detector and are calibrated according to the procedures described in Refs.~\cite{EGAM-2018-01,EGAM-2021-02}. Candidates are required to have $\pt > \SI{7}{\GeV}$ and $|\eta| < 2.47$, excluding the transition region between the barrel and endcap LAr calorimeters ($1.37 < |\eta| < 1.52$). Signal electron candidates are further required to satisfy the \texttt{TightLH} likelihood-based identification criteria~\cite{EGAM-2021-01} and the \texttt{Loose\_VarRad} isolation criteria~\cite{EGAM-2021-01}, which combine calorimeter- and track-based isolation variables. Electrons are associated with the primary vertex through a requirement on the longitudinal impact parameter: $|z_0 \sin\theta| < \SI{0.5}{mm}$. No selection on the transverse impact parameter ($d_{0}$) is applied to maximise the acceptance to leptonically decaying $\tau$-leptons.

Muons within $|\eta| < 2.5$ are reconstructed by combining tracks in the inner detector and muon spectrometer through a combined track fit, accounting for energy loss in the calorimeters. In the region beyond the inner-detector acceptance, they are reconstructed from the muon spectrometer alone~\cite{MUON-2018-03}. Muon candidates are required to have $\pt > \SI{7}{\GeV}$ and $|\eta| < 2.7$, and must satisfy the \texttt{Loose} identification working point for the baseline selection. As for electrons, selected muons are also required to satisfy $|z_0 \sin\theta| < \SI{0.5}{mm}$. Signal muon candidates are further required to satisfy the \texttt{Medium} identification working point and the \texttt{PflowLoose\_VarRad} isolation criteria~\cite{MUON-2018-03}, which combine track-based and particle-flow-based calorimeter isolation variables.

Jets are reconstructed from particle-flow objects~\cite{PERF-2015-09}, formed from energy deposits in topological clusters of calorimeter cells and tracks in the inner detector, using the anti-$k_t$ algorithm~\cite{Cacciari:2008gp,Fastjet} with a radius parameter of $R = 0.4$. Jets are calibrated using the procedures described in Ref.~\cite{PERF-2016-04}, including corrections to the jet energy scale and resolution. Central jets in the $|\eta| < 2.5$ region are required to have $\pt > \SI{20}{\GeV}$, and those with $20 < \pt < \SI{50}{\GeV}$ must additionally satisfy the \texttt{Tight} jet-vertex-tagging (JVT) working point~\cite{PERF-2014-03} to suppress pile-up contributions. Forward jets ($2.5 \leq |\eta| \leq 4.5$) are required to have $\pt > \SI{30}{\GeV}$ and to satisfy the \texttt{Tight} forward JVT working point~\cite{ATL-PHYS-PUB-2015-034,ATL-PHYS-PUB-2019-026}.

Simulated jets are labelled as $b$-jets by matching them to a $b$-hadron with $\pt \geq 5$~\GeV\ within a cone of radius $\Delta R = 0.3$ around the jet axis. The flavour of reconstructed jets is determined by the ATLAS GN2 flavour tagging algorithm~\cite{FTAG-2023-05}, which exploits charged-track information to separate $b$-jets, $c$-jets, and light-flavour jets, and is applied identically to data and simulation. Only central jets are considered for \btagging. Jets satisfying the 85\% \btagging efficiency working point, corresponding to a $c$-jet (light-flavour jet) misidentification rate of approximately 20\% (1\%), are classified as $b$-tagged jets and used to identify $b$-jet candidates from the $H\rightarrow \bb$ decay. These loose requirements are chosen to maximise the signal efficiency and correspond to the optimal
analysis sensitivity. In addition, to provide a richer representation of $b$-tagging information for the analysis MVA, jets are assigned to discrete categories based on the tightest efficiency working point they satisfy. This classification uses quantile bins corresponding to efficiency ranges of 100--90\%, 90--85\%, 85--77\%, 77--70\%, 70--65\%, and below 65\%. The tagging efficiencies and misidentification rates are calibrated in \ttbar events for $b$- and $c$-jets, and in $Z$+jets events for light-quark and gluon jets~\cite{FTAG-2018-01, FTAG-2019-02, FTAG-2020-08, FTAG-2023-10, FTAG-2024-07}.

Approximately 15\% of $b$-jets from $b$-hadron decays contain a muon from the semileptonic decay of the $b$-hadron. Dedicated muon-in-jet and \pt-reco corrections are applied to $b$-tagged jets to improve the jet energy resolution. The muon-in-jet correction adds the four-momentum of a non-isolated \texttt{Medium} muon found within the jet cone to the jet four-momentum~\cite{HIGG-2020-20}. The \pt-reco correction accounts for the missing energy from neutrinos in semileptonic $b$-hadron decays and out-of-cone effects where particles from the $b$-hadron decay are not clustered into the jet; it is derived from simulated \ttbar\ events~\cite{HIGG-2020-20} in bins of truth $b$-jet \pt and applied to all $b$-tagged jets. Together, these corrections yield a 30--36\% improvement in the di-$b$-jet invariant mass (\mbb) resolution.

The reconstruction of the visible components of hadronically decaying $\tau$-leptons is seeded from jets reconstructed using topological clusters of calorimeter deposits, as described in Ref.~\cite{ATL-PHYS-PUB-2022-044}. A recurrent neural network (RNN) classifies the tracks associated with each \tauhad candidate into four categories; $\tau$-tracks, photon-conversion tracks, tracks from jet fragmentation, and fake tracks, with the number of $\tau$-tracks defining the candidate's track multiplicity~\cite{ATL-PHYS-PUB-2022-044}. The \tauhad candidate energy is calibrated using a weighted average of the jet energy and \tauhad substructure information, with a boosted decision tree providing the final energy scale, and is subsequently corrected in situ using $Z\to\tau^{+}\tau^{-}$ events to match the energy response in data. The \tauhad candidates are required to have $\pt > \SI{20}{\GeV}$ and $|\eta| < 2.5$, excluding the transition region $1.37 < |\eta| < 1.52$, have one or three associated tracks, and carry unit absolute charge. To reject misidentified jets, signal \tauhad candidates must further satisfy the \texttt{Loose} RNN-based identification working point~\cite{ATL-PHYS-PUB-2019-033}, corresponding to identification efficiencies of 85\% and 75\% for 1-track and 3-track \tauhad candidates, respectively. In the \lephad channel, \tauhad candidates are additionally required to pass the \texttt{Loose} working point of the RNN-based electron veto (eVeto)~\cite{ATL-PHYS-PUB-2022-044}, corresponding to an efficiency of 95\% for 1-track candidates, to suppress electrons misidentified as 1-track \tauhad candidates. This eVeto requirement is applied only in the analysis regions where it provides a net improvement in sensitivity, to limit signal loss. Anti-\tauhad candidates are defined to estimate backgrounds with fake-\tauhad contributions, as described in Section~\ref{sec:background}; they satisfy the same reconstruction and selection criteria as signal \tauhad candidates but fail the \texttt{Loose} RNN identification requirement. At most one anti-\tauhad candidate is selected per event in regions where this background estimation is performed.

The missing transverse momentum, \metV, is reconstructed in each event as the negative vector sum of the transverse momenta of electrons, muons, \tauhadvis and jets, with an additional soft term from tracks originating from the primary vertex but not associated with any reconstructed physics object~\cite{JETM-2020-03}. The magnitude of \metV~is referred to as \met and it accounts for energies of neutrinos from $\tau$-lepton decays, semileptonic $b$-hadron decays, and other undetected particles. Its reconstruction and performance are described in Ref.~\cite{JETM-2020-03}.

The reconstruction of the di-$\tau$ invariant mass is complicated by the presence of multiple neutrinos that escape ATLAS undetected. The analysis employs the Missing Mass Calculator (MMC)~\cite{Elagin:2010aw}, a likelihood-based technique that uses the visible $\tau$-lepton decay products and \metV to constrain the %
system of neutrino four-momenta, yielding an estimate of the di-$\tau$ invariant mass \mMMC. The probability density functions used as input to the MMC are obtained from simulated $Z\to\tau^+\tau^-$ events separately for the different di-$\tau$ decay modes. %
In simulated \HH events, this technique improves the reconstructed Higgs boson mass resolution by $53$\%--$57$\% with respect to using only the visible decay products, depending on the di-$\tau$ decay mode.

To avoid double-counting of reconstructed physics objects, an overlap removal procedure is applied sequentially to electrons, muons, \tauhad
candidates, and jets, based on their angular separation.\footnote{When anti-\tauhad candidates are used for background estimation, they are given priority over jets except for $b$-tagged jets.}
The \tauhad candidates within $\Delta R < 0.2$ of an electron or muon are rejected. Jets within $\Delta R < 0.2$ of an electron are removed, afterwards electrons and muons within $\Delta R < 0.4$ of a remaining jet are rejected. Finally, jets with a small number of associated tracks that are ghost-associated~\cite{Cacciari:2008gn} with a muon are removed, to suppress jets arising from  muon energy depositions.

%


\section{Event reconstruction}

\label{sec:event}
\begin{figure}[t!]
\centering
\includegraphics[width=0.99\textwidth]{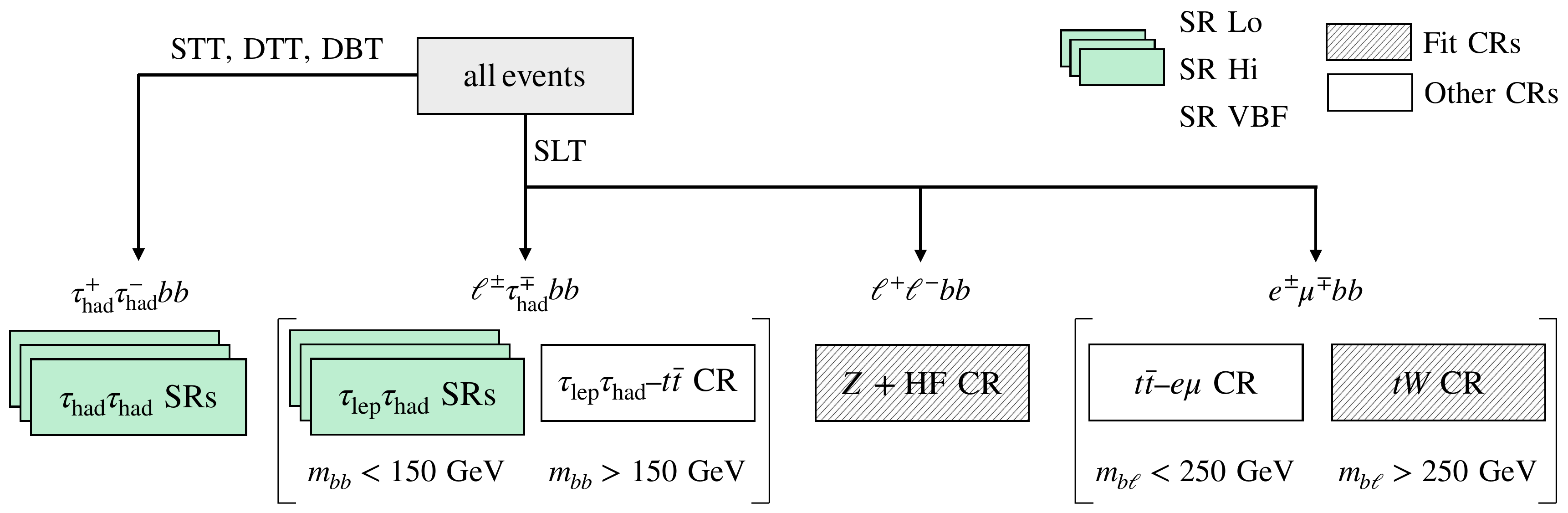}
\caption{Schematic view of the signal and control regions used in the analysis. In the \hadhad channel, events are selected using single-\tauhad (STT), di-\tauhad (DTT), and di-$b$-jet triggers (DBT). In all other regions, single-lepton triggers (SLT) are used. Regions are grouped according to the targeted final state. In addition to the signal regions (shown as filled boxes), the control regions indicated by the hatched boxes are included in the final statistical fit, while the remaining control regions are used for background estimation.}
\label{fig:regions}
\end{figure}

Events are classified into different channels depending on the number of signal electrons, muons, and $\tauhad$ candidates present, with a dedicated trigger strategy used in each channel to target the corresponding final-state object multiplicity. Within each channel, events are further categorised into signal regions (SRs), which provide the main phase space for signal extraction, and control regions (CRs), which are used to constrain the leading background predictions and support the estimation of specific background contributions. The regions that enter the final statistical interpretation are indicated in Figure~\ref{fig:regions}; detailed definitions are provided in the following subsections.

Unless otherwise specified, all regions require a leading central jet with $\pt > 45$~\GeV, exactly two $b$-tagged jets, and an opposite-sign charge requirement between the two selected lepton candidates.
The selected events in the \hadhad and \lephad signal regions are further categorised into three mutually exclusive regions, \SRHi, \SRLo, and \SRVBF, based on the reconstructed invariant mass of the \HH system, \mhh, and on the output scores of the neural network classifiers used to separate \HH ggF, \HH VBF, and background events, as described in Section~\ref{sec:mva}.

\subsection{\hadhad channel}
\label{subsec:SRhadhad}

Events in the \hadhad channel are required to have exactly two \tauhad candidates of opposite electric charge and no additional electrons or muons. The visible invariant mass of the di-$\tau$ system, \mttvis, must exceed 40~\GeV, and \mMMC\ is required to be greater than 60~\GeV\ to reject low-mass Drell--Yan background. The invariant mass of the two $b$-tagged jets must satisfy $40 < \mbb < 210$~\GeV.

Several trigger selections are used to collect events in this channel, each involving a different set of physics objects and multiplicities. For each trigger, offline \pt thresholds are chosen to select the phase space where the trigger is highly efficient. Offline objects are required to be geometrically matched to the corresponding trigger objects that fired the event. A trigger bucketing strategy is employed, in which each trigger category defines a separate offline selection region and events are assigned exclusively to one bucket. Exclusive offline selection regions are applied due to the challenges of calibrating multi-object trigger decisions conditional on an event passing one trigger while failing another in a given phase space. The buckets are evaluated in an order dictated by both selection complexity and analysis sensitivity, as detailed in the sequence below.

Single-\tauhad triggers (STT) require one HLT \tauhad candidate, with the \pt threshold raised progressively from 80~\GeV\ to 160~\GeV\ during 2016 and kept constant at 160~\GeV\ for the remainder of Run~2 and Run~3 data-taking. Events are assigned to the STT bucket if they pass the corresponding STT and contain a reconstructed \tauhad candidate with offline \pt at least 20~\GeV\ above the corresponding trigger threshold.

Di-\tauhad triggers (DTT) require a pair of \tauhad candidates at the HLT with \pt thresholds of 35~\GeV\ and 25~\GeV, and constitute the most sensitive trigger category for the \hadhad channel. From the 2016 data-taking period onward, additional L1 requirements were applied to reduce the DTT rate~\cite{HDBS-2018-40}. In 2016, these consisted of requiring one additional jet. From 2017 onward, the requirements were either a single additional jet together with an angular separation requirement on the two \tauhad candidates, or two additional central jets. Events are considered for the DTT bucket if they pass at least one of the DTTs and fail the offline selection criteria of the STT bucket. The offline selection criteria of the DTT bucket include requiring two reconstructed \tauhad candidates with offline \pt thresholds of 40~\GeV\ and 30~\GeV\ for the leading and subleading \tauhad candidates, respectively. A single jet with $\pt > 70$~\GeV\ or two jets with $\pt > 50$~\GeV\ is additionally required. In 2023, starting with the data-taking period based on 2400 proton bunches, DTT chains with HLT \tauhad \pt thresholds lowered by 5~\GeV\ were added to the ATLAS delayed trigger stream~\cite{TRIG-2022-01}, recovering events with offline \pt above 35~(25)~\GeV\ for the leading (subleading) \tauhad candidate.

Di-$b$-jet triggers (DBT) are used only in Run~3 and are recorded in the delayed stream. These triggers require four central jets at the HLT, two of which are $b$-tagged~\cite{TRIG-2022-01}. Events are assigned to the DBT bucket if they pass the DBT and fail the offline selections of both the STT and DTT buckets. The offline DBT selection requires that the \pt-ordered collection of the two \tauhad candidates and all central jets satisfies \pt thresholds of 85, 60, 50, and 25~\GeV, with the leading object required to be within $|\eta| < 2.1$, consistent with the L1 seed requirement.\footnote{HLT jets in DBT can include $\tau$-lepton signatures.}

All events satisfying the above selection enter the \hadhad SR. The overall acceptance times efficiency for SM $HH\to b\bar{b}\tauhad\tauhad$ events is approximately 5.3\% for the Run~2 configuration, about 32\% higher than reported in the previous ATLAS result~\cite{HDBS-2019-27}, primarily owing to the higher-efficiency $b$-tagging working point and the removal of the eVeto requirement in this channel. For the Run~3 dataset, the acceptance times efficiency increases to approximately 6.3\% from the inclusion of the additional triggers.

\subsection{\lephad channel}
Events in the \lephad channel are required to contain exactly one electron or muon and one \tauhad candidate of opposite electric charge. The electron or muon must fulfil the stricter signal lepton selection described in Section~\ref{sec:object}. Data are collected using single-lepton triggers (SLT). Events with an electron are selected using single-electron triggers~\cite{TRIG-2018-05}, with the electron required to have an offline \pt greater than 27~\GeV\ to ensure high trigger efficiency.\footnote{The threshold is lowered to 25~\GeV\ for 2015 data.} Events with a muon are selected using single-muon triggers~\cite{TRIG-2018-01}, with the muon required to have an offline \pt at least 1~\GeV\ above the corresponding trigger thresholds, which vary between 20~\GeV and 26~\GeV depending on the data-taking period. The lepton-plus-\tauhad triggers used in previous ATLAS analyses~\cite{HDBS-2019-27,HDBS-2018-40} are not used, as they were found to provide little improvement in sensitivity.

The \tauhad candidate must satisfy the eVeto requirement, suppressing background from $Z\to e^+e^-$ events. The $H\rightarrow \tau^+\tau^-$ system is reconstructed from the lepton and the \tauhad candidate. Events are required to have $\mttvis > 40$~\GeV\ and $\mMMC > 60$~\GeV. The invariant mass of the $b$-tagged jet pair is used to separate the signal and control regions. The \lephad SR requires $40 < \mbb < 150$~\GeV. The overall acceptance times efficiency for SM $HH\to b\bar{b}\lephad$ events is approximately 4.7\% for Run~2 and 4.5\% for Run~3, a relative improvement of about 20\% for Run~2 when compared to Ref.~\cite{HDBS-2019-27}.

The \lephad--\ttbar CR is defined by the same preselection as the SR but with an alternative requirement of $\mbb > 150$~\GeV. It is used in the data-driven determination of the background with misidentified \tauhad candidates, as described in Section~\ref{sec:background}.

\subsection{The $ee/\mu\mu/e\mu$ channels}
\label{subsec:CR_preselection}

For the $ee/\mu\mu/e\mu$ channels, events are required to have a pair of opposite-sign and either same-flavour ($\ell^+\ell^-$) or opposite-flavour ($e^\pm\mu^\mp$) signal leptons and no \tauhad candidate or additional leptons. The same SLTs used in the \lephad channel are used to collect these events. No signal regions are defined in these channels, as topologies where both $\tau$-leptons decay leptonically have a small branching fraction and suboptimal Higgs boson mass reconstruction due to the presence of four neutrinos in the final state. Instead, dedicated regions are built to target some of the leading backgrounds in the analysis. These regions share as much of the SR selection as possible: exactly two $b$-tagged jets are required, and the two leptons must have $\pt > 40$~\GeV. Three regions are defined:

The \textbf{\ZHF CR} targets the $Z$+jets process by requiring a pair of same-flavour leptons. Together with the common two-$b$-tagged-jet requirement, this selection provides a sample enriched in $Z$+heavy flavour (HF) jets production.
The dilepton invariant mass must satisfy $75~\GeV < \mll < 110~\GeV$, consistent with the $Z$ boson mass, and the $b$-tagged jet pair invariant mass must satisfy $40~\GeV < \mbb < 210~\GeV$, the same requirement as in the \hadhad SR.

The \textbf{\temu CR} requires an $e\mu$ pair with invariant mass $m_{e\mu} > 110$~\GeV. The lepton--$b$-jet invariant mass must satisfy $\mbl < 250$~\GeV, where $\mbl = \min\!\left[\max(m_{11}, m_{22}),\,\max(m_{12}, m_{21})\right]$ with $m_{ij} = m(b_i, \ell_j)$. In \ttbar\ events, \mbl\ is kinematically bounded from above by the top-quark mass, so this requirement retains the bulk of the \ttbar\ contribution. The \temu CR is used to derive data-driven corrections to the \ttbar\ background simulation through a reweighting procedure described in Section~\ref{sec:background}.

The \textbf{$tW$ CR} has a similar selection to the \temu CR but requires $\mbl > 250$~\GeV\ to increase the purity of $tW$ events and provide sensitivity to their normalisation and modelling uncertainties.

%


\section{Background estimation}

\label{sec:background}

The main sources of background come from top-quark, $Z$+jets, diboson, single Higgs boson, and multijet production. While most of these backgrounds genuinely produce a final state similar to the signal, reducible backgrounds containing misidentified objects have non-negligible contributions. In particular, a reconstructed \tauhad candidate in background events can originate from a genuine hadronic $\tau$-lepton decay (true-\tauhad), from an electron or muon misidentified as a \tauhad (lepton-fake-\tauhad), or from a misidentified quark- or gluon-initiated jet (fake-\tauhad). The \ttbar and multijet production processes are the leading sources of fake-\tauhad candidates. Simulated event samples, described in \Sect{\ref{sec:samples}}, are used to model the majority of the background processes. The exceptions are backgrounds containing fake-\tauhad candidates, which are estimated either with data-driven methods or by applying data-driven corrections to simulation. In this section, a brief overview of the composition of the leading backgrounds and the techniques used to estimate them is given.

\textbf{\ttbar background}\\
Top-quark pair production is the main background in the analysis and is the leading contribution in all signal and control regions. Different sub-processes are considered depending on the decay mode of the two $W$ bosons produced in \ttbar decays. Dileptonic decays produce events with one (two) true-\tauhad in the \lephad (\hadhad) channel, as well as events in the $ee$, $\mu\mu$, and $e\mu$ channels. They represent the leading component of the \ttbar contribution in all regions, with a larger contribution in the \lephad channel due to the larger branching fraction to different-flavour leptons. Dileptonic \ttbar decays are also responsible for events with lepton-fake-\tauhad. In the \hadhad channel, the rate of events with at least one lepton-fake-\tauhad is approximately half that of events with two true-\tauhad; in the \lephad channel, the rate of lepton-fake-\tauhad drops to 2\% due to the eVeto requirement. Semileptonic and all-hadronic \ttbar events require a fake-\tauhad to fulfil the event selection in the SRs and they contribute approximately 45\% and 60\% of the total \ttbar background in the \lephad and \hadhad channels, respectively.

Predictions for the \ttbar process are reweighted to data using the \temu CR, which has a \ttbar purity of 96\%. The reweighting factors are computed as a function of two quantities. The first is the scalar sum of the transverse momenta of all selected central jets, electrons, muons, and \tauhad candidates; the second is the number of central jets.\footnote{While \tauhad candidates are not present in the \temu CR, they are included when computing this quantity in the SRs.} The reweighting is then applied to the simulated \ttbar predictions in every analysis region where they are used; this includes contributions in the \ZHF and $tW$ CRs, events with true- and lepton-fake-\tauhad in all the SRs, and the contribution from events with fake-\tauhad in the \hadhad channel. A common free-floating normalisation factor is applied to all of these contributions in the final likelihood fit. The fake-\tauhad background in the \lephad channel is not included, as it is estimated with an independent, fully data-driven method.

The \ttbar background with at least one fake-\tauhad candidate in the \hadhad channel originates primarily (85\%) from semileptonic \ttbar decays, where one true-\tauhad and one fake-\tauhad are selected. The remaining 15\% arises from fully hadronic \ttbar decays with two fake-\tauhad candidates. Following the approach detailed in Ref.~\cite{HDBS-2018-40}, the background is estimated using simulation with dedicated correction factors derived from data and applied per fake-\tauhad object. The factors are extracted via a binned maximum-likelihood fit to data in a modified \lephad--\ttbar CR where the \tauhad candidate has the same selection as in the \hadhad channel; in particular, no eVeto is applied.
The fit uses the lepton \mtw distribution, which discriminates between \ttbar events with true- and fake-\tauhad, and is performed simultaneously in bins of \tauhad \pt. Separate corrections are computed for different \tauhad track multiplicities, and the measurements are repeated for the different trigger-matching criteria that each reconstructed \tauhad candidate must satisfy in the \hadhad channel, as described in Section~\ref{subsec:SRhadhad}.
These factors are designed to correct the shape and normalisation of the \ttbar background with fake-\tauhad relative to other \ttbar events.

In the \lephad channel, the \ttbar background from events with a fake-\tauhad candidate is the second largest background in the signal regions.
This background is estimated with a fake-factor method~\cite{HDBS-2018-40}. Fake factors are derived in the \lephad--\ttbar CR as the ratio of the number of events containing a fake-\tauhad that passes the \texttt{Loose} RNN identification to the number of events
containing an anti-\tauhad instead, as a function of \pt, $\eta$, and track multiplicity of the \tauhad candidate. These factors are then applied to events satisfying the \lephad SR selection but containing an anti-\tauhad instead of a \texttt{Loose} \tauhad, thereby obtaining the fake-\tauhad background template in the SR.\footnote{Both in the fake-factor derivation and in the fake factor application, the number of fake-\tauhad candidates is obtained by subtracting the simulated backgrounds containing true-\tauhad from the number of reconstructed \tauhad candidates observed in data.} The same approach also accounts for the contribution of events with a fake-\tauhad from $tW$, $W$+jets, and multijet production, although these processes provide a subdominant contribution with respect to \ttbar.

\textbf{Single top-quark background}\\
Associated production of a top quark and a $W$ boson ($tW$) is the dominant component of the single top-quark background in the analysis. The features of this process are similar to those of \ttbar, except for the origin of one of the two $b$-tagged jets. The overall contribution is roughly 4\% of that of \ttbar in the SRs, while it matches the \ttbar yields in the $tW$ CR. The impact of $tW$ is particularly relevant in the \lephad channel, where it represents 24\% of the background contribution in the most sensitive classifier score bin.
The normalisation of $tW$ is a free-floating parameter in the final likelihood fit to data, primarily constrained by the $tW$ CR.

\textbf{$Z$+jets background}\\
The dominant contribution in the \ZHF CR arises from $Z$+jets events in which the $Z$ boson decays into a pair of electrons or muons; the $Z\to \taulep\taulep$ contribution is heavily suppressed by the leptonic $\tau$ decay branching ratio and the \pt requirement on the leading lepton. Conversely, the $Z\to\tautau$ process dominates the $Z$+jets background within the signal regions, accounting for 98\% of the events in the \hadhad SRs and 70\% in the \lephad SRs. Simulated events are categorised according to the true flavour of the $b$-tagged jets. The \ZHF background comprises $bb$, $bc$, $bl$, and $cc$ contributions, where $l$ denotes any light-flavour jet. These account for approximately 90\% of the total number of $Z$+jets events in the signal and control regions, with the $bb$ component being the leading process owing to the performance of the flavour-tagging algorithms. The \ZHF background is particularly important in the most sensitive SR due to its irreducible nature and its similarity to the signal topology; its normalisation is left free-floating in the fit to data. The contribution from $Z$+jets is negligible in the \temu CR and $tW$ CR due to the different-flavour lepton requirement.

\textbf{$W$+jets background}\\
$W$+jets events represent a subdominant background, as they require at least one misidentified physics object to populate the signal and control regions. In the \hadhad channel, the selected events contain a true-\tauhad candidate from a $W$ boson decay and a second fake-\tauhad candidate. The same scale factors applied to the \ttbar events with fake-\tauhad objects are also used to correct the $W$+jets events, although their contribution is only 2\% of the \ttbar process. In the \lephad channel, the contribution originates almost exclusively from $W$ bosons decaying into electrons and muons, produced together with one additional fake-\tauhad. This process is therefore absorbed into the fake-factor method described earlier. MC studies indicate that the yield from the $W$+jets process is expected to be approximately 20 times smaller than that of the \ttbar component with fake-\tauhad candidates.

\textbf{Multijet background}\\
The production of events with multiple jets has a large cross-section and thus, despite not being a source of genuine prompt electrons, muons, or $\tau$-leptons, can contribute to the background in the analysis regions. The multijet contribution, referred to as MJ in the figures, is small or negligible in the \lephad channel and in the \ZHF, \temu, and $tW$ CRs owing to the low rates for jets to be misidentified as electrons or muons. The misidentification rate for \tauhad is significantly larger, making multijet production an important background in the \hadhad channel.
A modified fake-factor method is used to estimate the multijet contribution, following the approach of Ref.~\cite{HDBS-2018-40}. Fake factors are derived in a dedicated set of control regions requiring a pair of same-sign \tauhad candidates: regions where both candidates fulfil the \texttt{Loose} RNN selection, and regions with one \texttt{Loose} \tauhad candidate and one anti-\tauhad candidate. In all regions used to extract the fake factors and to obtain the final multijet background template, contributions from non-multijet backgrounds are subtracted from the observed data. The fake factors are extracted separately for each of the three trigger buckets and as a function of \pt, $\eta$, and track multiplicity of the \tauhad candidate. For events in the STT and DTT buckets, the extraction is performed in events with exactly one $b$-tagged jet, and \pt- and $\eta$-inclusive transfer factors are then applied to extrapolate the result to the two-$b$-tagged-jet selection. A selection requiring two $b$-tagged jets is used for events in the DBT bucket.
These fake factors are then applied to events satisfying the \hadhad SR selection but containing one anti-\tauhad and one \texttt{Loose} RNN \tauhad instead of two \texttt{Loose} RNN \tauhad candidates, thereby obtaining the multijet template in the SR.

In Run~2, the multijet contribution is of the same magnitude as that of the \ttbar process with fake-\tauhad. Since the fake rate rises sharply as the \pt of the \tauhad candidate decreases, delayed-stream events exhibit a larger multijet contamination than the rest of the SRs. %
In Run~3, the multijet yields are twice as high as the \ttbar contribution with a fake-\tauhad candidate.

\textbf{Diboson background}\\
The diboson background is completely dominated by the $ZZ$ process. The configuration where one $Z$ boson decays into a $b$-quark pair and the other into a pair of $\tau$-leptons is the leading contribution, accounting for 88\% and 68\% of $ZZ$ events in the \hadhad and \lephad SRs, respectively. The lower fraction in the \lephad channel is explained by the presence of electrons or muons in the final state, which naturally increases the acceptance for $Z\to e^{+}e^{-}/\mu^{+}\mu^{-}$. While $ZZ$ is a subleading contribution to the total background, it has a topology similar to that of the $HH$ signal and is used to validate the analysis methodology. The predicted $ZZ$ yield, summed over all signal regions, is 10 times that of SM $HH$ production, although a smaller fraction of events (43\%) than for the $HH$ signal falls in the most sensitive SR, owing to the softer kinematics of diboson events.

\textbf{Single Higgs boson background}\\
Processes containing a single Higgs boson have a subdominant contribution to the total background in the SRs, but they display signal-like features that increase their importance in the phase space relevant to the signal extraction. The single-Higgs contribution is dominated by associated production with a top-quark pair (40\%--60\%), followed by $ZH$ production (22\%--36\%) and ggF production in association with additional heavy-flavour jets (\ggHHF, 11\%--23\%). The \ggHHF component displays the most signal-like behaviour. %
The $ZH$ process, with resonant features in both \mbb and \mMMC, represents another key process for validating the analysis methodology. Its yield is 4.9 times that of SM $HH$ production, summed over all SRs. The final state with a pair of $b$-quarks and a pair of $\tau$-leptons represents 90\% and 88\% of the $ZH$ yield in the \hadhad and \lephad SRs, respectively, with a ratio of approximately 2:3 between \ZHbbtt and \ZHttbb.

%


\section{Multivariate analysis}

\label{sec:mva}

A transformer MVA strategy is developed to simultaneously categorise events into signal regions and discriminate signal from background within each region, superseding the approach based on boosted decision trees used in the previous ATLAS $HH\rightarrow \bbtt$ analysis~\cite{HDBS-2019-27}. Two independent models are trained for the \hadhad and \lephad channels, sharing the same architecture and hyperparameters.

\subsection{Architecture}

The transformer~\cite{vaswani2023attentionneed} architecture is chosen for its ability to capture complex correlations among reconstructed objects in an event and to handle variable-length input sequences, which is well suited to the \bbtt analysis with diverse final-state kinematic objects. The architecture follows that of the GN2 flavour tagging algorithm~\cite{FTAG-2023-05}, adapted to the analysis-specific input features and classification tasks described below.

Each model receives two types of input. The first is a set of per-object features for the reconstructed jets, \tauhad candidates, \metV, the di-$\tau$ four-momentum reconstructed by the MMC algorithm, and additionally the lepton in the \lephad channel. The second is a set of global event-level features, including the kinematic properties of the $H\to\bbbar$ and $H\to\tau^+\tau^-$ candidates, the reconstructed \HH system, and observables describing the VBF-jet topology. The full list of input features, given in Table~\ref{tab:kinematic_features}, is obtained by starting from the significantly larger set used in the previous ATLAS analysis~\cite{HDBS-2019-27} and iteratively removing variables until the model performance began to degrade.

\begin{table}[htbp!]
\centering
\footnotesize
\caption{Summary of input features used in the MVA. Event-level features are listed in the upper portion of the table, separated by channel, and object-level features common to all objects in the graph are listed in the lower portion. Since the number of reconstructed objects varies from event to event, each object additionally carries a validity flag, and only valid objects contribute to the network. VBF jets are defined as non-$b$-tagged central jets with $\pt > 20$~\GeV\ and $|\eta| < 2.5$, together with all forward jets with $\pt > 30$~\GeV\ and $2.5 < |\eta| < 4.5$; in their ordering, forward jets are placed ahead of central jets regardless of \pt. The indices 0 and 1 denote the \pt-leading and \pt-subleading object of each type.}
\label{tab:kinematic_features}
\centering
\begin{tabular}{ll}
\toprule
\textbf{Variable} & \textbf{Description} \\
\midrule
\multicolumn{2}{c}{\textit{Global event-level features: both channels}} \\
\midrule
\mhh           & Invariant mass of the reconstructed \HH system using the di-$\tau$ four-momentum after the MMC algorithm \\
\mbb           & Invariant mass of the $b$-jet pair \\
\dRbb          & $\Delta R$ between the two $b$-jets \\
\mjj           & Invariant mass of the two leading VBF jets \\
\dRjj          & $\Delta R$ between the two leading VBF jets \\
\EtaOneEtaTwoVBF & Product of pseudorapidities of the two leading VBF jets \\
\midrule
\multicolumn{2}{c}{\textit{Global event-level features: \hadhad only}} \\
\midrule
\dRtautau      & $\Delta R$ between the two \tauhad candidates \\
\midrule
\multicolumn{2}{c}{\textit{Global event-level features: \lephad only}} \\
\midrule
\dRTauLep      & $\Delta R$ between the \tauhad candidate and the electron or muon \\
\pTHH          & Transverse momentum of the reconstructed \HH system \\
\etaHH         & Pseudorapidity of the reconstructed \HH system \\
\met           & Missing transverse energy \\
\dRbZerotau    & $\Delta R$ between the leading $b$-jet and the \tauhad candidate \\
\dRbOnetau     & $\Delta R$ between the subleading $b$-jet and the \tauhad candidate\\
\mbl           & $\min(x,y)$, where $x = \max(m_{b_1,\tauhad}, m_{b_2,\ell})$ and
$y = \max(m_{b_1,\ell}, m_{b_2,\tauhad})$ \\
\mttvis        & Invariant mass of the visible di-$\tau$ system \\
\pTttvis       & Transverse momentum of the visible di-$\tau$ system \\
$T_1$          & Topness~\cite{Kim:2018cxf}, with $\sigma_t = \sigma_W = 5$~\GeV \\
\htsum         & Scalar sum of transverse momenta of all reconstructed objects \\
\dEtaHH        & $\Delta\eta$ between the $b$-jet pair and the MMC di-$\tau$ systems \\
\dPhiTauTauMET & $\Delta\phi$ between \metV and the MMC di-$\tau$ system \\  %
\dRbOnelep     & $\Delta R$ between the subleading $b$-jet and the lepton \\
\mTTauZero     & Transverse mass of the \tauhad candidate and \metV system\\
\midrule
\multicolumn{2}{c}{\textit{Object-level features (per-object inputs to the transformer network)}} \\
\midrule
$\eta$         & Pseudorapidity; set to 0 for \metV \\
$\phi$         & Azimuthal angle \\
$m$            & Mass; set to 0 for \metV, leptons, and $\tauhad$ candidates \\
$\pt$          & Transverse momentum \\
\multirow{2}{*}{PCBT} & GN2 pseudo-continuous $b$-tagging score (bins: 100\%--90\%--85\%--77\%--70\%--65\%) \\
& set to $-2$ for non-jet objects and $-1$ for forward jets \\
isTau          & Boolean flag: object is a \tauhad candidate \\
isJet          & Boolean flag: object is a non-$b$-tagged jet \\
isBjet         & Boolean flag: object is a $b$-tagged jet \\
isMET          & Boolean flag: object is \metV \\
isMMC          & Boolean flag: object is the MMC di-$\tau$ system \\
isLepton       & Boolean flag: object is a lepton (only in \lephad channel) \\
\bottomrule
\end{tabular}
\end{table}

The per-object and event-level features are concatenated for each object and passed to an initialiser network following a Deep Sets architecture~\cite{zaheer2018deepsets}, consisting of two dense layers of sizes (32, 64) with rectified linear unit (ReLU) activations~\cite{nair2010relu}, which maps the inputs to a 256-dimensional latent space. The embedded representations are then processed by a transformer encoder with four layers and four attention heads per layer, employing scaled dot-product self-attention~\cite{vaswani2023attentionneed} to capture correlations among objects. The sequence of encoded object representations is aggregated using a learned global attention pooling layer, yielding a 128-dimensional event representation. This representation is passed simultaneously to four separate three-layer feedforward networks with hidden-layer sizes [128, 64, 32], ReLU activations, and a dropout rate of 2.6\%, each corresponding to a distinct classification task.

\subsection{Training}

The training sample consists of simulated SM ggF and VBF \HH signal events, simulated ggF \HH events with $\kl = 5$, simulated background events, and, for the \hadhad channel, the data-driven fake-\tauhad background estimate. In the \lephad channel, given the limited statistical precision on the data-driven fake-\tauhad estimate, the prediction from simulation is used instead for training. The inclusion of $\kl = 5$ events is motivated by the need to train the model on kinematic configurations that yield a softer \mhh spectrum, which differs substantially from the SM, allowing this information to be partially exploited when setting limits on \kl. The same event weights applied to simulated events in the analysis to determine the signal and background yields are also applied during training, to account for cross-section and other differences. The MVA strategy is optimised to maximise the expected signal significance, evaluated with a metric similar to that of the likelihood fit described in Section~\ref{sec:fit} but without systematic uncertainties.

Each network has four output heads corresponding to the following classification tasks:

\begin{itemize}
\item \textbf{Multiclass classification}: Distinguishes four event categories, background, SM ggF signal, SM VBF signal, and BSM ggF signal with $\kl = 5$, yielding output scores \pB, \pGGF, \pVBF, and \pggFBSM, respectively.

\item \textbf{Binary classification in the high-\mhh region}: A dedicated binary classifier is trained on events with $\mhh > 350$~\GeV\ collected via the main-stream trigger to distinguish the SM ggF and VBF signal from the SM background. The classifier targets the kinematic regime that contributes most to the sensitivity to the SM \HH production cross-section. Events collected via delayed-stream triggers are not included in this training. The classifier outputs two scores, \pHighMassS\ and \pHighMassB, corresponding to the signal and background hypotheses, respectively. Adding this classifier on top of the multiclass head improves the sensitivity by about 10\% in both the \hadhad and \lephad channels.

\item \textbf{Auxiliary Higgs boson parentage classification}: An object-level auxiliary task classifying whether or not a reconstructed object originates from a Higgs boson decay, improving discrimination against single-Higgs-boson backgrounds. This task improves the sensitivity by 2--10\%, depending on the signal region, in both the \hadhad and \lephad channels.

\item \textbf{Auxiliary top-quark parentage classification}: An object-level auxiliary task classifying whether or not a reconstructed object originates from a top-quark decay, improving discrimination against \ttbar\ and single-top-quark backgrounds. This task improves the sensitivity by 2--8\%, depending on the signal region, in both the \hadhad and \lephad channels.
\end{itemize}

Each task is trained with a cross-entropy loss, and the total loss function is a weighted sum of the individual task losses:
\begin{equation}
L_{\text{total}} = L_{\text{Multiclass}}
+ 10\, L_{\text{Binary}}
+ 2\, L_{\text{Aux Top}}
+ 5\, L_{\text{Aux Higgs}},
\end{equation} where the weights are determined based on a dedicated study to maximise the analysis sensitivity.

Training is performed inclusively on Run~2 and Run~3 simulated events. A 5-fold cross-validation strategy is employed to mitigate overfitting and to allow every event to be used in the analysis while never being evaluated by a model that was trained on it. The dataset is divided into five disjoint subsets: for each fold, three subsets are used for training, one for validation, and one for testing, so that five independent models are trained. Each event is classified by the model for which it falls in the test subset, ensuring statistical independence between training and evaluation. Models are trained for up to 30 epochs with a batch size of 1024, and the checkpoint with the lowest validation loss is selected. To reduce the effect of statistical fluctuations from the random initialisation, a model is trained 20 times with different random seeds for each fold, and the one whose validation signal significance is closest to the ensemble mean is selected for the final evaluation.

\subsection{Event categorisation}

\begin{figure}[htbp!]
\centering
\includegraphics[width=0.99\textwidth]{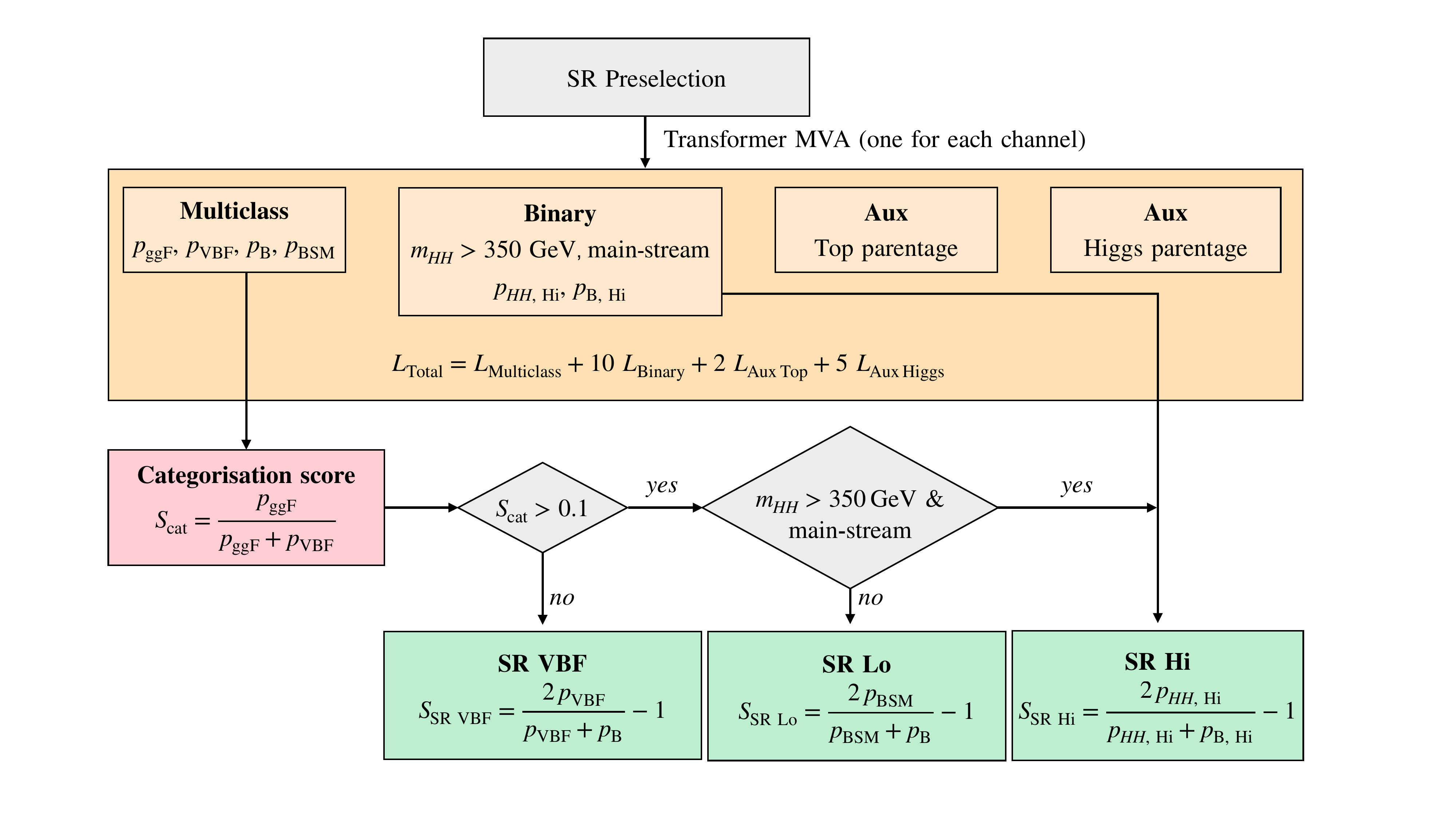}
\caption{Overview of the event categorisation scheme. The MVA produces multiclass, binary, and two auxiliary classification outputs, optimised jointly via the combined loss function. Events are assigned to one of three orthogonal signal regions (\SRVBF, \SRLo, \SRHi) based on the categorisation score $\Scat$ and the reconstructed $\mhh$. For the \hadhad channel, the DBT and DTT events in the delayed stream are designed to enter only \SRLo.}
\label{fig:event_categorization}
\end{figure}

The output scores from the multiclass and binary classification heads are used to define the signal regions and the signal--background discriminants, as illustrated in Figure~\ref{fig:event_categorization}. This categorisation is performed independently in the \lephad and \hadhad channels, using the scores from the respective channel's classifier. Preselected events are assigned to ggF-like or VBF-like regions based on a ggF--VBF categorisation score \Scat, defined as
\begin{equation}
\Scat = \frac{\pGGF}{\pGGF + \pVBF},
\end{equation}
where \pGGF and \pVBF are the ggF and VBF signal probabilities from the multiclass head. Events with $\Scat < 0.1$ are assigned to \SRVBF, with discriminant
\begin{equation}
\SVBF = \frac{2\,\pVBF}{\pVBF + \pB} - 1.
\end{equation}
The threshold $\Scat = 0.1$ is chosen such that more than 90\% of ggF events and more than 60\% of VBF events are assigned to their respective signal regions in both channels.
Among the remaining events, those with $\mhh > 350$~\GeV\ are assigned to the \SRHi signal region, with discriminant
\begin{equation}
\SHighMass = \frac{2\,\pHighMassS}{\pHighMassS + \pHighMassB} - 1.
\end{equation}
For the \hadhad channel, as the background composition of the delayed-stream trigger events differs, only main-stream trigger events are retained in \SRHi. All remaining events are assigned to \SRLo, with discriminant
\begin{equation}
\SggFLowMass = \frac{2\,\pggFBSM}{\pggFBSM + \pB} - 1.
\end{equation}

\begin{figure}[tb!]
\centering
\subfloat[]{
\includegraphics[width=0.4\textwidth]{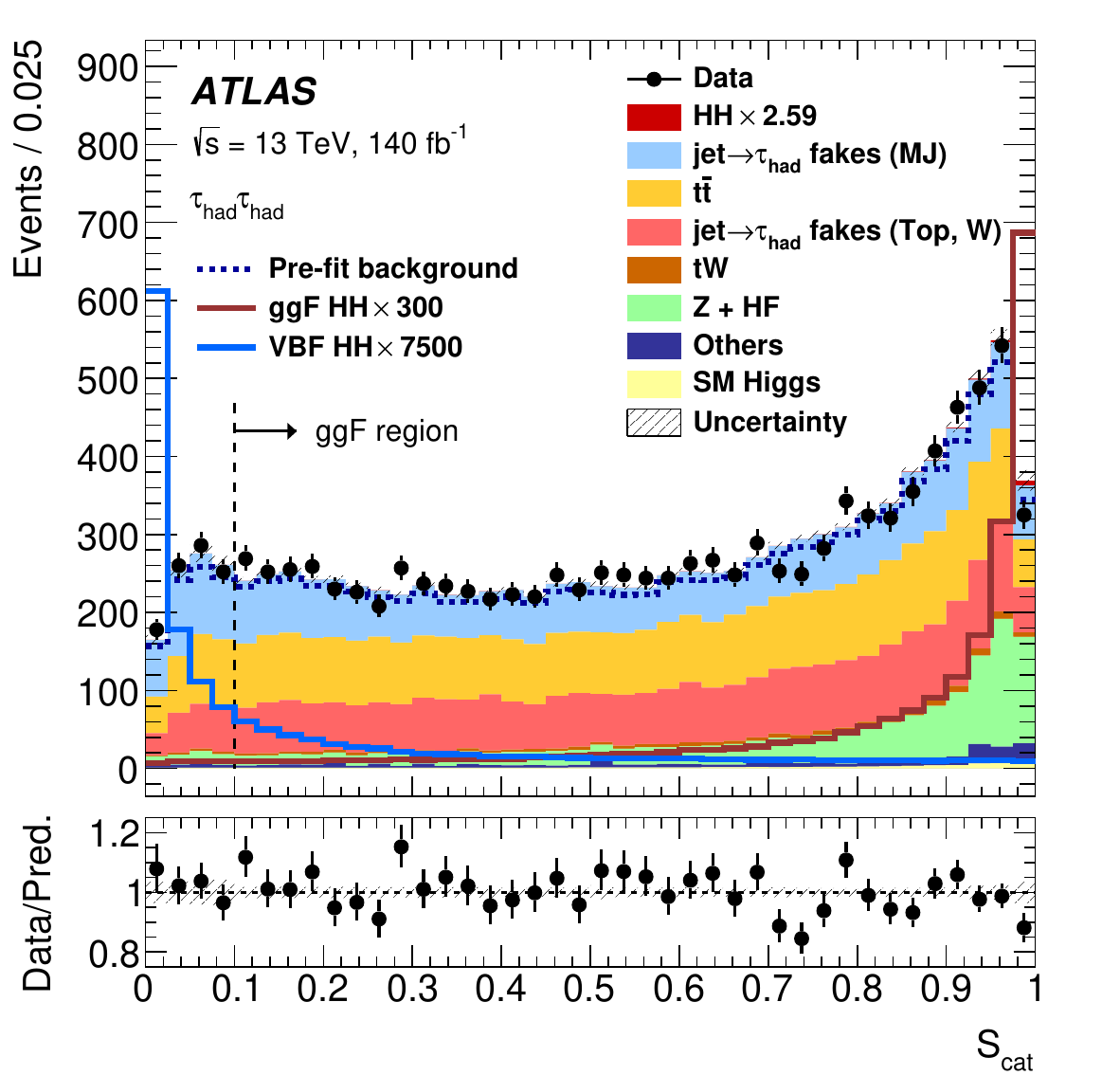}
\label{fig:Scat_Run2_HadHad}
}
\subfloat[]{
\includegraphics[width=0.4\textwidth]{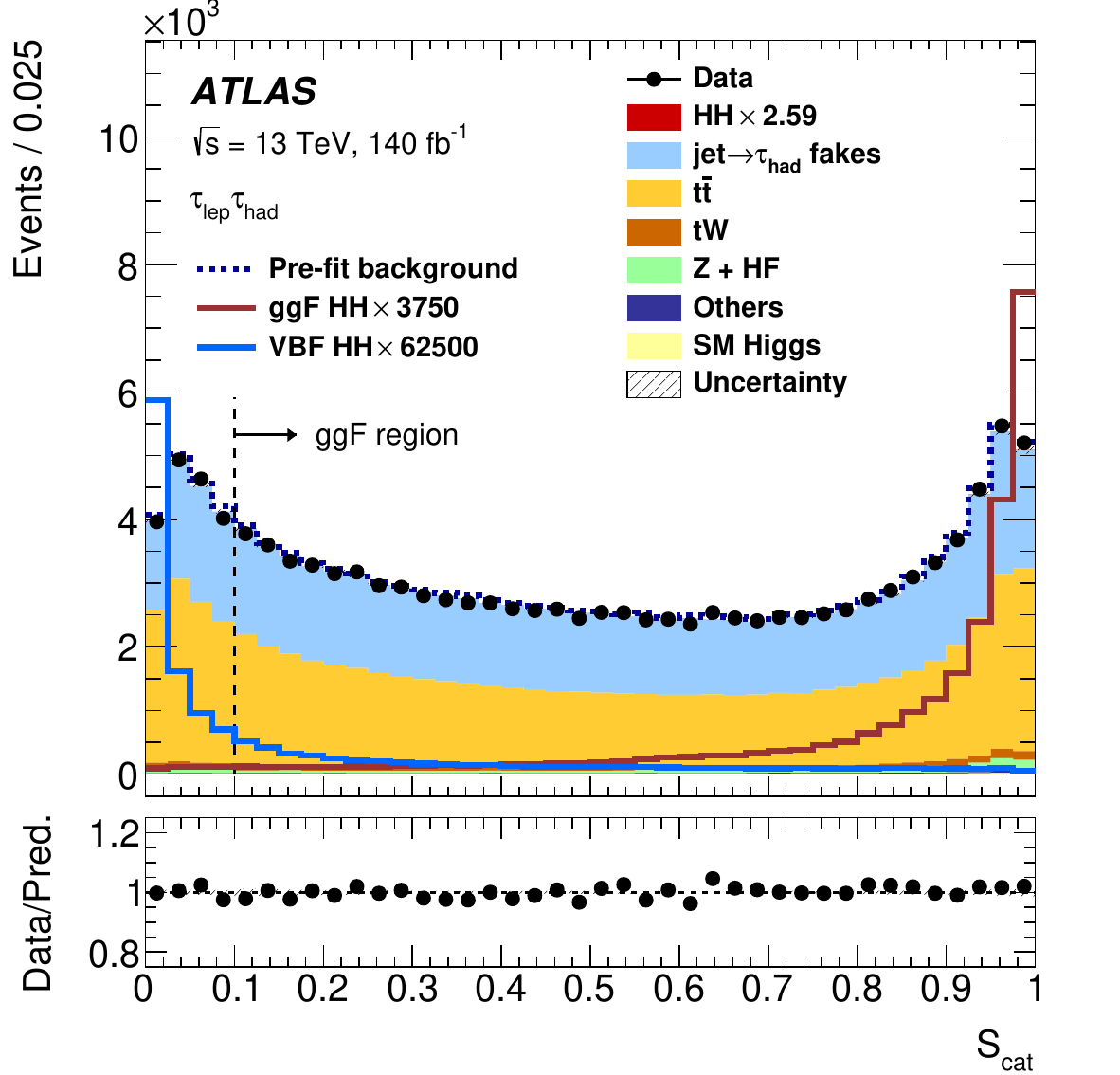}
\label{fig:Scat_Run2_LepHad}
} \\
\subfloat[]{
\includegraphics[width=0.4\textwidth]{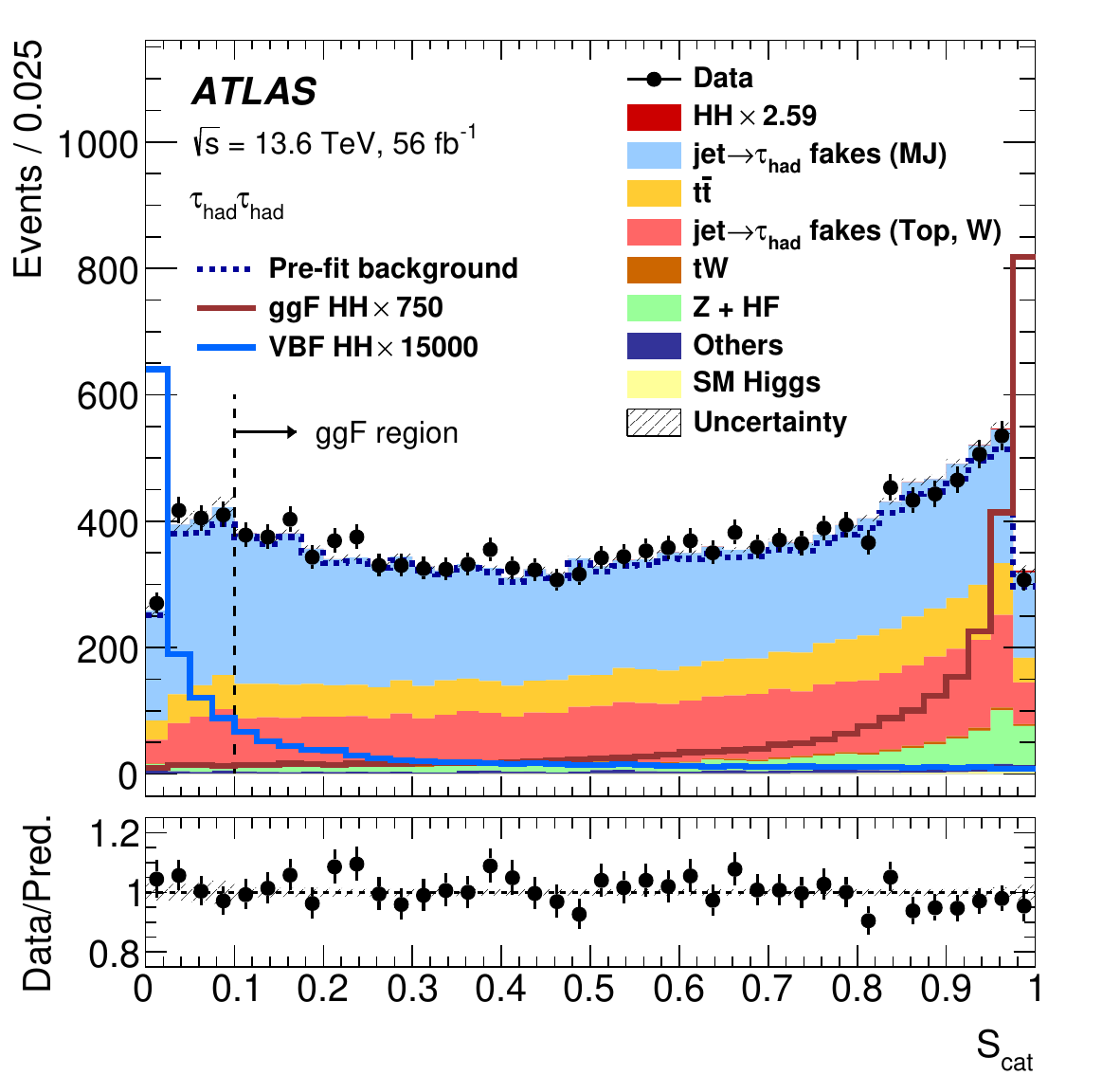}
\label{fig:Scat_Run3_HadHad}
}
\subfloat[]{
\includegraphics[width=0.4\textwidth]{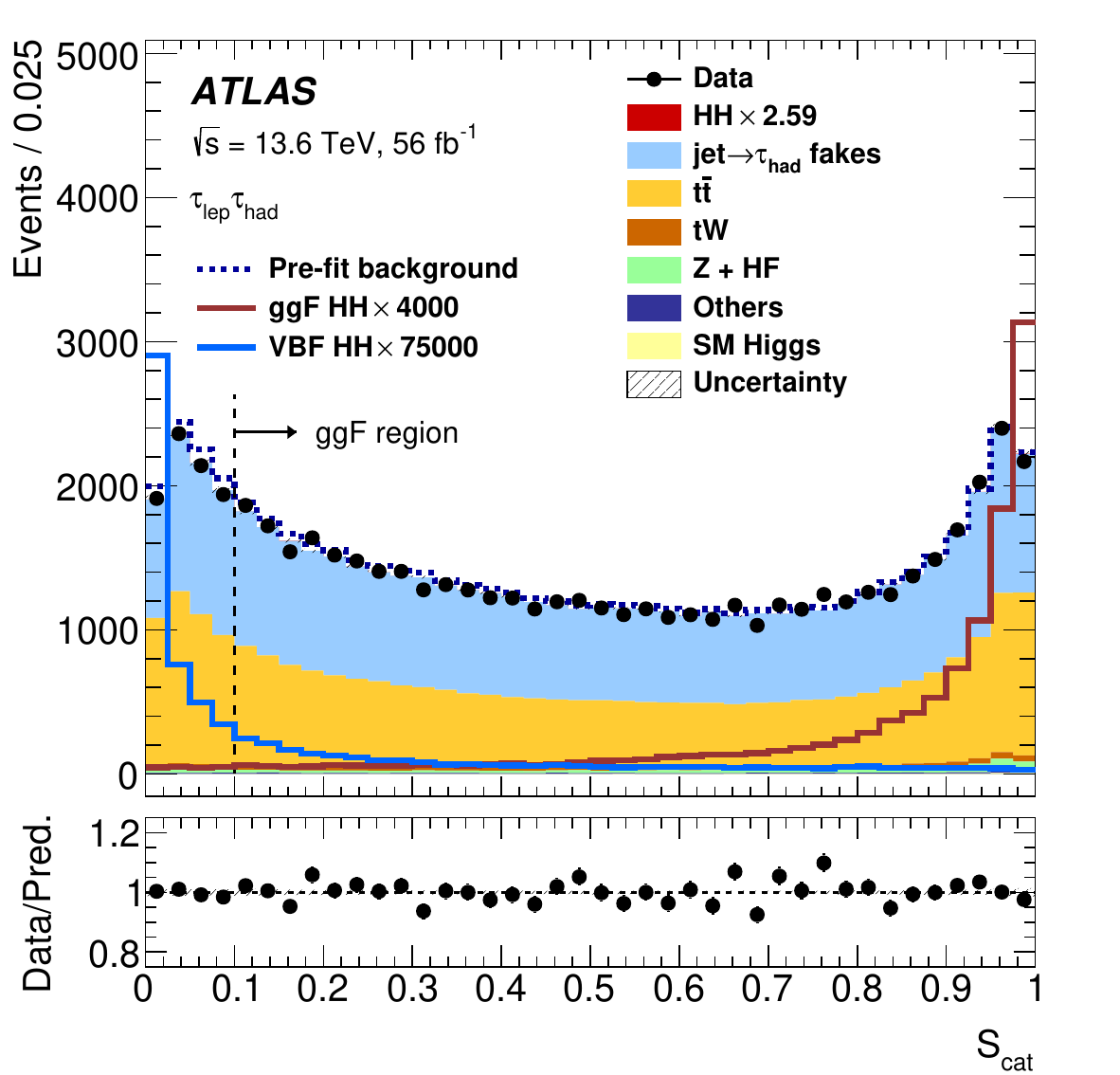}
\label{fig:Scat_Run3_LepHad}
}
\caption{Predicted and observed distributions of the categorisation score \Scat in the sum of the SRs for the \hadhad channel (left) and \lephad channel (right) in Run~2 (top) and Run~3 (bottom). The background distributions are shown as obtained from the combined likelihood fit to data described in Section~\ref{sec:fit}. The ggF and VBF \HH signal distributions are overlaid and scaled to the factor indicated in the legend times the SM expectation. The dashed histogram shows the total pre-fit background. The lower panels show the ratio of data to the total post-fit sum of signal and background. The vertical dashed line indicates the threshold value that defines the boundary between the ggF and VBF SRs.}
\label{fig:ggFVBFScores}
\end{figure}

Figure~\ref{fig:ggFVBFScores} shows the post-fit distribution of the ggF--VBF categorisation score \Scat for the \lephad and \hadhad channels in Run~2 and Run~3.

The three signal regions are designed to maximise the sensitivity to distinct kinematic regimes of \HH production. \SRVBF targets events with a topology consistent with VBF production, providing the primary sensitivity to anomalous values of \kvv. Together, \SRHi and \SRLo span the full range of ggF $HH$ production kinematics, where \SRHi provides the dominant contribution to the sensitivity to the SM \HH production cross-section and \SRLo provides complementary sensitivity to anomalous values of \kl.

%


\section{Systematic uncertainties}

\label{sec:sys}
The dominant source of uncertainty is the limited size of the available collision dataset. Systematic uncertainties provide a subdominant contribution to the total uncertainty and can be broadly categorised into three groups: experimental uncertainties, theoretical uncertainties related to the simulated signal and background processes, and uncertainties related to the estimation of the fake-\tauhad backgrounds. The estimation of these uncertainties builds upon the methodology outlined in Refs.~\cite{HDBS-2018-40,HDBS-2019-27} and is briefly summarised below. Unless otherwise specified, uncertainties are treated as correlated across channels and uncorrelated across data-taking periods.

\subsection{Experimental uncertainties}
Several sources of experimental uncertainty are considered, with those related to jets and \tauhad candidates being dominant. Jet-related uncertainties arise from the energy scale and resolution~\cite{JETM-2018-05}, and the efficiency of the JVT algorithm~\cite{PERF-2014-03}. A dedicated uncertainty arising from the energy calibration of $b$-jets is also included. As Run~2 data are used to calibrate the jet energy scale and resolution for the Run~3 simulation, the corresponding uncertainties are correlated between Run~2 and Run~3. Additional dedicated uncertainties are considered to cover the extrapolation to Run~3 conditions.
The uncertainties arising from the calibration of the $b$-tagging efficiencies and misidentification rates are decomposed into several independent and uncorrelated contributions. Uncertainties associated with the \tauhad candidate reconstruction, identification, trigger selection, and energy scale are estimated following the methodologies described in Refs.~\cite{ATLAS-CONF-2017-029,ATL-PHYS-PUB-2019-033,ATL-PHYS-PUB-2022-044}. Uncertainties associated with the eVeto algorithm used to distinguish \tauhad candidates from electrons are also considered. Dedicated uncertainties in the efficiency of the DBT triggers are included following the procedure described in Ref.~\cite{TRIG-2022-02}.

The subdominant experimental uncertainties include those related to the trigger, reconstruction, and identification of electrons~\cite{EGAM-2021-01} and muons~\cite{MUON-2018-03}. The uncertainties in the energy and momentum calibration of the physics objects are propagated to the estimation of \met, which also incorporates
contributions from the modelling of the underlying event and the reconstruction efficiency of tracks used to compute the soft term~\cite{JETM-2020-03}. Uncertainties of 0.83\%~\cite{DAPR-2021-01} and 2.0\%~\cite{ATL-DAPR-PUB-2023-001,ATL-DAPR-PUB-2024-001} are assigned to the integrated luminosities in Run~2 and Run~3 data, respectively. An uncertainty arising from the correction of the pile-up distribution in simulation compared with that in data based on the measurement of the visible cross-section in minimum-bias events~\cite{STDM-2015-05} is also included. Experimental uncertainties in individual physics objects are propagated to uncertainties in the predicted normalisations and shapes of signal and background distributions used in the final statistical analysis.

\subsection{Theoretical modelling uncertainties}
\label{subsec:sigmod}

Various sources of uncertainty are considered in the modelling of the simulated backgrounds and grouped according to their effect into three categories: absolute normalisation uncertainties, relative acceptance uncertainties between regions of phase space, and shape uncertainties in the predicted distribution of the fit discriminant. Relative acceptance uncertainties are used for samples whose overall normalisation is determined in the fit to data. Overall cross-section uncertainties are applied to all background processes except \ZHF, \ttbar, and $tW$ production, whose normalisations are free parameters in the likelihood fit to data. Unless stated otherwise, acceptance shape and normalisation uncertainties are treated as uncorrelated.

The \ttbar normalisation is determined from the fit to data and is primarily constrained by the \ZHF CR and the low-classifier-score parts of the \lephad signal regions. Relative acceptance uncertainties are defined for each fit region with respect to the \ZHF CR, in addition to all statistically significant shape uncertainties affecting the fitted observables in each region. Uncertainties associated with the choice of parton-shower model and with the matching of the matrix element to the parton shower are evaluated by comparing the nominal samples with alternative samples generated with different configurations. \HERWIG[7]~\cite{Bahr:2008pv,Bellm:2015jjp,Bellm:2019zci} is used as an alternative parton shower in place of \PYTHIA[8]. The matching uncertainty is probed through variations of the \hdamp parameter, which controls the transverse momentum of the first emission in the \POWHEG matching, and of the \pthard setting, which governs the scale at which the parton shower is matched to the matrix-element emission. An additional uncertainty in the treatment of recoil in top-quark decays is evaluated using an alternative recoil scheme~\cite{TOPQ-2018-19}. Uncertainties due to the modelling of initial- and final-state radiation, the choice of the QCD scales, and PDF uncertainties are also considered.

The contributions from the $s$- and $t$-channels of single top-quark production are small in the analysis phase space, and therefore only overall cross-section uncertainties are applied. Dedicated modelling uncertainties are derived for the $tW$ background process. As for the \ttbar background, only relative acceptance and shape uncertainties are assigned. Uncertainties in the choice of matrix-element and parton-shower generator, as well as the choice of the \hdamp parameter, are evaluated from comparisons with alternative samples, while those from the choice of QCD scales, PDFs, and the modelling of initial- and final-state radiation are taken from internal weight variations and found to be small. A further uncertainty arises from the modelling of the interference between $tW$ and \ttbar production, and is evaluated by comparing the nominal diagram-removal scheme with an alternative diagram-subtraction scheme~\cite{Frixione:2008yi}. This interference uncertainty is the largest individual source of uncertainty in the modelling of $tW$ production.

For \ZHF processes, only relative acceptance and shape uncertainties are assigned. In addition to variations of the QCD scales and the PDF uncertainties, the nominal \SHERPA sample is compared to an alternative \MGNLO~+~\PYTHIA[8] sample generated using FxFx merging~\cite{Frederix:2012ps}. This is the only \ZHF modelling uncertainty with a significant impact on the shape of the classifier score in the signal regions. A flavour-composition uncertainty, derived from the same generator comparison, covers the differences in $Z+bc$, $Z+bl$, and $Z+cc$ contributions relative to $Z+bb$. A further uncertainty covers mismodelling in the prediction of the $Z$ boson \pt spectrum and is derived from data in the \ZHF CR.

For single-Higgs-boson production, the dominant contributions are $t\bar{t}H$ and $ZH$ production. Their modelling uncertainties from the QCD scale variations and the PDF choice are considered, and additional uncertainties are evaluated from comparisons with alternative samples, using \HERWIG[7] in place of \PYTHIA[8] for the parton shower and \MGNLO for the hard-scatter calculation in the $ttH$ case. For the $ZH$ process, uncertainties are evaluated separately for the quark- and gluon-initiated production and for the $H\to\bbbar$ and $H\to\tau^+\tau^-$ decay configurations.

An uncertainty of 100\% is assigned to the normalisation of the ggF, VBF, and $WH$ processes with $H\to\tau^+\tau^-$ and additional heavy-flavour jets.
Due to the modelling challenges associated with production in conjunction with heavy-flavour jets, a conservative approach is adopted with the magnitude of the uncertainties motivated by the observations from several ATLAS measurements~\cite{HIGG-2019-13,HIGG-2018-29}. The \ggHHF component is the leading background modelling uncertainty in the analysis. Only overall normalisation uncertainties are considered for the remaining minor backgrounds, including $Z$+light-flavour, $W$+jets, and the subleading $WW$ and $WZ$ diboson processes.

Systematic uncertainties that affect the modelling of the signal are estimated with procedures that closely follow those outlined in Ref.~\cite{HDBS-2018-40}. The uncertainties in the calculations of the \HH production cross-sections in the ggF and VBF modes, as well as in the $H\rightarrow\bbbar$ and $H\rightarrow\tau^+\tau^-$ branching ratios, are assigned following the recommendations in Refs.~\cite{Baglio:2020wgt,deFlorian:2016spz}. Uncertainties in the acceptance and in the shape of the classifier score distributions are estimated from QCD scale variations, as well as from propagating PDF uncertainties, and by comparing \PYTHIA[8] with \HERWIG[7] for parton showering. All signal theory uncertainties are correlated between Run~2 and Run~3, but uncorrelated between processes with the exception of the branching ratio uncertainties.

\subsection{Uncertainties in the fake-\tauhad background estimates}

Uncertainties in the data-driven fake-\tauhad background estimate account for differences in the composition of fake-\tauhad objects between the control regions, where the fake factors are derived, and the signal regions. In the \hadhad channel, an additional uncertainty is assigned to account for the charge asymmetry between the event selections applied in the control and signal regions. Both sets of uncertainties are estimated by varying the event selection used to derive the fake factors or by comparing the origin of fake-\tauhad candidates in simulated samples. Finally, uncertainties in the normalisation of other simulated processes, which are subtracted from the data when deriving fake factors or in the template region, are propagated as shape uncertainties to the fake-\tauhad background template. The statistical uncertainties in the fake factors are also propagated to the fit. None of these systematic uncertainties has a significant impact on the final results. The largest effect arises from the statistical uncertainty on the background template due to the limited number of data events available for its construction.

The \ttbar background with a fake-\tauhad candidate in the \hadhad channel, by contrast, is taken from simulation corrected by data-derived scale factors. As described in Section~\ref{sec:background}, these scale factors are obtained from a binned maximum-likelihood fit to data in a modified \lephad--\ttbar CR, and their uncertainties are propagated to the background estimate. As the scale factors measured across the \tauhad candidate \pt and track-multiplicity bins are strongly correlated, a principal-component analysis is used to decompose their combined uncertainty into a set of independent components that are propagated to the final likelihood fit.

%


\section{Statistical interpretation}

\label{sec:fit}

Signal and background yields in the selected event sample are determined using fits based on a likelihood function $\mathcal{L}(\mu, \vec{\theta})$ constructed as the product of Poisson probability terms over the bins of the input distributions, with parameters of interest (POIs) $\mu$ extracted by maximising the likelihood. The effects of systematic uncertainties enter the likelihood as nuisance parameters (NPs), $\vec{\theta}$. The normalisations of the largest backgrounds, \ZHF, \ttbar without misidentified \tauhad candidates, and $tW$ processes in several analysis regions are determined by the fit, so they are left unconstrained in the likelihood and floated independently for Run~2 and Run~3 regions. Experimental, theoretical, and fake-\tauhad background estimation uncertainties, discussed in Section~\ref{sec:sys}, are described by NPs constrained with Gaussian or log-normal probability density functions. The uncertainties due to the limited number of events in the background templates %
are included using the ``light'' Beeston--Barlow technique~\cite{Barlow:1993dm}. Systematic variations that are subject to large statistical fluctuations are smoothed following the procedure detailed in \Refn{\cite{HDBS-2019-27}}. Systematic uncertainties that have a negligible impact on the final results are pruned away region-by-region, with the signal and control regions treated separately. Signal strength upper limits are derived using a modified frequentist CL$_\mathrm{s}$ method~\cite{Read:2002hq}, with the profile-likelihood ratio as the test statistic~\cite{Cowan:2010js}.

The global likelihood fit comprises 12 signal regions, corresponding to \SRLo, \SRHi, and \SRVBF in each of the \hadhad and \lephad channels for both run periods, and four control regions, two for each LHC run periods. In the signal regions, the classifier scores are used in the fit, whereas in the control regions the fitted observables are the invariant mass of the two leptons, \mll, in the \ZHF CR, and the \mbl variable, defined in Section~\ref{subsec:CR_preselection}, in the $tW$ CR. The choice of \mll is motivated by the different distributions expected for $Z$+jets and \ttbar events, while \mbl is chosen to better constrain the modelling of the $tW$ background.

The binning of the classifier score is determined following the algorithm defined in Ref.~\cite{HIGG-2016-29} to obtain a smoother distribution for the background processes and finer binning in the regions with the largest signal contribution, while ensuring that the statistical uncertainty of the background estimation is less than 20\% in each bin, leading to an enhanced sensitivity for the signal. In addition, bins are required to have at least three expected background events to support the validity of asymptotic formulae~\cite{Cowan:2010js} in the statistical interpretation of the data. The final number of bins for each distribution varies from 9 to 15, depending on the expected number of background events.

Multiple sets of POIs are probed by the analysis and extracted in different fits to data. Initially, a single signal strength, $\mu_{HH}$, is considered as a multiplicative factor that scales all the expected $HH$ predictions. Next, a 2-POI fit is performed to extract simultaneously the signal strength for the ggF and VBF production modes ($\mu_{\mathrm{ggF}}$, $\mu_{\mathrm{VBF}}$). The extraction of confidence intervals for \kl and \kvv is performed by considering the full dependence of the cross-section and event kinematics of the ggF and VBF $HH$ production on these parameters. The dependence of the single-Higgs boson production cross-section and $H\rightarrow b\bar{b}$ and $H\rightarrow\tau^+\tau^-$ branching fractions on these parameters is also taken into account, but it constitutes only a very small effect.

\subsection{$ZZ$ and $ZH$ validation measurements}
To validate the full analysis chain, before extracting the \HH signal strength(s), alternative fits are performed to extract signal strengths for the diboson process ($\mu_{ZZ}$) and for the associated production of a Higgs boson and a $Z$ boson ($\mu_{ZH}$). The configuration of the fit is determined as a compromise between maximum sensitivity for these processes and the need to minimise the differences with respect to the $HH$ signal extraction setup. The binary classifier scores used in \SRHi are replaced with alternative discriminants used to separate $ZZ$ and $ZH$ processes from the total background. These discriminants are obtained analogously to those used to separate the $HH$ signal from the background in the \SRHi regions, employing the same network architecture, training strategy, hyperparameters, and input features described in \Sect{\ref{sec:mva}}. The categorisation of the SRs as well as the discriminants in \SRLo and \SRVBF remain unchanged. This approach allows the same acceptance uncertainties to be maintained across analysis regions. In addition, while both $ZZ$ and $ZH$ processes have non-negligible expected yields in the \ZHF CR, such contributions are kept fixed to their SM prediction so that the $\mu_{ZZ}$ and $\mu_{ZH}$ POIs only scale the expected predictions in the \hadhad and \lephad final states. Finally, in both fits, the process that is not treated as signal is fixed to its SM prediction including the full set of uncertainties described in \Sect{\ref{subsec:sigmod}}, and the same approach is used for the $HH$ process.
%


\section{Results}

\label{sec:result}

\begin{figure}[ht!b]
\centering
\subfloat[]{
\includegraphics[width=0.4\textwidth]{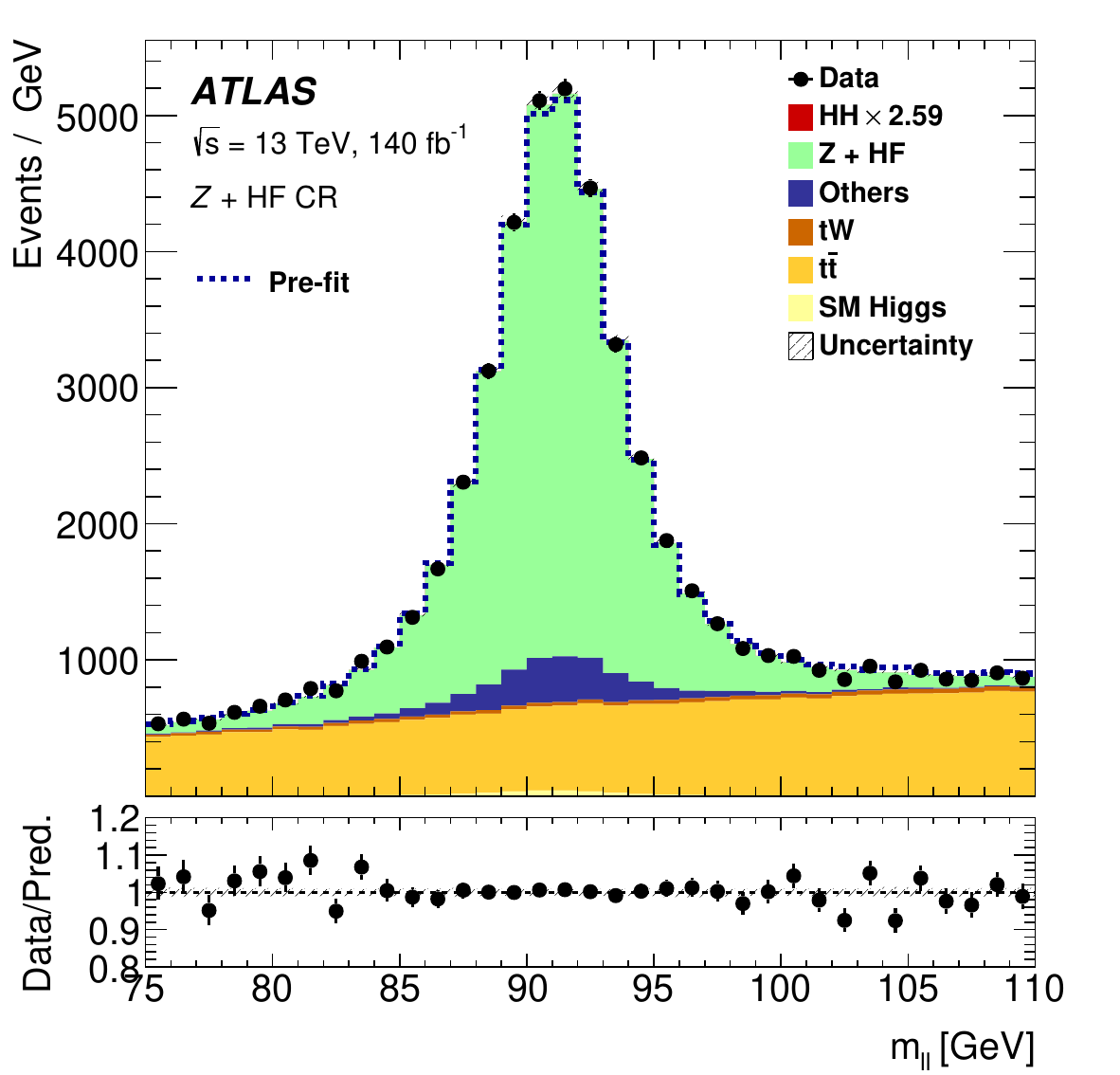}
\label{fig:CR_a}
}
\subfloat[]{
\includegraphics[width=0.4\textwidth]{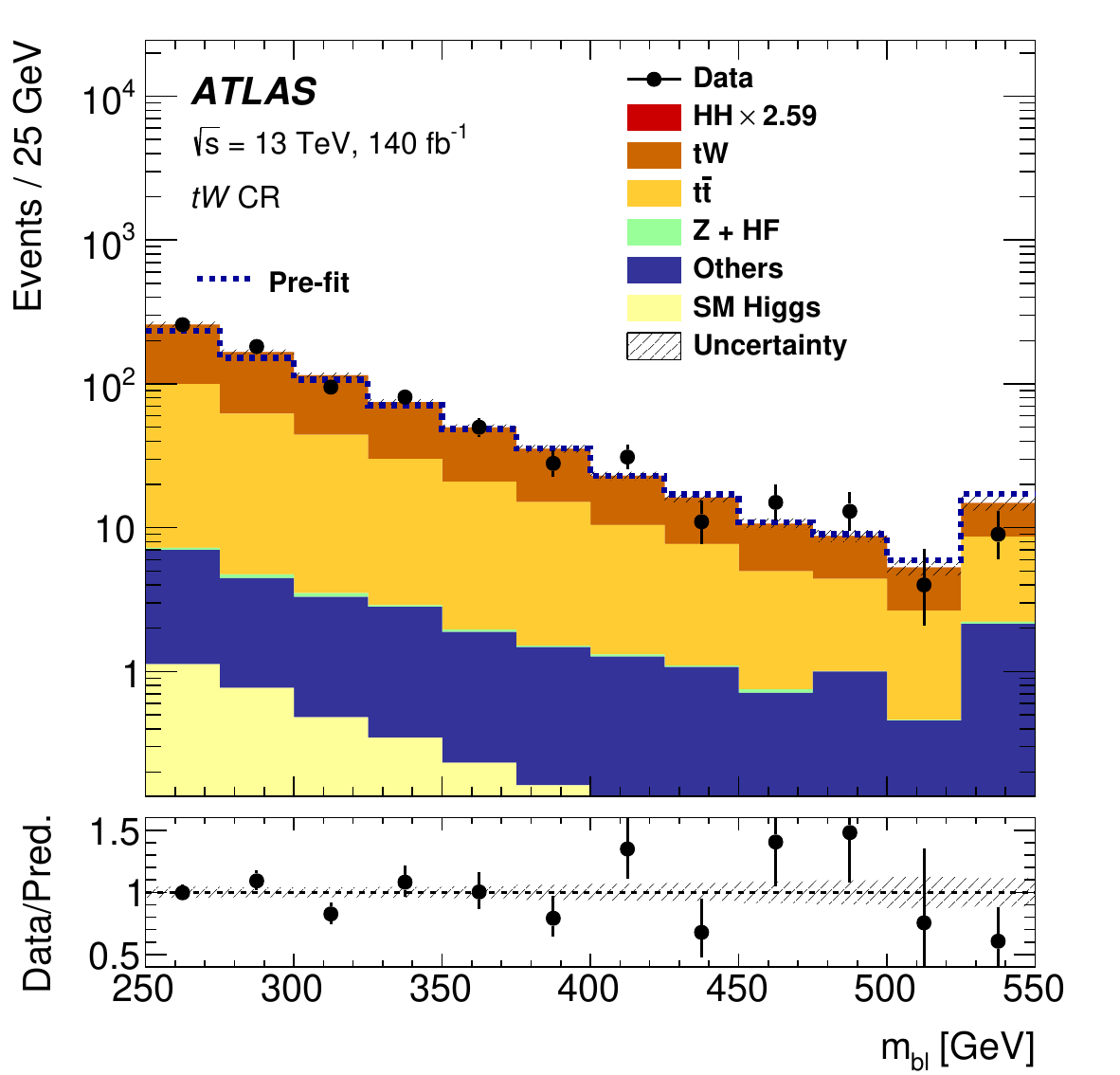}
\label{fig:CR_b}
} \\
\subfloat[]{
\includegraphics[width=0.4\textwidth]{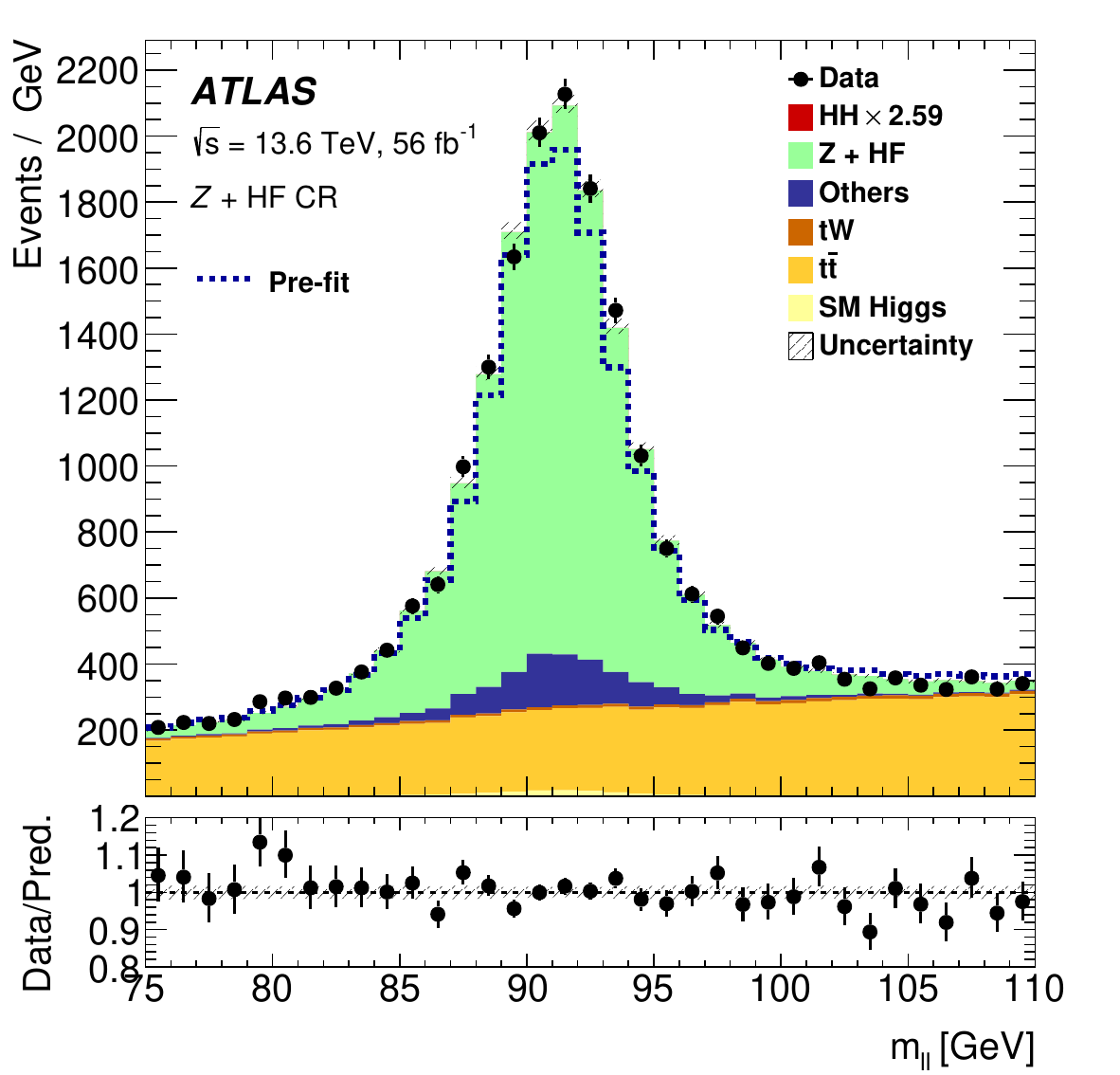}
\label{fig:CR_c}
}
\subfloat[]{
\includegraphics[width=0.4\textwidth]{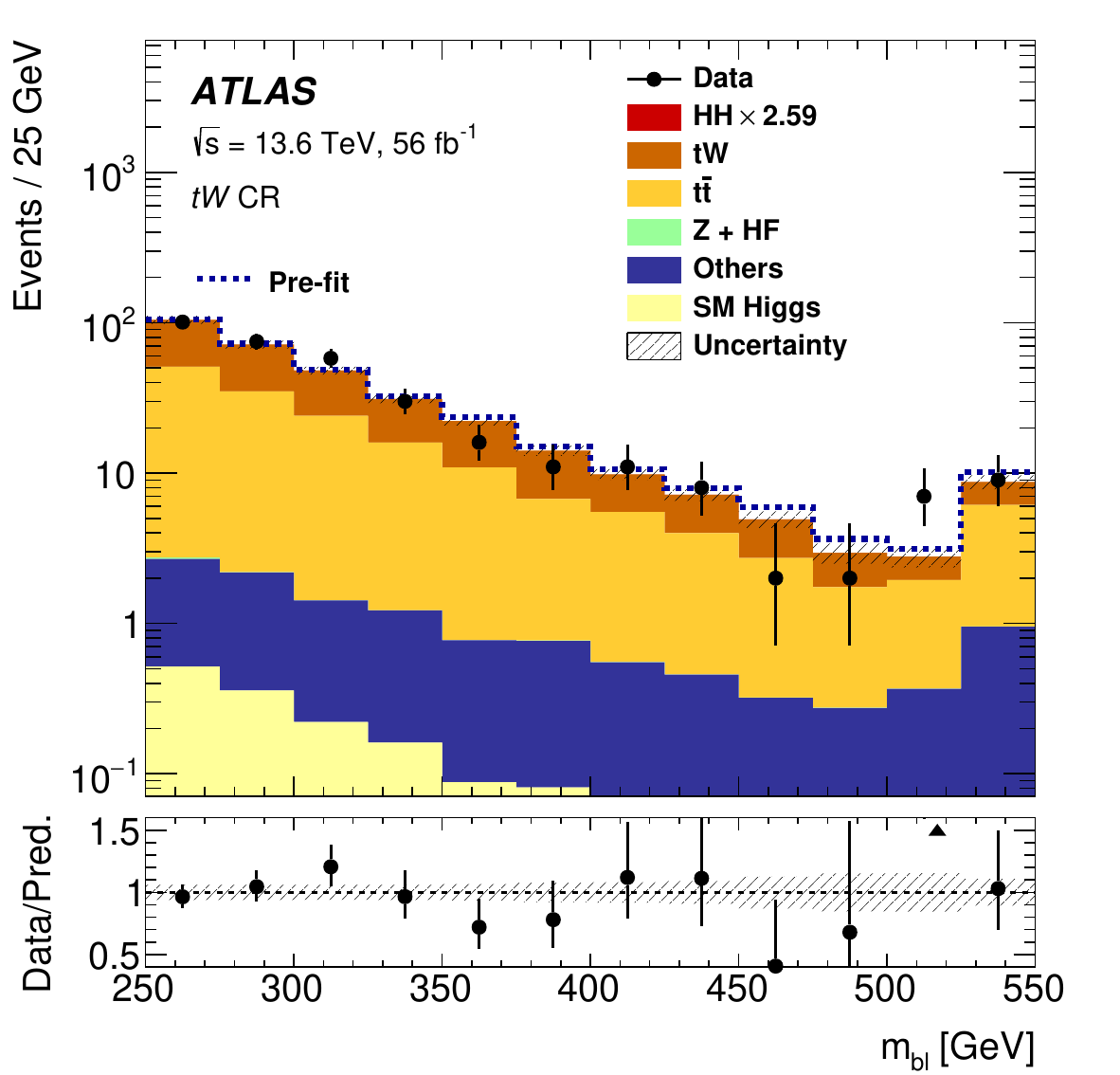}
\label{fig:CR_d}
} \\
\caption[]{Post-fit distributions of fit variables in the $\ZHF$ CR (left) and $tW$ CR (right) for Run~2 (top) and Run~3 (bottom). The background contributions after the fit are shown as filled histograms. The size of the combined statistical and systematic uncertainties for the sum of the fitted signal and background are indicated by the hatched band. The ratios of the data to the sum of the fitted signal and background are shown in the lower panel. For the $tW$ CR plots, the overflow events are included in the last displayed bin.
}
\label{fig:CR}
\end{figure}

The post-fit normalisation factors of the unconstrained backgrounds in the $\mu_{HH}$ fit are shown in Table~\ref{tab:NF}.\footnote{The normalisation factors do not reflect the overall post-fit normalisation of the backgrounds in the CRs and SRs due to normalisation effects of systematic uncertainties that can be modified in the likelihood fit.} Good agreement is observed in the values for these background processes between Run~2 and Run~3. Figure~\ref{fig:CR} shows the fitted distribution in the CRs while the data/MC agreement for the classifier scores in all the SRs is reported in Figure~\ref{fig:postfit1} for Run~2 and in Figure~\ref{fig:postfit2} for Run~3. The goodness-of-fit evaluated using a saturated model~\cite{Baker:1983tu,Saturated} yields a probability of 6.1\%, with this modest compatibility driven primarily by the test statistic contributions from the \SRLo regions.

\begin{table}[htb!]
\caption{The normalisation factors (NF) applied to the key background processes in the analysis as obtained from a fit to data.
The errors represent the combined statistical and systematic uncertainties.}
\centering
\scalebox{0.95}{
\begin{tabular}{c  c  c  c }
\toprule
& NF$_{\ZHF}$      & NF$_{\ttbar}$     &  NF$_{tW}$ \\
\midrule
Run ~2		  & 1.11 $\pm$ 0.08  & 0.97 $\pm$ 0.04  & 1.34 $\pm$ 0.16 \\
Run ~3		  & 1.21 $\pm$ 0.11  & 0.96 $\pm$ 0.05  & 1.05 $\pm$ 0.19 \\
\bottomrule
\end{tabular}
}
\label{tab:NF}
\end{table}

\begin{figure}[ht!b]
\centering
\subfloat[]{
\includegraphics[width=0.45\textwidth]{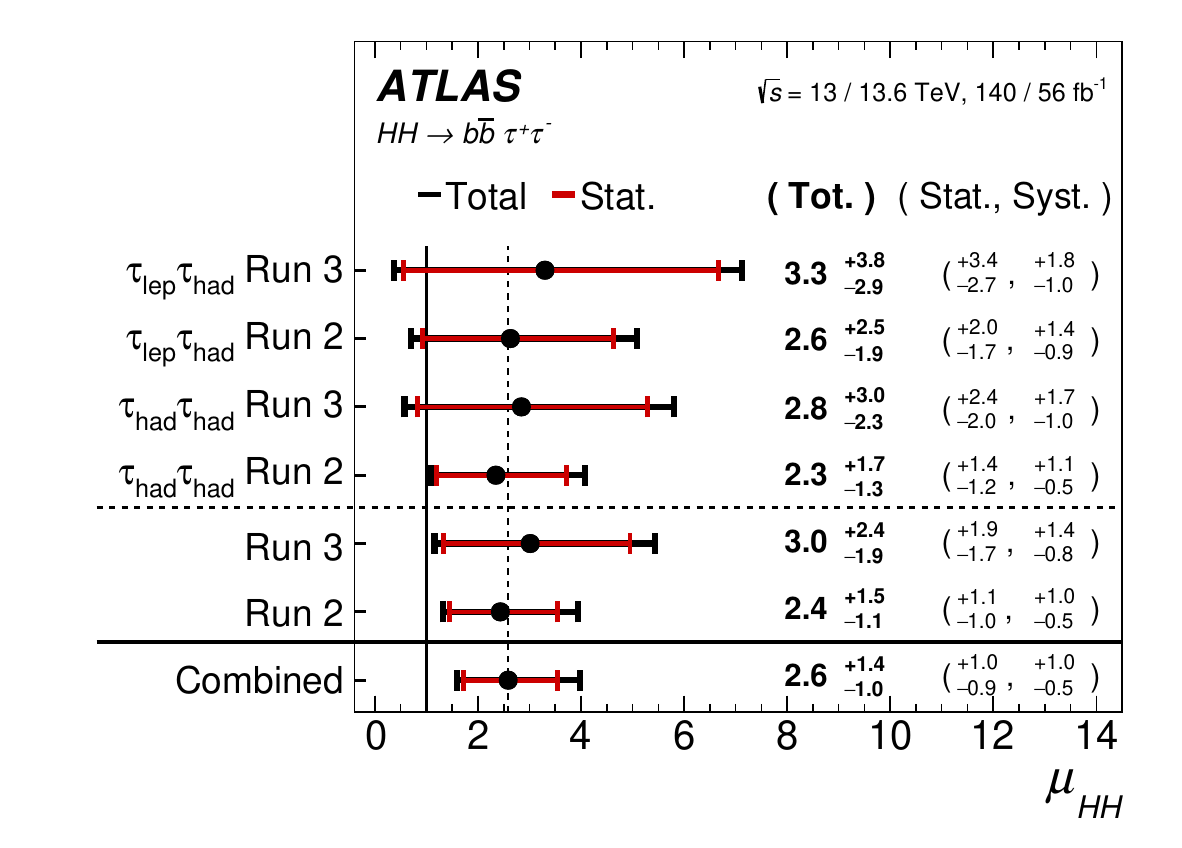}
\label{fig:multimu}
}
\subfloat[]{
\includegraphics[width=0.45\textwidth]{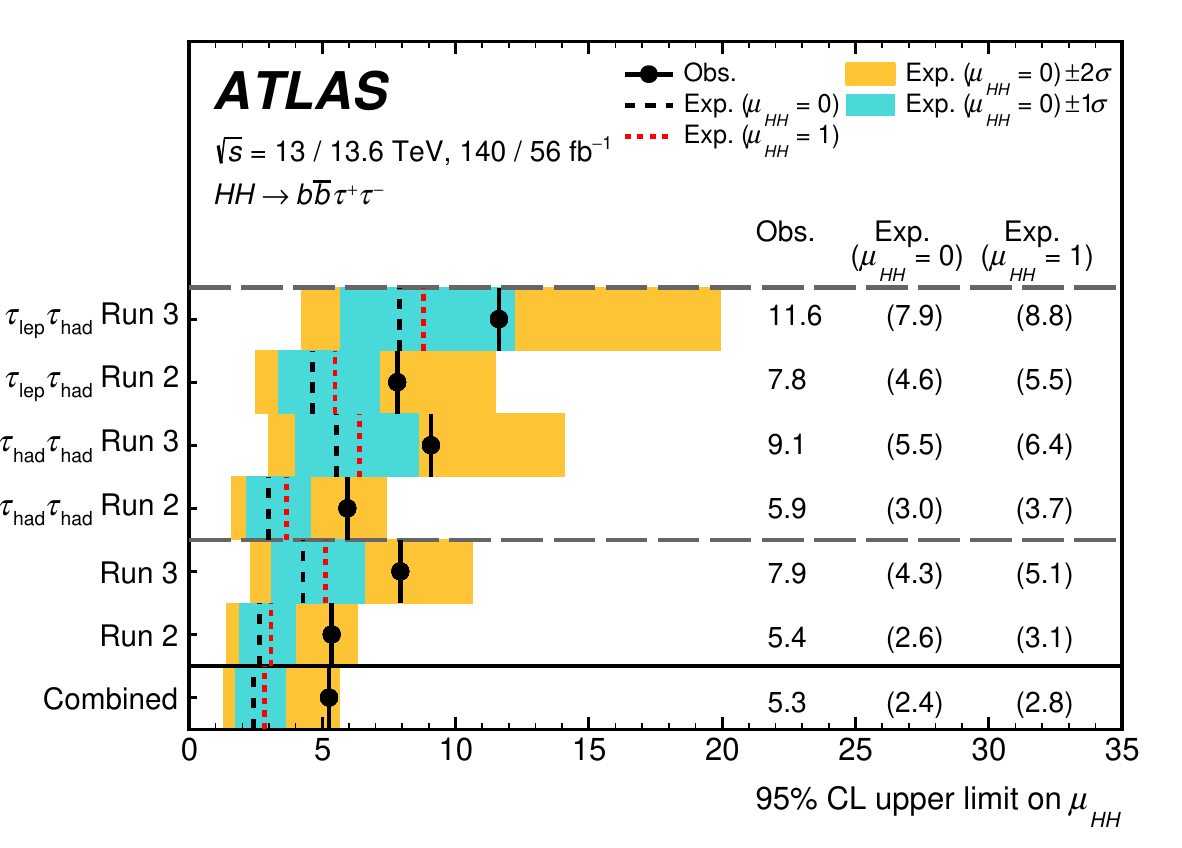}
\label{fig:limitSummary}
}
\caption{\protect\subref{fig:multimu} Fitted values of the signal strengths for the $HH$ process from the combined fit to Run~2 and Run~3 data. The values for the individual signal regions and for Run~2 and Run~3 are obtained from a simultaneous fit in which the signal strength for each region and run period floats independently. The solid vertical line represents the SM prediction, and the dashed vertical line indicates the observed best-fit value from the final likelihood fit. \protect\subref{fig:limitSummary} Summary of observed (filled circles) and expected (short vertical dashes) 95\% CL upper limits on $\mu_{HH}$ from the fit to each individual channel and from the combined fit, under the background-only ($\mu_{HH}=0$) hypothesis. A set of dotted lines indicates the expected 95\%~CL upper limits on $\mu_{HH}$ under the SM hypothesis ($\mu_{HH}=1$). The inner and outer bands indicate the $\pm 1\sigma$ and $\pm 2\sigma$ variations of the expected limit under the background-only hypothesis.
}
\end{figure}

\begin{figure}[ht!b]
\centering
\includegraphics[width=0.5\textwidth]{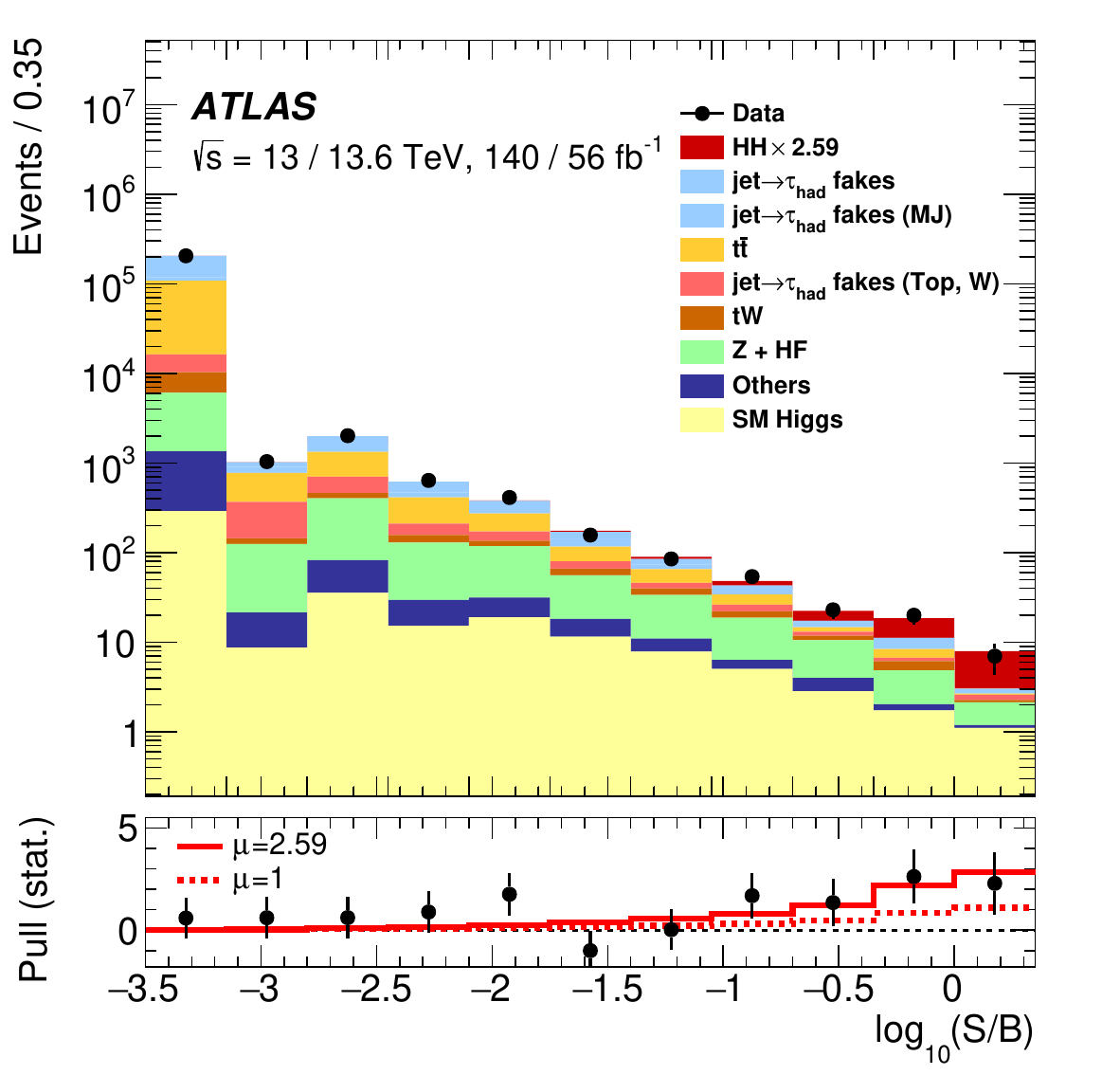}
\caption{Event yields as a function of $\log_{10}(S/B)$ for data, background, and signal, for all channels and run periods combined. Final-discriminant bins in the signal regions are combined according to their $\log_{10}(S/B)$ values, where $S$ and $B$ denote the fitted signal and background yields, respectively. Signal contributions in the upper panel are shown after rescaling the SM signal cross-section by the signal strength extracted from data. The lower panel shows the pull of the data relative to the fitted background, defined as the statistical significance of the difference between the data and the fitted background, with statistical uncertainties only. In the lower panel, the solid (dashed) line represents the pull expected from the sum of the fitted background and signal scaled to the fitted (SM) signal strength.}
\label{fig:HH_SoverB}
\end{figure}

The measured signal strength for the $\HH\to \bbtt$ is:
\begin{align*}
\mu_{HH} 	& = \muHHRes^{+ 1.4}_{- 1.0} = \muHHRes ^{+1.0}_{-0.9}\:(\textrm{stat.})^{+1.0}_{-0.5}\:(\textrm{syst.}).
\end{align*}
The result has an observed (expected) significance of \hhSigObs~(\hhSigExp) standard deviations with respect to the background-only hypothesis and a 1.65 standard deviation excess over the SM prediction. Compatible signal strength values are extracted in both the \hadhad and \lephad channels for both Run~2 and Run~3, as shown in Figure~\ref{fig:multimu}. The \hadhad channel provides greater sensitivity than the \lephad channel, while Run~2 SRs are more sensitive than the Run~3 ones due to the larger integrated luminosity. The larger-than-one observed signal strength is primarily driven by an excess of data in the most sensitive bin of the classifier score in \SRHi of both channels. Figure~\ref{fig:HH_SoverB} shows the data, background $B$ and signal $S$ yields, where the bins of the classifier score in SRs are combined into bins of $\log_{10} (S/B)$, with $S$ being the fitted $HH$ contribution and $B$ the fitted background yields.

\newlength{\myfigwidth}
\setlength{\myfigwidth}{0.40\textwidth} %

\begin{figure}[htb!]
\centering
\subfloat[]{
\includegraphics[width=\myfigwidth]{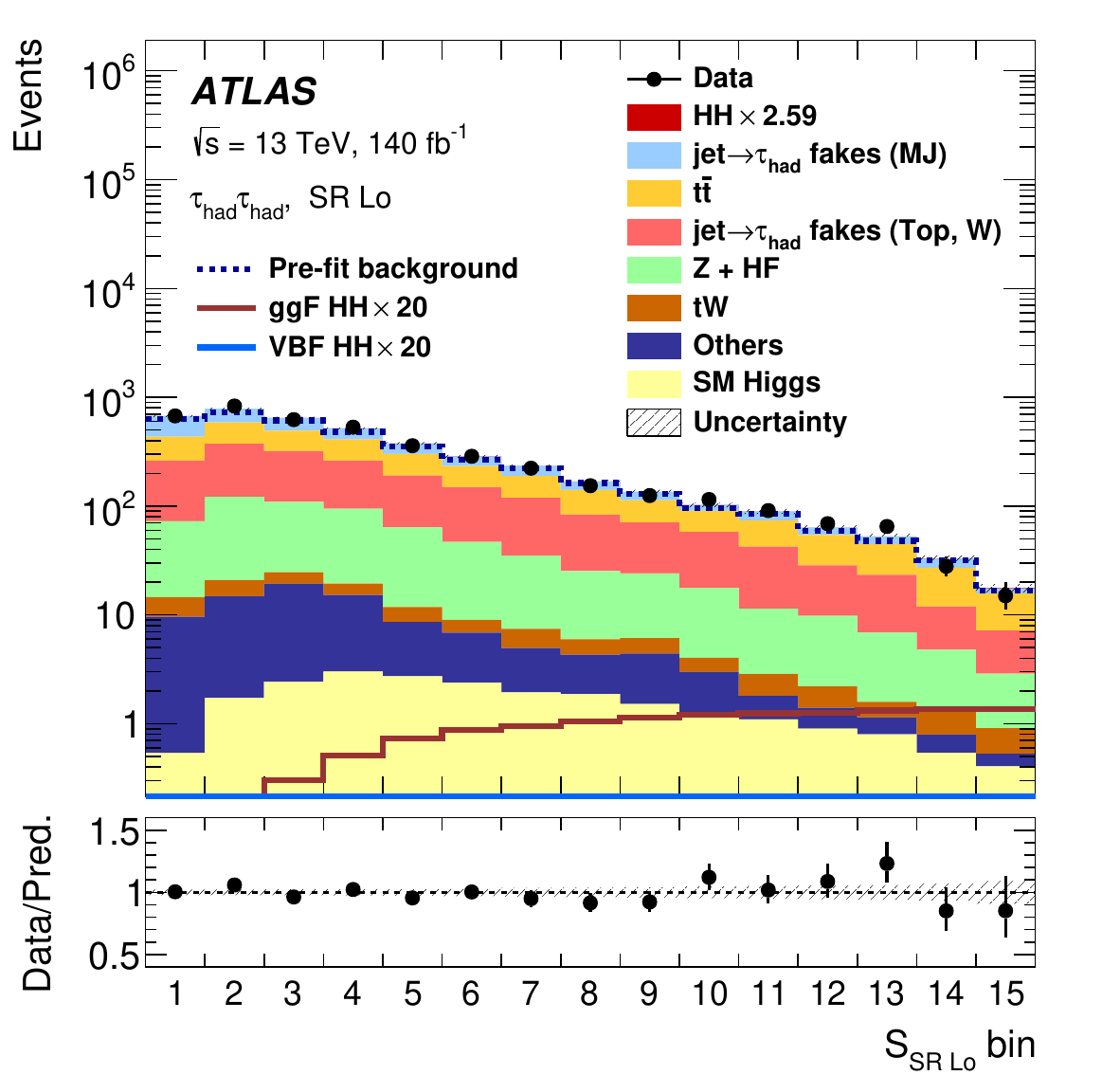}
\label{fig:BDT_1a}
\vspace{-15pt}
}
\subfloat[]{
\includegraphics[width=\myfigwidth]{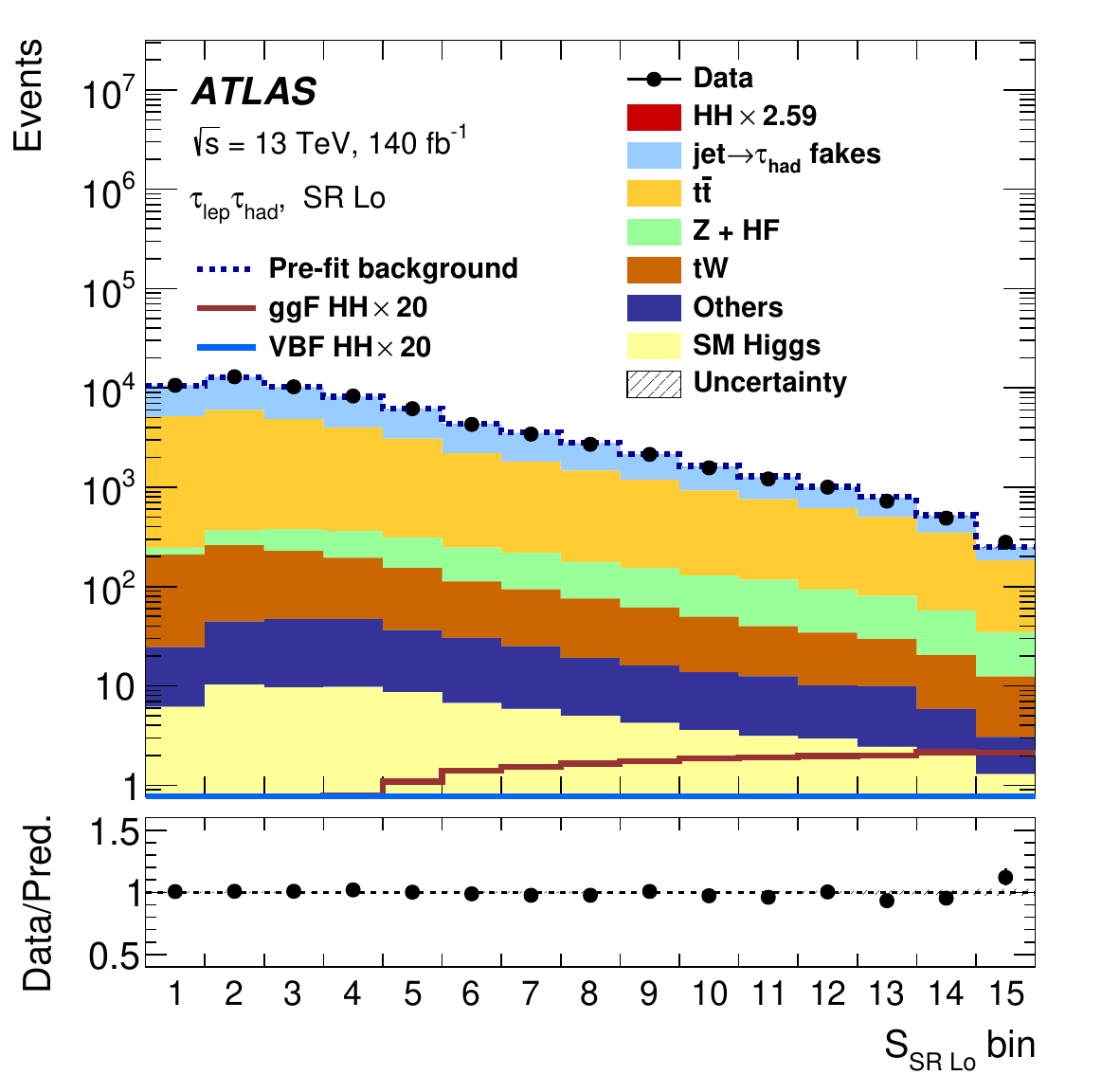}
\label{fig:BDT_1b}
} \\ \vspace{-0.9em}%

\subfloat[]{
\includegraphics[width=\myfigwidth]{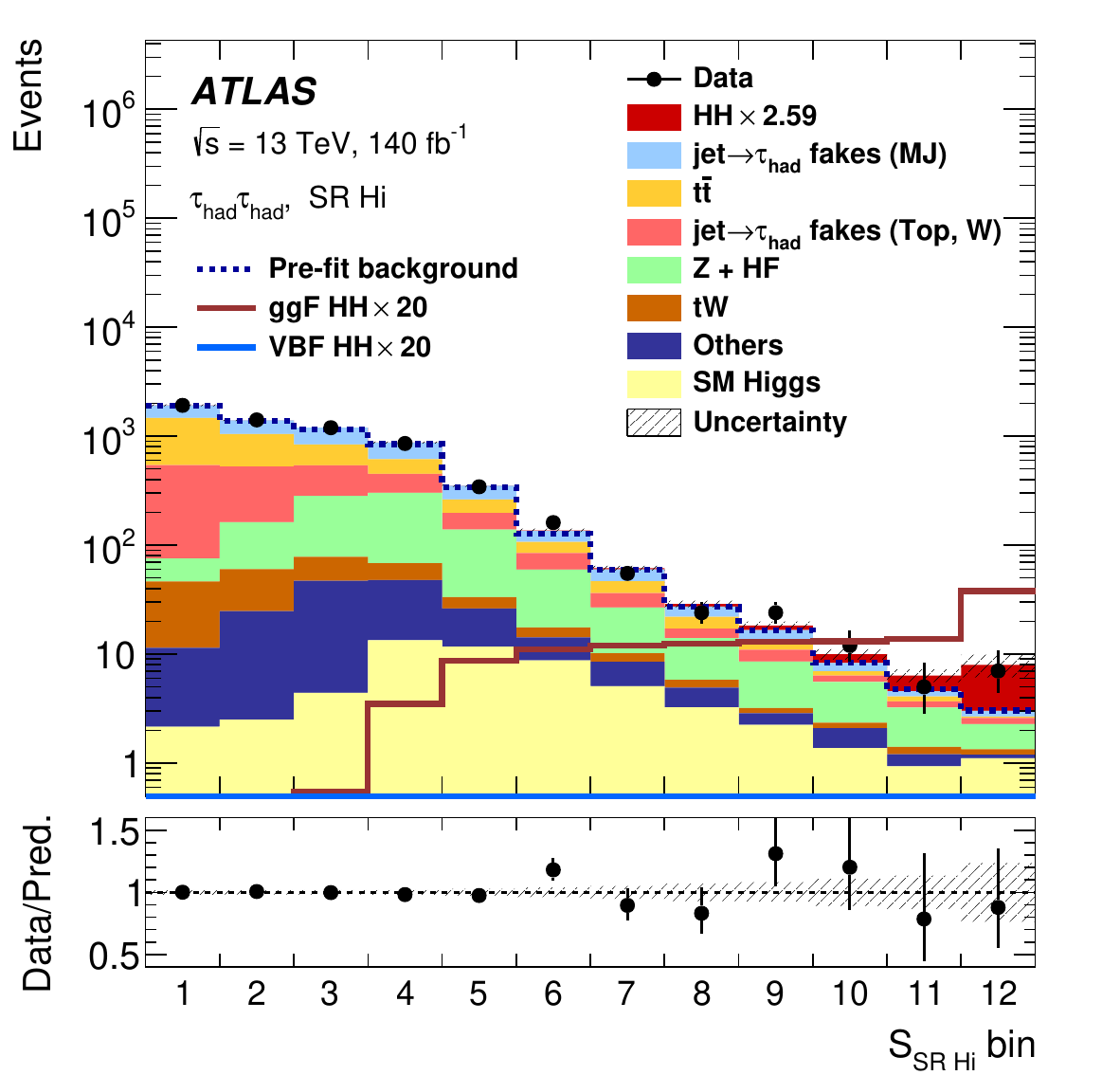}
\label{fig:BDT_1c}
}
\subfloat[]{
\includegraphics[width=\myfigwidth]{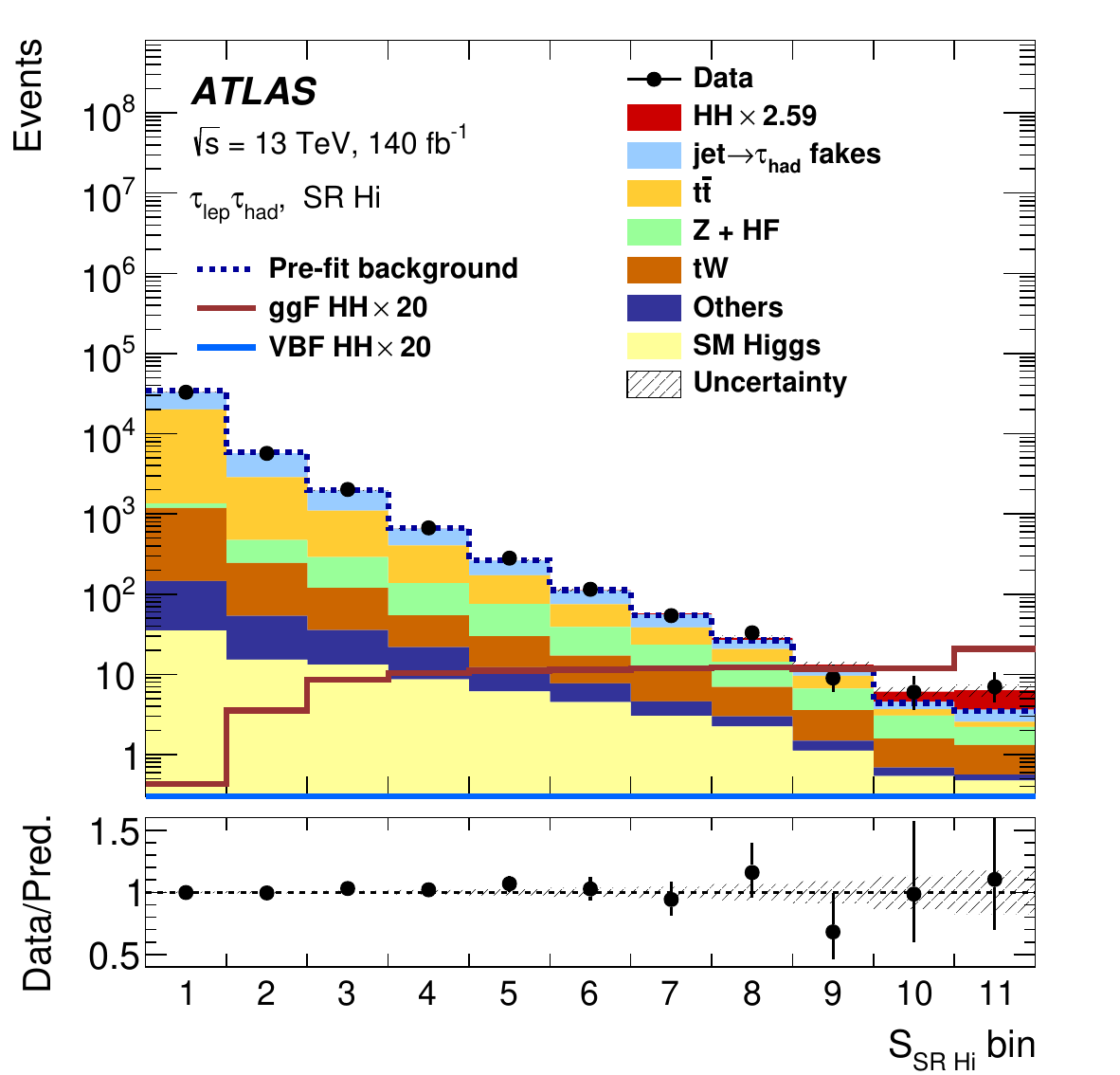}
\label{fig:BDT_1d}
}  \vspace{-0.9em}%

\subfloat[]{
\includegraphics[width=\myfigwidth]{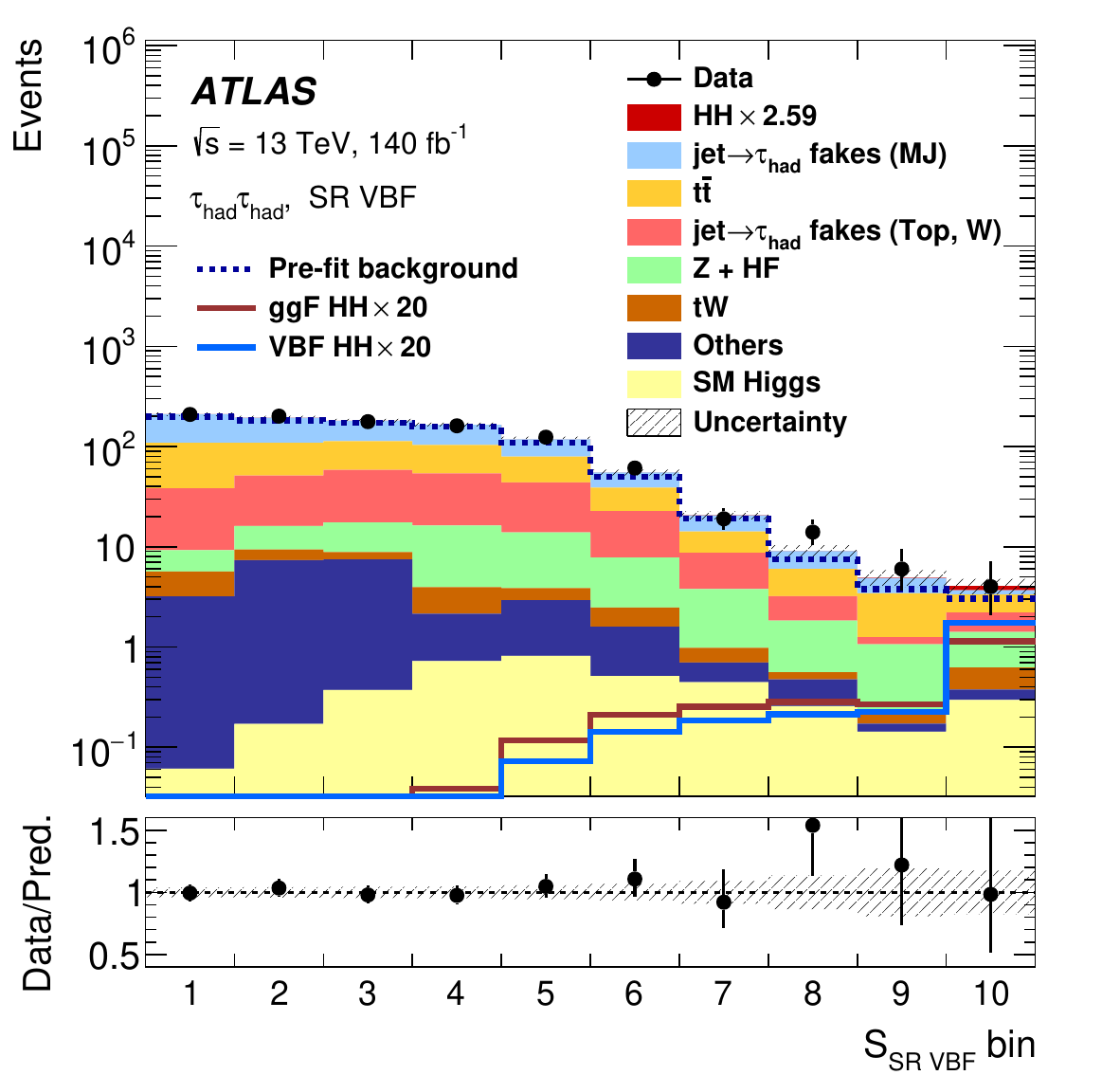}
\label{fig:BDT_1e}
}
\subfloat[]{
\includegraphics[width=\myfigwidth]{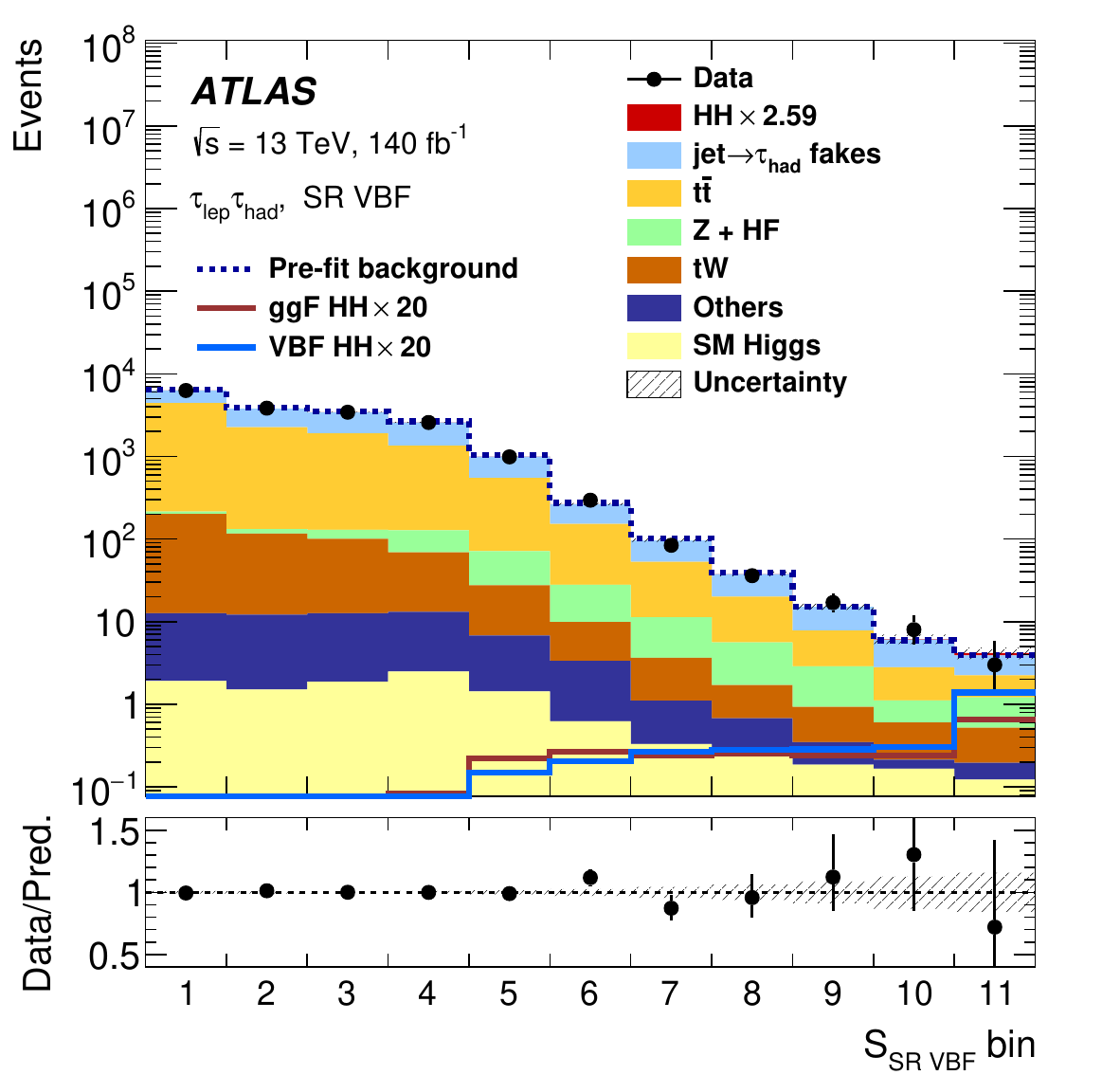}
\label{fig:BDT_1f}
}\\ \vspace{-0.3em}
\caption[]{Post-fit classifier score distributions for the Run~2 \hadhad SRs (left column), and the Run~2 \lephad SRs (right column). Rows from top to bottom show the \SRLo, \SRHi and \SRVBF regions. The background contributions are shown as filled histograms. The $HH$ signal is shown both stacked (normalised to the extracted $\mu_{HH}$) and unstacked (unfilled, scaled by a factor of 20). The hatched band indicates the combined statistical and systematic uncertainty on the signal plus background prediction, with data-to-prediction ratios shown in the lower panels.
}
\label{fig:postfit1}
\end{figure}

\begin{figure}[htb!]
\centering
\subfloat[]{
\includegraphics[width=\myfigwidth]{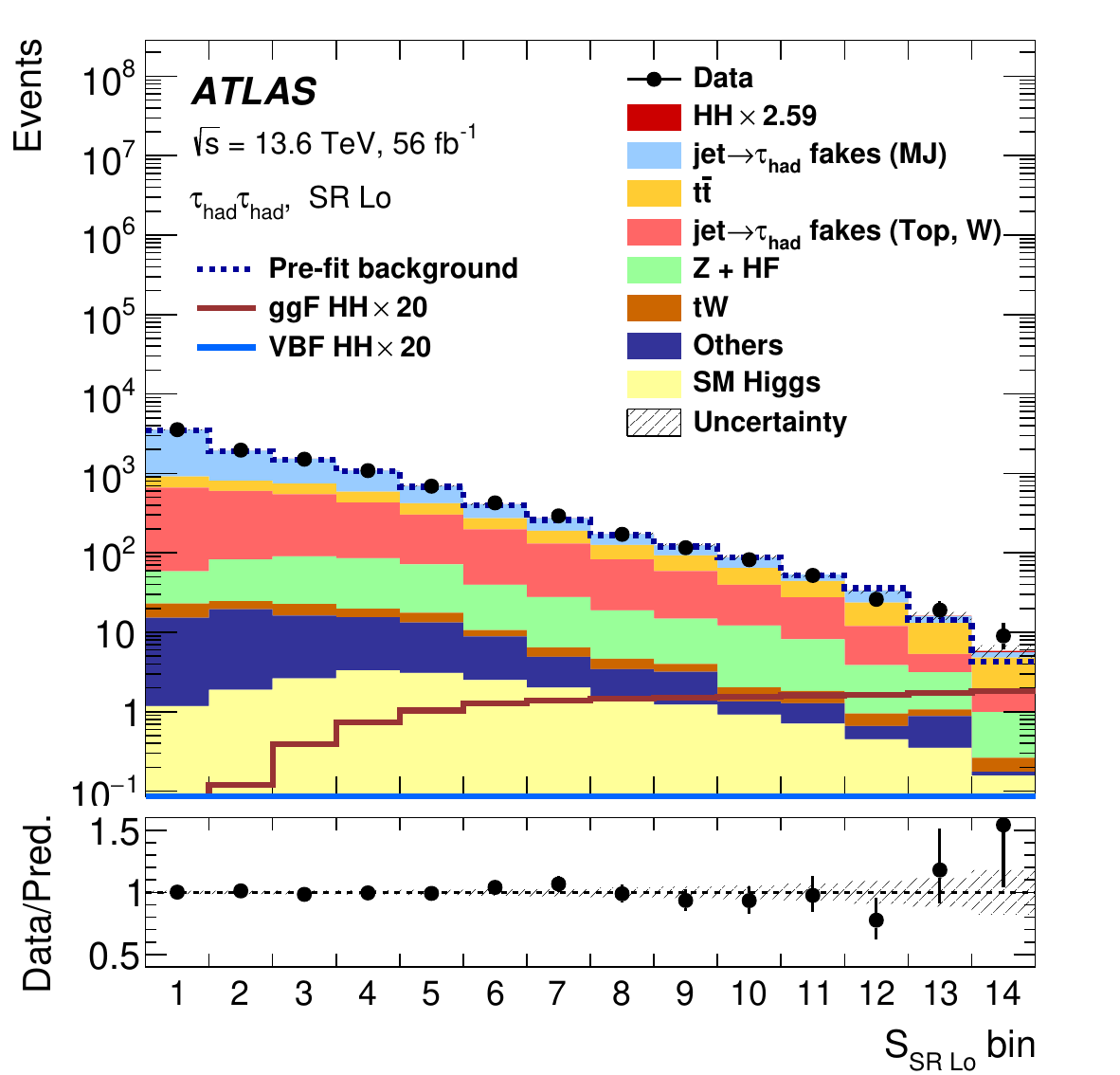}
\label{fig:BDT_2a}
\vspace{-15pt}
}
\subfloat[]{
\includegraphics[width=\myfigwidth]{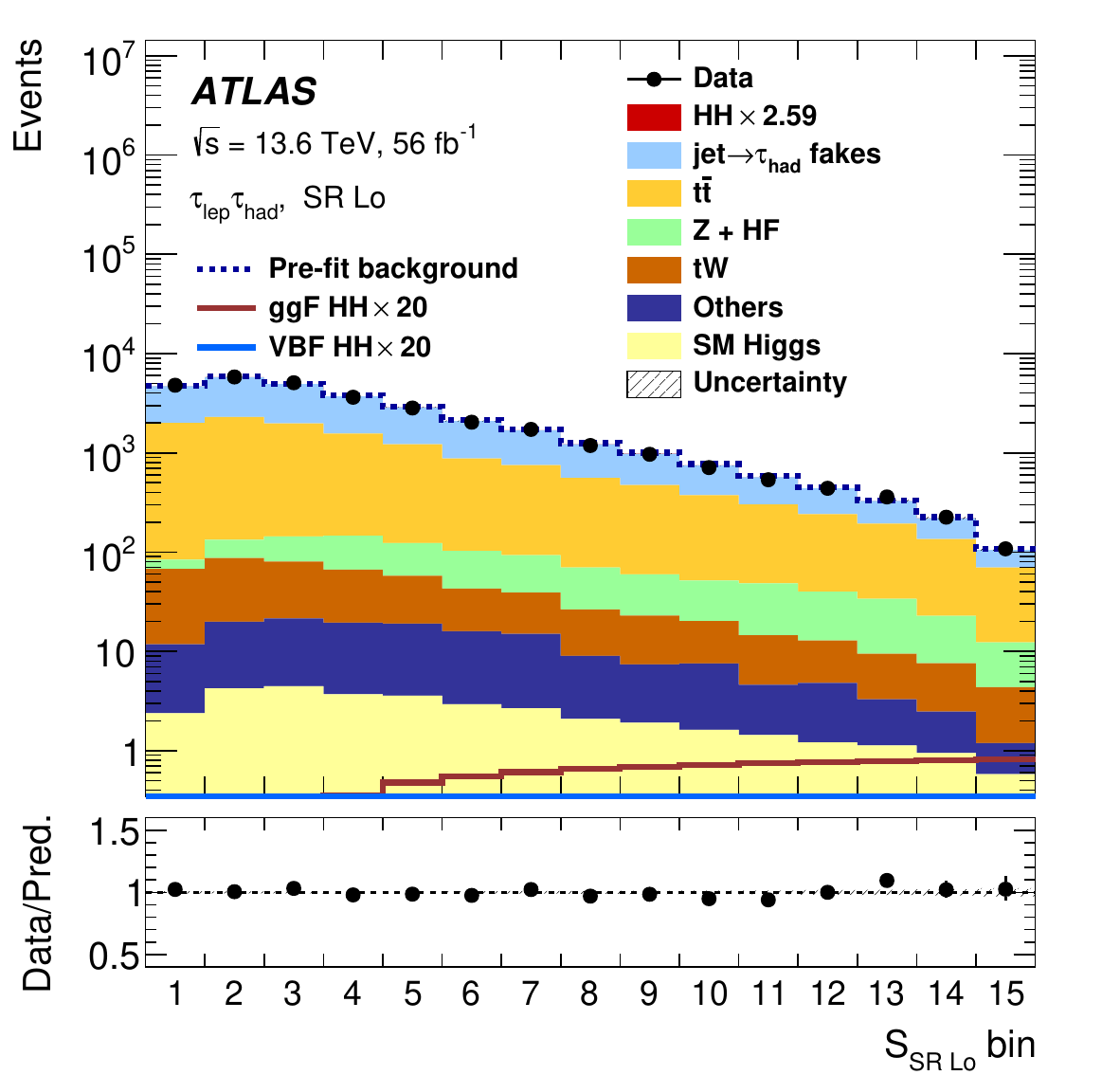}
\label{fig:BDT_2b}
} \\ \vspace{-0.9em}%

\subfloat[]{
\includegraphics[width=\myfigwidth]{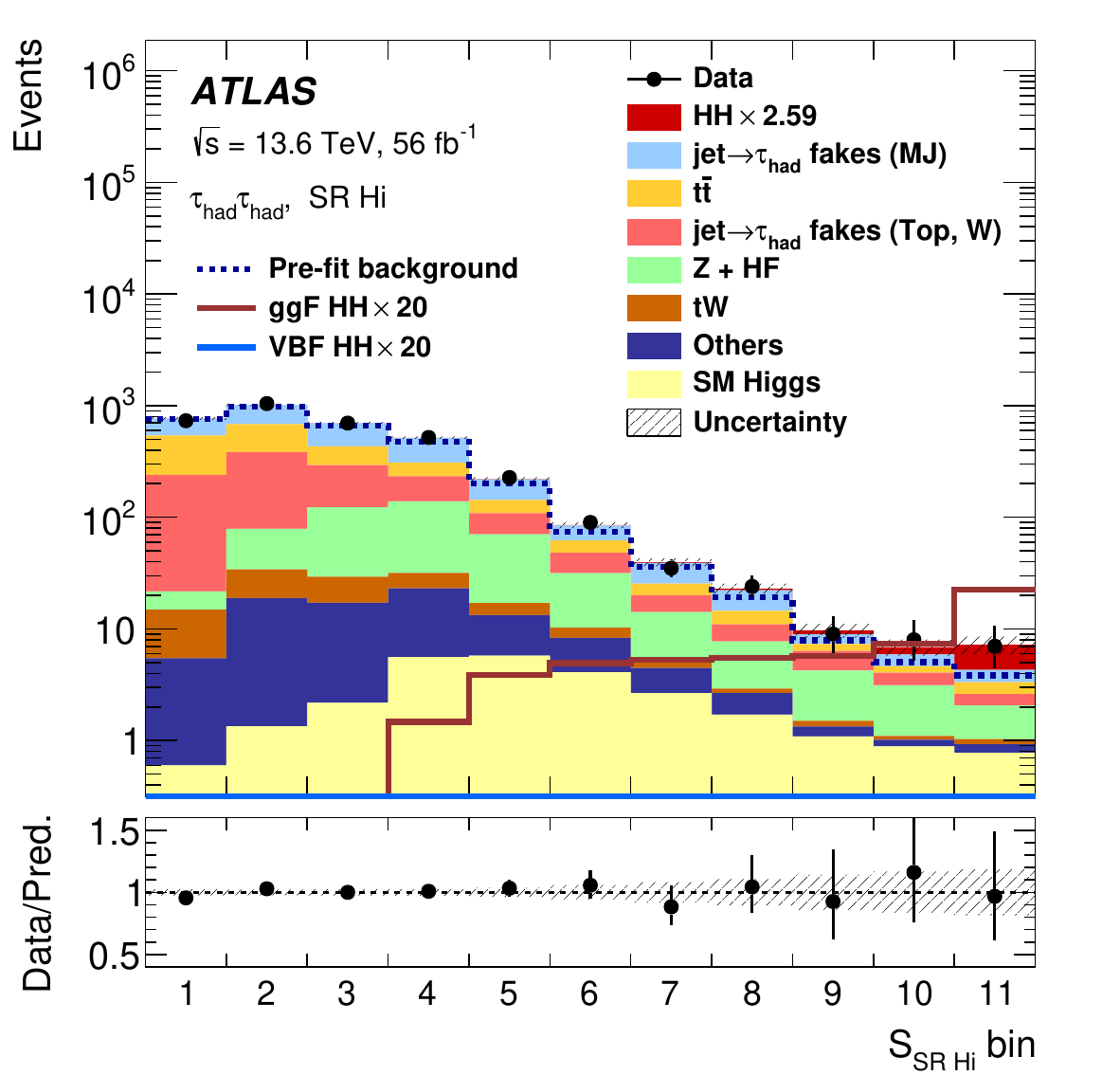}
\label{fig:BDT_2c}
}
\subfloat[]{
\includegraphics[width=\myfigwidth]{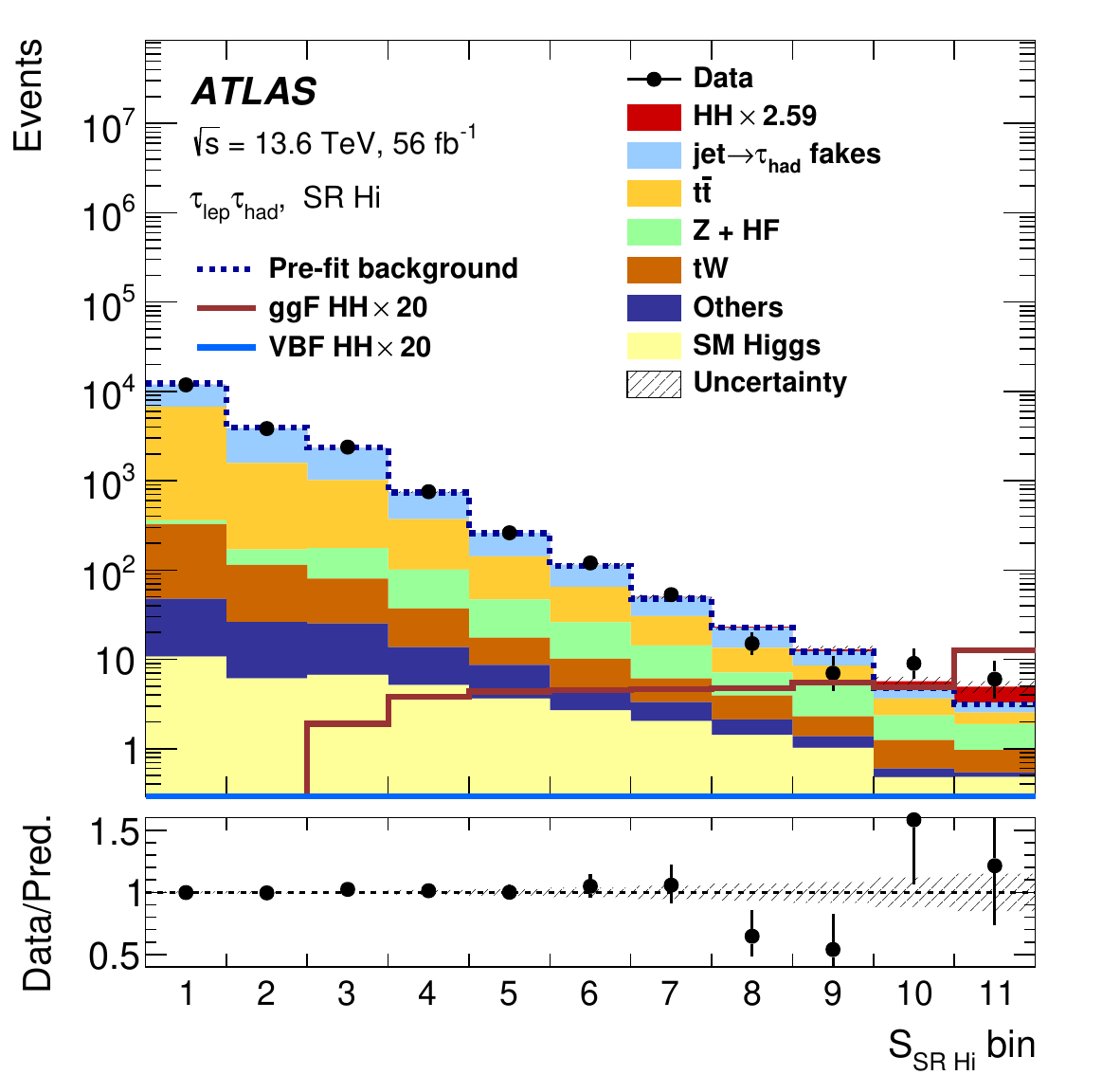}
\label{fig:BDT_2d}
}  \vspace{-0.9em}%

\subfloat[]{
\includegraphics[width=\myfigwidth]{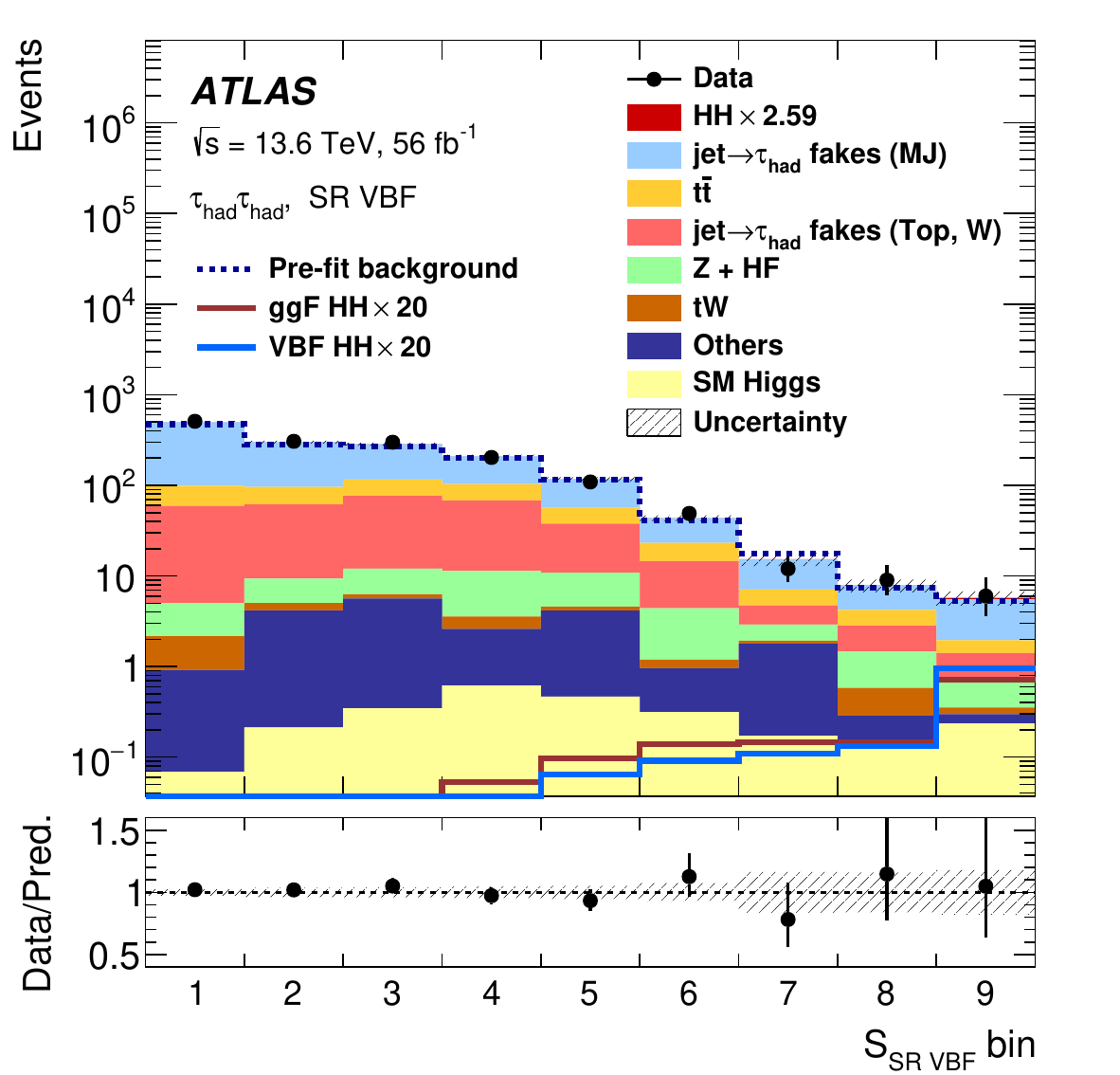}
\label{fig:BDT_2e}
}
\subfloat[]{
\includegraphics[width=\myfigwidth]{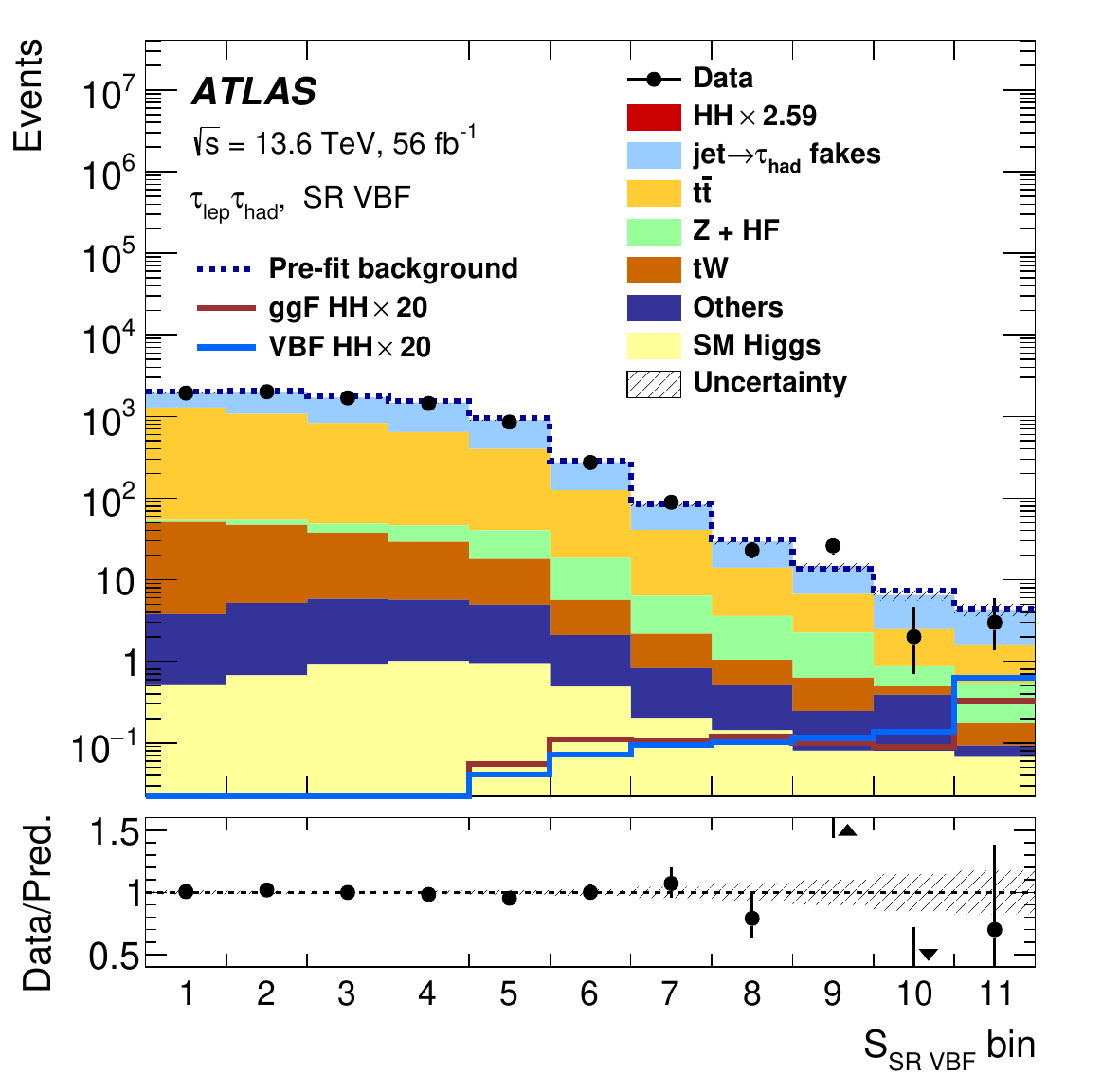}
\label{fig:BDT_2f}
}\\ \vspace{-0.3em}
\caption[]{Post-fit classifier score distributions for the Run~3 \hadhad SRs (left column), and the Run~3 \lephad SRs (right column). Rows from top to bottom show the \SRLo, \SRHi and \SRVBF regions. The background contributions are shown as filled histograms. The $HH$ signal is shown both stacked (normalised to the extracted $\mu_{HH}$) and unstacked (unfilled, scaled by a factor of 20). The hatched band indicates the combined statistical and systematic uncertainty on the signal plus background prediction, with data-to-prediction ratios shown in the lower panels.
}
\label{fig:postfit2}
\end{figure}

The signal strength upper limits are summarised in Table~\ref{tab:limits} and Figure~\ref{fig:limitSummary}, by channel and data-taking campaign along with the combined fit.
The observed 95\% CL upper limit on $\mu_{\HH}$ is \hhLimObs, to be compared with an expected limit of \hhLimExp~for the background-only hypothesis ($\mu_{HH}=0$).
The expected limit is obtained using an Asimov dataset where the background nuisance parameters are obtained from a conditional $\mu_{HH}=0$ fit to data. As mentioned above, this hypothesis is disfavoured by 2.6 standard deviations. In the presence of a data excess, conditioning $\mu_{HH}=0$ in the fit will lead to pulls that increase the background yields in the most sensitive bins to partially compensate for this excess, resulting in a larger expected limit compared to the pre-fit expectation of $\mu_{HH}<1.9$.

\FloatBarrier

From the simultaneous fit of $\mu_{\mathrm{ggF}}$ and $\mu_{\mathrm{VBF}}$, observed (expected) 95\% CL upper limits are $\mu_{\mathrm{ggF}}<5.5~(2.5)$ and $\mu_{\mathrm{VBF}}<41~(39)$, respectively, assuming that the signal strength parameters can vary independently for each production mode.
When $\mu_{\mathrm{VBF}}$ is fixed to the SM prediction, the observed (expected) 95\% CL upper limit is $\mu_{\mathrm{ggF}}<5.4~(2.4)$.
Similarly, if $\mu_{\mathrm{ggF}}$ is fixed to the SM prediction, the observed (expected) 95\% CL upper limit is $\mu_{\mathrm{VBF}}<43~(39)$.
These results confirm the capability of the analysis to determine the two processes independently as the upper limits are minimally affected by the free-floating normalisation of the other production mode.

\begin{table}[thb]
\footnotesize
\setlength
\extrarowheight{2.55pt}
\centering
\caption{Observed and expected 95\% CL upper limits on $\mu_{HH}$, $\mu_{\mathrm{ggF}}$ and $\mu_{\mathrm{VBF}}$ from the individual SR likelihood fits as well as the combined results, under the background-only hypothesis ($\mu_{HH}=0$).
The $\mu_{\mathrm{ggF}}$ and $\mu_{\mathrm{VBF}}$ limits are quoted both from the results of the simultaneous fit of both signal strengths (central column), and from independent fits for the individual production modes, assuming the other to be as predicted by the SM.
The uncertainties quoted on the combined expected upper limits correspond to the $1\sigma$ uncertainty band. \protect\label{tab:limits} }
\begin{tabular}{ c c  c  cc  cc }
\toprule
& & $\mu_{HH}$ & $\mu_{\mathrm{ggF}}$ & $\mu_{\mathrm{VBF}}$ & $\mu_{\mathrm{ggF}}$ ($\mu_{\mathrm{VBF}}$=1) & $\mu_{\mathrm{VBF}}$ ($\mu_{\mathrm{ggF}}$=1)\\
\midrule
\multirow{2}{*}{\lephad Run~3}  & observed & 11.6 & 12.5 & 135 & 12.1 & 143 \\[0.4pt]
& expected &  7.9 &  8.1 & 176 &  8.0 & 169 \\[0.4pt]
\multirow{2}{*}{\lephad Run~2}  & observed & 7.8 & 8.2 & 69 & 8.1 & 73 \\[0.4pt]
& expected & 4.6 & 4.8 & 76 & 4.7 & 74 \\[0.4pt]
\multirow{2}{*}{\hadhad Run~3} & observed & 9.1 & 9.2 & 134 &  9.2 & 140 \\[0.4pt]
& expected & 5.5 & 5.7 & 127 &  5.6 & 123 \\[0.4pt]
\multirow{2}{*}{\hadhad Run~2} & observed & 5.9 & 6.0 & 70 & 6.0 & 75 \\[0.4pt]
& expected & 3.0 & 3.1 & 61 & 3.0 & 58 \\[0.4pt]
\midrule

\multirow{2}{*}{Run~3} & observed & 7.9 & 8.4 &  89  & 8.1 & 96 \\[0.4pt]
& expected & 4.3 & 4.4 &  99  & 4.3 & 96 \\[0.4pt]
\multirow{2}{*}{Run~2} & observed & 5.4 & 5.5 & 46 & 5.5 & 51 \\[0.4pt]
& expected & 2.6 & 2.7 & 45 & 2.7 & 43 \\[0.4pt]

\midrule

\multirow{2}{*}{Combined} & observed & \hhLimObs 	         	    & 5.5 		              & 41			    & 5.4 			        & 43                     \\[1.5pt]
& expected & $\hhLimExp^{+1.2}_{-0.7}$        & $2.5^{+1.2}_{-0.7}$             & $39^{+18}_{-11}$            & $2.4^{+1.2}_{-0.7}$               & $39^{+17}_{-11}$ \\[1.5pt]
\bottomrule
\end{tabular}
\end{table}

The breakdown of the systematic uncertainties and their impact on the inclusive signal strength extraction is presented in Table~\ref{tab:uncert_breakdown}. The measurement precision is primarily limited by the statistical uncertainty on $\mu_{HH}$, resulting from the limited data sample size, which contributes 60\% to the total uncertainty.\footnote{The relative contributions are defined relative to the total uncertainty added in quadrature, $\sigma_i^2 / \sigma_{\mathrm{tot}}^2 $.} The leading systematic uncertainty in the measurement of $\mu_{HH}$ is the uncertainty in the ggF $\HH$ production cross-section arising from variations of the QCD scales and the top-quark mass scheme. The next leading sources are the uncertainty on the \ggHHF production, the uncertainty related to the interference between the $tW$ and \ttbar processes, and the statistical precision of the background templates in the most signal-like discriminant bins. Signal and background modelling uncertainties contribute 35\% to the total uncertainty, while experimental uncertainties contribute only 3\%.
The combined impact of all sources of systematic uncertainties leads to an increase in the expected upper limits on the signal strength $\mu_{HH}$ by 21\% with respect to the case in which systematic uncertainties are neglected (excluding the floating normalisations and background template statistics).

\begin{table}[htbp]
\caption{The breakdown of contributions to the uncertainty in the fitted value $\mu_{HH} = 2.6$ of the signal strength ($\sigma_{\mu_{HH}}$). The statistical uncertainty is the uncertainty of a fit where only the signal strength and the floating normalisations are free to vary, while all other nuisance parameters are fixed to their post-fit values. The impact of each other uncertainty group is evaluated by performing the combined fit with all nuisance parameters floating and a second fit in which the nuisance parameters belonging to the corresponding group are fixed to their post-fit values obtained from the combined fit. The contribution of the uncertainty group is then computed as the quadratic difference between the total uncertainty from the nominal combined fit and the uncertainty obtained from the fit with the group fixed. The sum in quadrature of uncertainties from different sources may differ from the total due to correlations. In cases where the upward and downward systematic variations have different values, the mean of the absolute values is shown.}
\label{tab:uncert_breakdown}
\centering
\small
\begin{tabular}{lr}
\\
\\
\toprule
Source of uncertainty & $\sigma_{\mu_{HH}}$ \\
\midrule
Total & 1.20 \\
\\
Statistical & 0.91 \\
Systematic & 0.76 \\
\\
Experimental uncertainties & 0.20 \\
\hspace{1.5em} Electron and muon & 0.01 \\
\hspace{1.5em} $\tau_{\mathrm{had}}$ & 0.13 \\
\hspace{1.5em} $E_{\mathrm{T}}^{\mathrm{miss}}$ & 0.03 \\
\hspace{1.5em} Jet & 0.14 \\
\hspace{1.5em} Flavour tagging & 0.04 \\
\hspace{1.5em} Pile-up & 0.01 \\
\hspace{1.5em} Luminosity & 0.03 \\
\\
Modelling uncertainties & 0.71 \\
\hspace{1.5em} Background template statistical & 0.21 \\
\hspace{1.5em} $HH$ & 0.54 \\
\hspace{1.5em} $ZH$ & 0.02 \\
\hspace{1.5em} $ZZ$ & $<$0.01 \\
\hspace{1.5em} Other SM Higgs & 0.25 \\
\hspace{1.5em} $Z$+jets & 0.07 \\
\hspace{1.5em} $tW$ & 0.21 \\
\hspace{1.5em} Top & 0.12 \\
\hspace{1.5em} jet $\rightarrow \tau_{\mathrm{had}}$ fakes & 0.07 \\
\hspace{1.5em} Others & 0.01 \\
\bottomrule
\end{tabular}
\end{table}

Compared to the previous analysis in the same final state~\cite{HDBS-2019-27}, the current result represents an improvement of the expected significance of 65\%. The largest single gain comes from the new machine learning classifier, which yields a 15--20\% boost depending on the region. Upgrades to the flavour-tagging algorithm increase signal efficiency at a comparable misidentification rate; since the primary backgrounds also contain $b$-jets, this enhances overall sensitivity by about 10\%. Similarly, removing the eVeto algorithm in the \hadhad channel provides an approximately 5\% gain in that channel. Finally, incorporating the Run~3 dataset expands sensitivity by an additional 15\%. The remaining improvement in sensitivity arises from the combined effects of the individual changes and other smaller improvements in the event selection and background modelling. For $\mu_{\mathrm{VBF}}$, the above-mentioned improvements lead to a reduction of the expected upper limit by 45\%. In particular, the new triggers in Run~3 contribute to a 20\% increase in the precision of $\mu_{\mathrm{VBF}}$ for that data-taking period.

\begin{figure}[th!b]
\centering
\subfloat[]{
\includegraphics[width=0.45\textwidth]{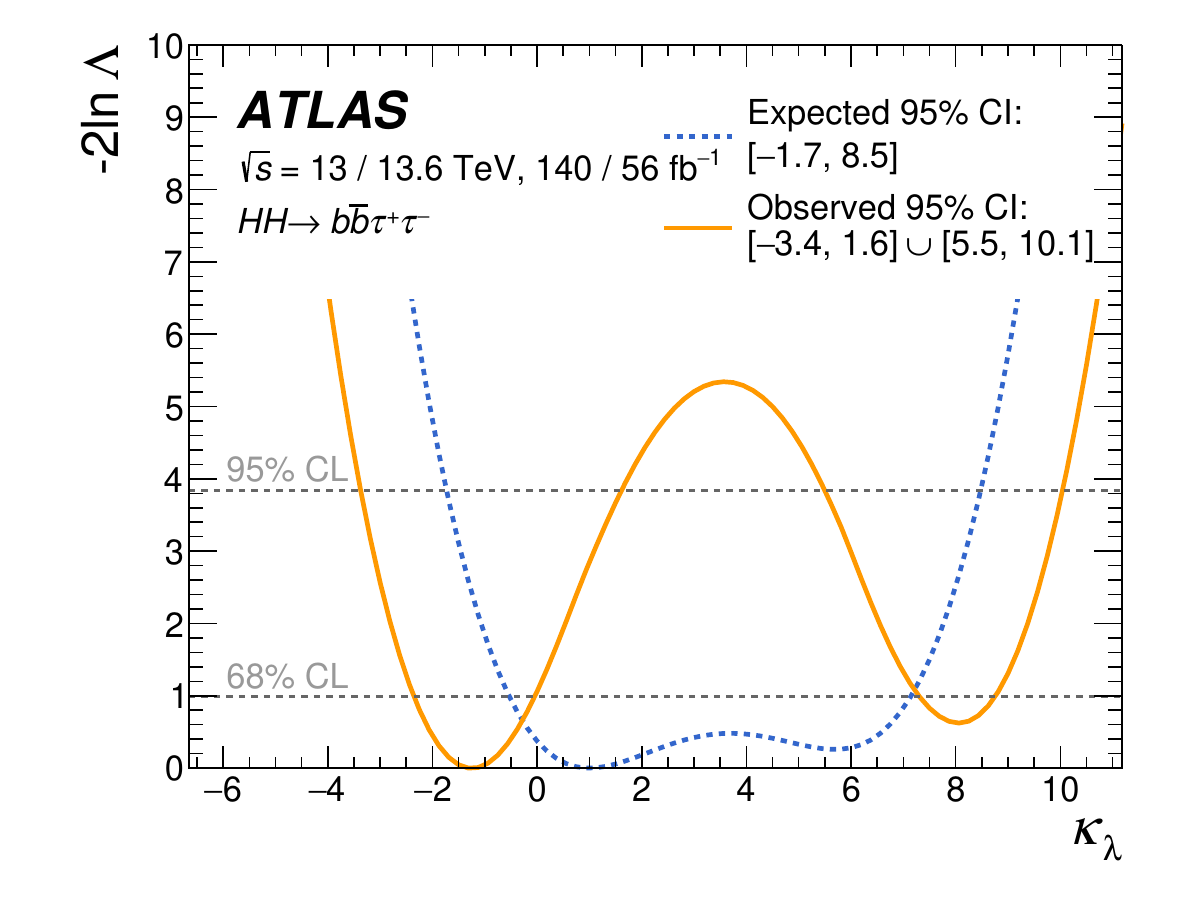}
\label{fig:kl}
}
\subfloat[]{
\includegraphics[width=0.45\textwidth]{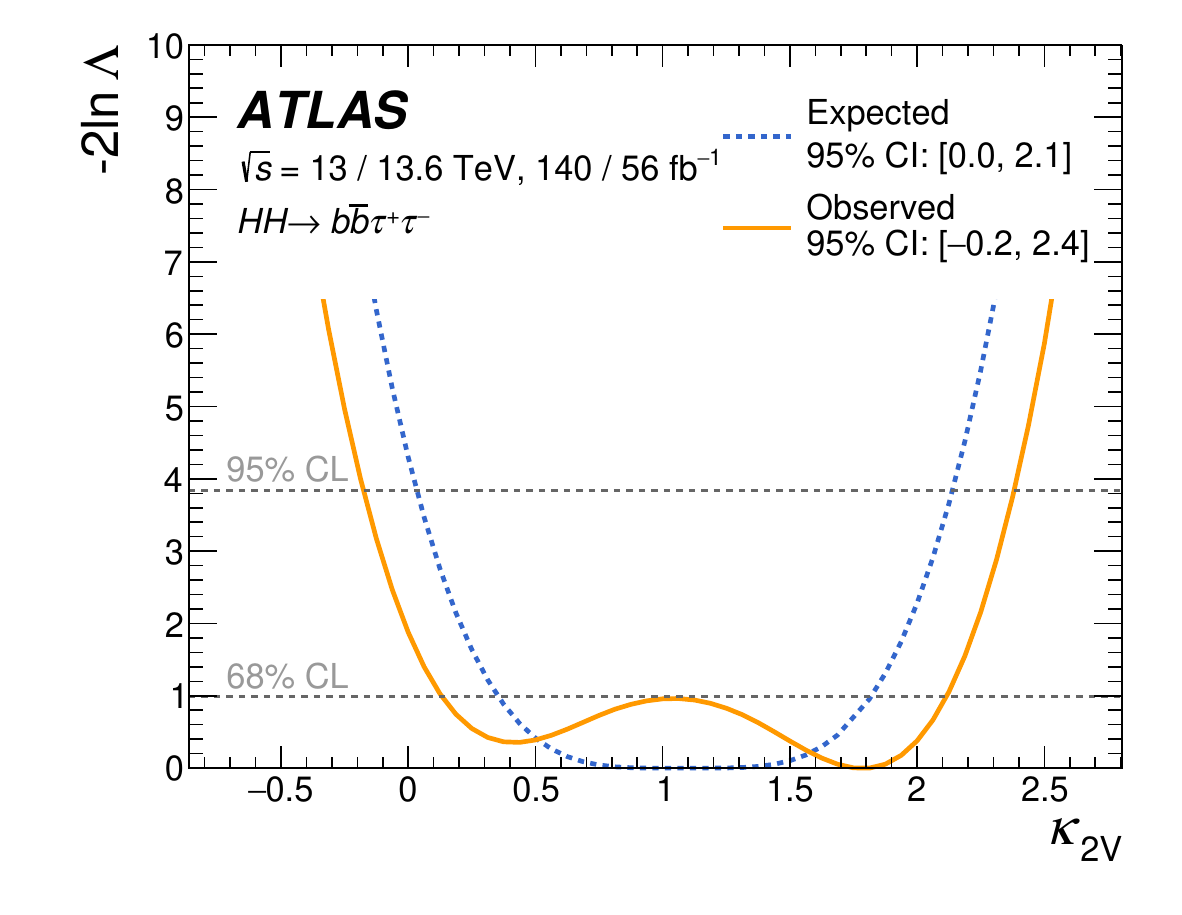}
\label{fig:kvv}
}
\caption{Values of the negative profile log-likelihood ratio for different \protect\subref{fig:kl}~\kl and \protect\subref{fig:kvv}~\kvv hypotheses obtained from fits to the data (solid curve) and to an Asimov dataset (dashed curve) constructed under the SM hypothesis. In each case, all coupling modifiers other than the scanned parameter are fixed to their SM values. The values of the 95\% confidence interval (CI) are also reported.
}
\label{fig:kappa1d}
\end{figure}

The compatibility, considering statistical and systematic uncertainties, between the signal strength $\mu_{HH}$ from this result and that of Ref.~\cite{HDBS-2019-27} is evaluated using a bootstrap technique~\cite{PhysRevD.39.274} considering a standalone fit to only Run~2 data separately for the independent SR fits and for the combined fit.
The compatibility between the two results is at the level of 1.6 standard deviations for the individual fit of the \hadhad SR and of 1.7 standard deviations for the individual fit of the \lephad SR. The compatibility is within 0.3 standard deviations for the combined Run~2 fit driven by the fact that the signal strengths change in opposite directions between the two channels.

Profile-likelihood fits are performed separately for the two coupling modifiers \kl and \kvv, with all others fixed to their SM values. The likelihood scans are shown in Figure~\ref{fig:kappa1d}. The combined fit sets observed (expected) 95\% CL intervals of $\kl\in[\kLdoObsA, \kLupObsA] \cup [\kLdoObsB, \kLupObsB]$ ($[\kLdoExp, \kLupExp]$) (assuming \kvv = 1) and $\kvv \in [\kVVdoObs, \kVVupObs]$ ($[\kVVdoExp, \kVVupExp]$) (assuming \kl = 1). The observed constraints on \kl are looser than the expected primarily due to the excess in the fitted signal strength with respect to the SM predictions. Owing to the observed combined $\mu_{HH}$ signal strength, the observed best-fit value of \kl is close to the point where the product of the production cross-section and acceptance times efficiency is approximately 2.6 times the corresponding SM prediction. The exact position of the best fit is also influenced by the similarity of the signal shapes among different \kl values, although this effect is smaller. The second minimum corresponds to the \kl value with the production cross-section times acceptance closest to that of the first minimum. The difference between the two minima is driven by the differences in the shapes of the two signal hypotheses and their relative contribution to the three SRs. The best-fit value of \kvv is larger than the SM prediction, and the difference between the expected and observed results can be understood in terms of the same effects as for \kl. Compared to Ref.~\cite{HDBS-2019-27}, the expected width of the 95\% confidence interval decreases by approximately 15\% for \kl and 20\% for \kvv, respectively.

\FloatBarrier
\subsection{$ZZ$, $ZH$ signal strength measurements}
Figures~\ref{fig:postfitzzzh} and \ref{fig:postfitzzzh2} show the dedicated classifier score distributions for the $ZZ$ and $ZH$ fits in the \SRHi regions for Run~2 and Run~3, respectively. Similar goodness of fit as in the main \HH fit is obtained. Measurements of $ZZ$ and $ZH$ production yield signal strengths of
\begin{align*}
\mu_{ZZ} & =  0.64^{+ 0.39}_{- 0.36} = 0.64\pm 0.28\:(\textrm{stat.})^{+0.25}_{-0.28}\:(\textrm{syst.}), \\
\mu_{ZH} & =  1.52^{+ 0.52}_{- 0.48} = 1.52\pm 0.40\:(\textrm{stat.})^{+0.33}_{-0.28}\:(\textrm{syst.}),
\end{align*}
respectively, in good agreement with the SM predictions, thus providing a strong validation of the analysis methodology.

In both results, the measurement precision is primarily limited by the statistical uncertainty on the signal strength, resulting from the limited data sample size.
For the $ZZ$ process, the observed (expected) significance with respect to the background-only prediction is 1.8~(2.7) standard deviations.
Evidence for the $ZH\rightarrow \bbtt$ process is found for the first time %
with an observed (expected) significance of 3.5 (2.4) standard deviations.

\begin{figure}[ht!b]
\centering
\subfloat[]{
\includegraphics[width=0.43\textwidth]{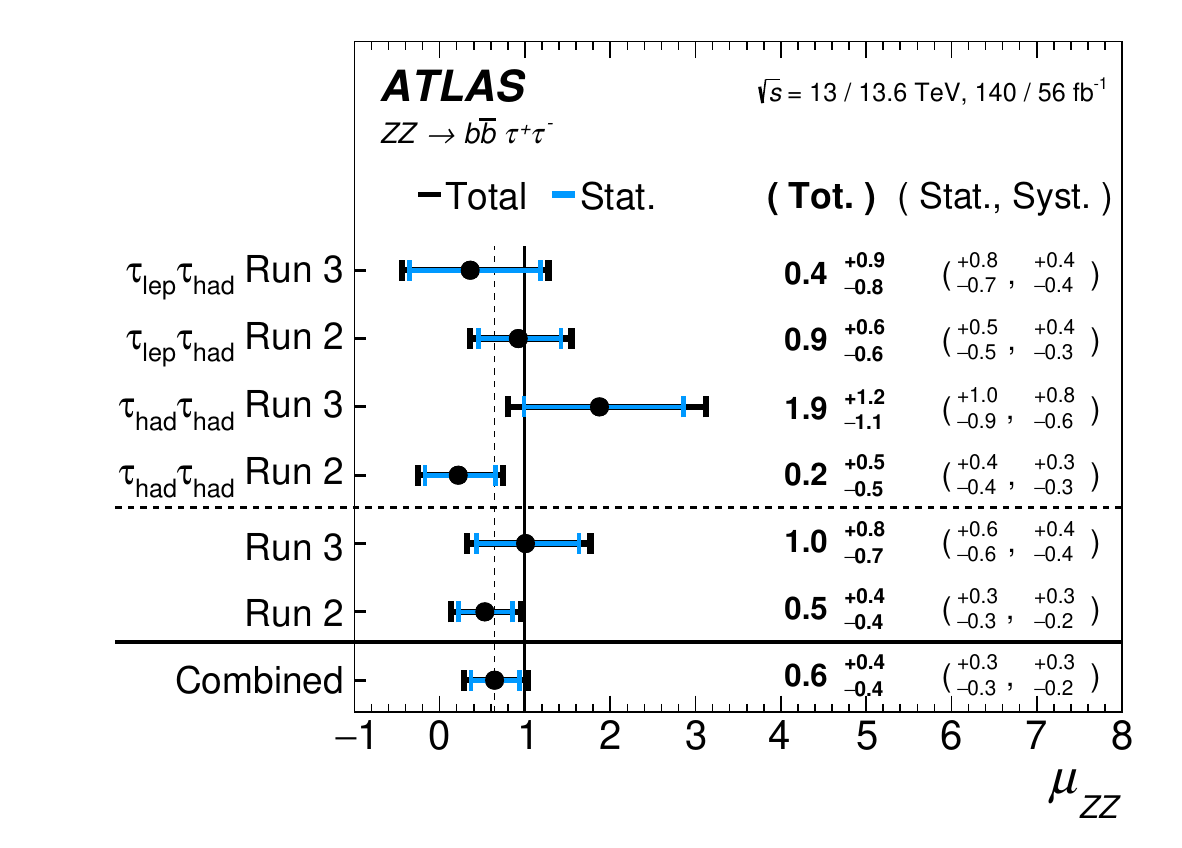}
\label{fig:zz}
}
\subfloat[]{
\includegraphics[width=0.43\textwidth]{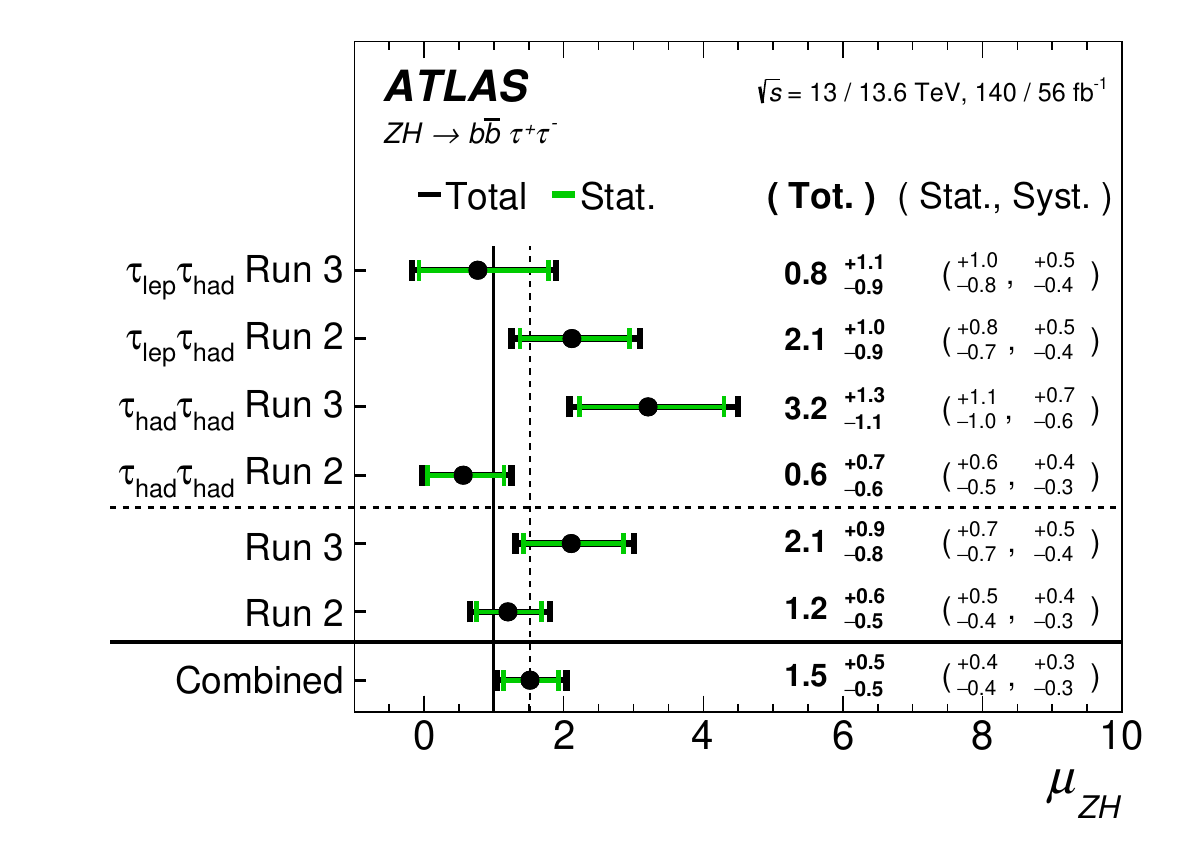}
\label{fig:zh}
}
\caption{Fitted values of the signal strengths for the \protect\subref{fig:zz} $ZZ$ and \protect\subref{fig:zh} $ZH$ processes from combined fits to Run~2 and Run~3 data.
The values for the signal strengths in the various SRs as well as for Run~2 and Run~3 are obtained from a simultaneous fit with the signal strengths for each region and run period floating independently. The solid and dashed vertical lines represent the SM prediction and the observed best-fit value from the corresponding likelihood fit, respectively.
}
\label{fig:zzzh}
\end{figure}

Additional fits are performed to extract the signal strength in each analysis SR as well as separately for Run~2 and Run~3 data. The results are reported in \Fig{\ref{fig:zzzh}}, showing good compatibility
among the different configurations.

\newlength{\myfigwidthB}
\setlength{\myfigwidthB}{0.40\textwidth} %

\begin{figure}[ht!b]
\centering
\subfloat[]{
\includegraphics[width=\myfigwidthB]{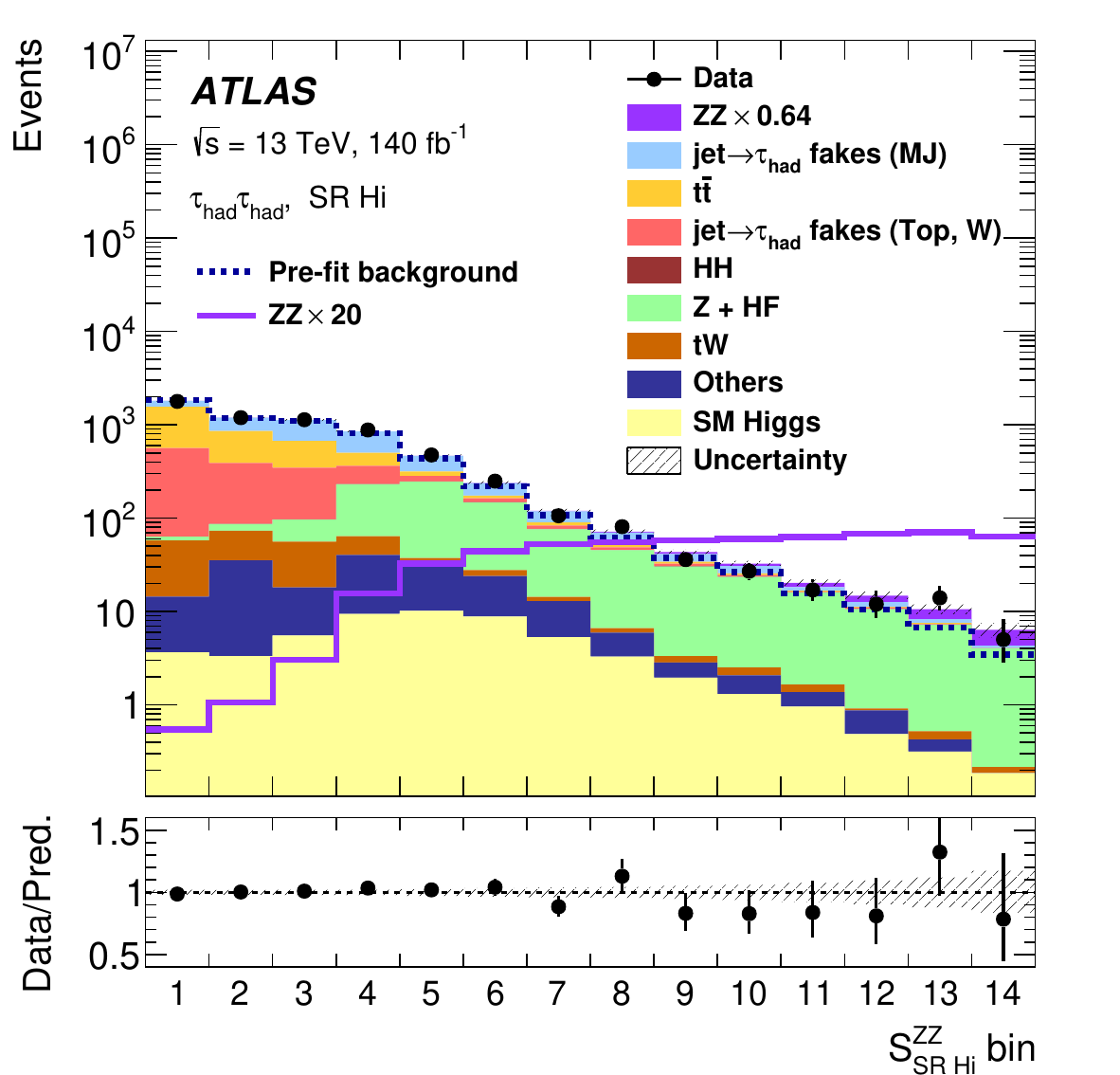}
\label{fig:BDT_3a}
}
\subfloat[]{
\includegraphics[width=\myfigwidthB]{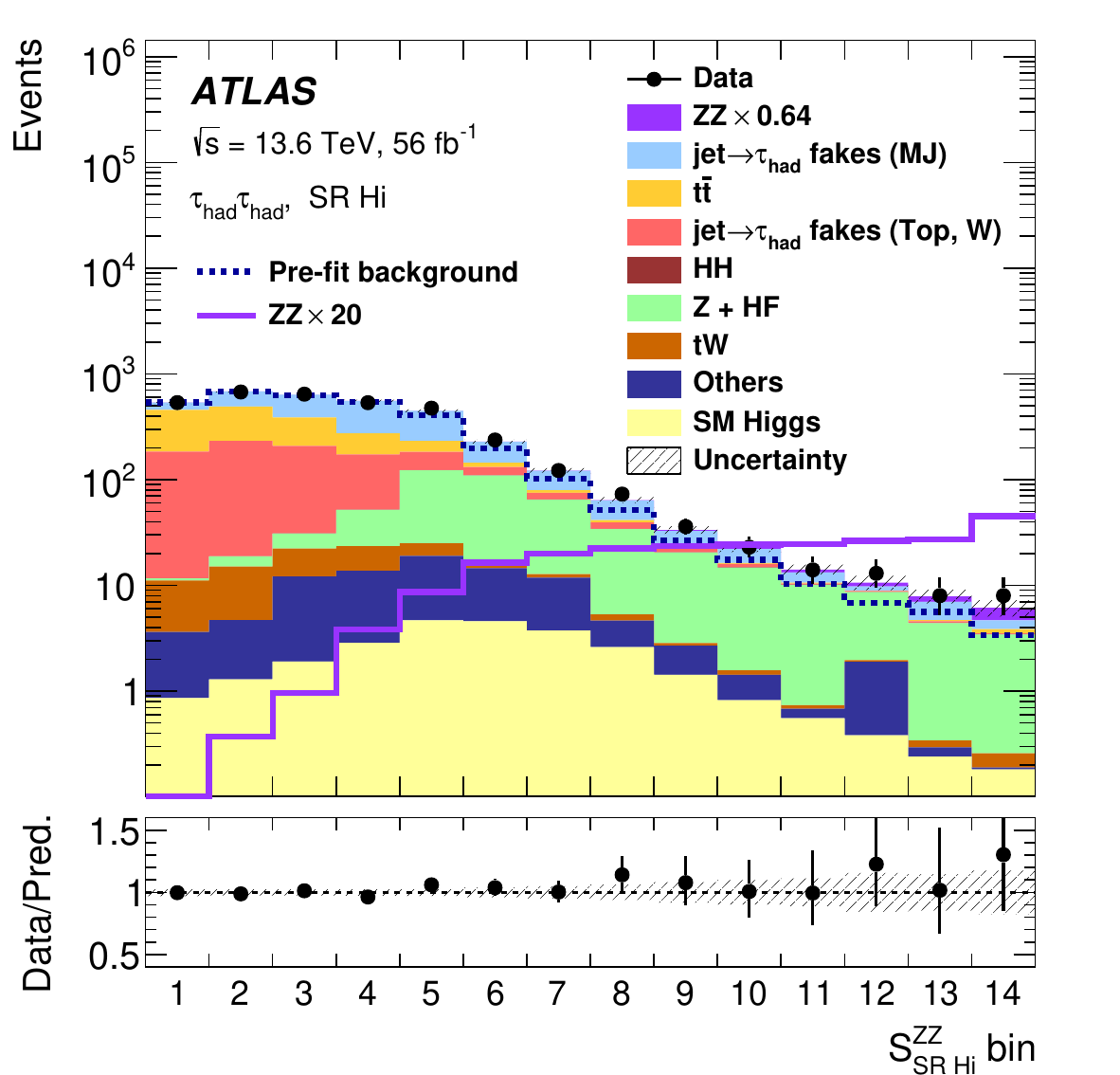}
\label{fig:BDT_3b}
}
\\    \vspace{-0.9em} %
\subfloat[]{
\includegraphics[width=\myfigwidthB]{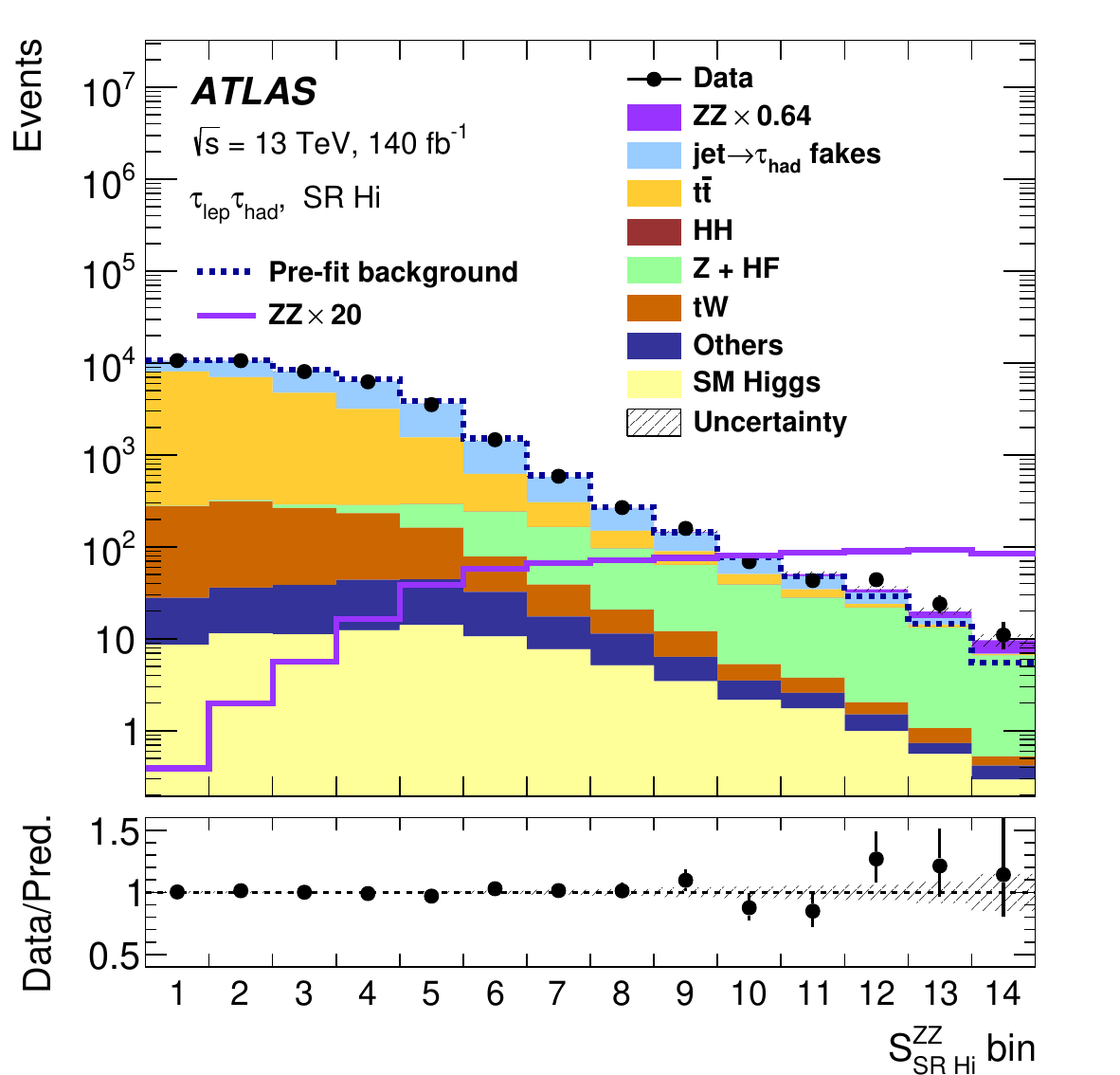}
\label{fig:BDT_3c}
}
\subfloat[]{
\includegraphics[width=\myfigwidthB]{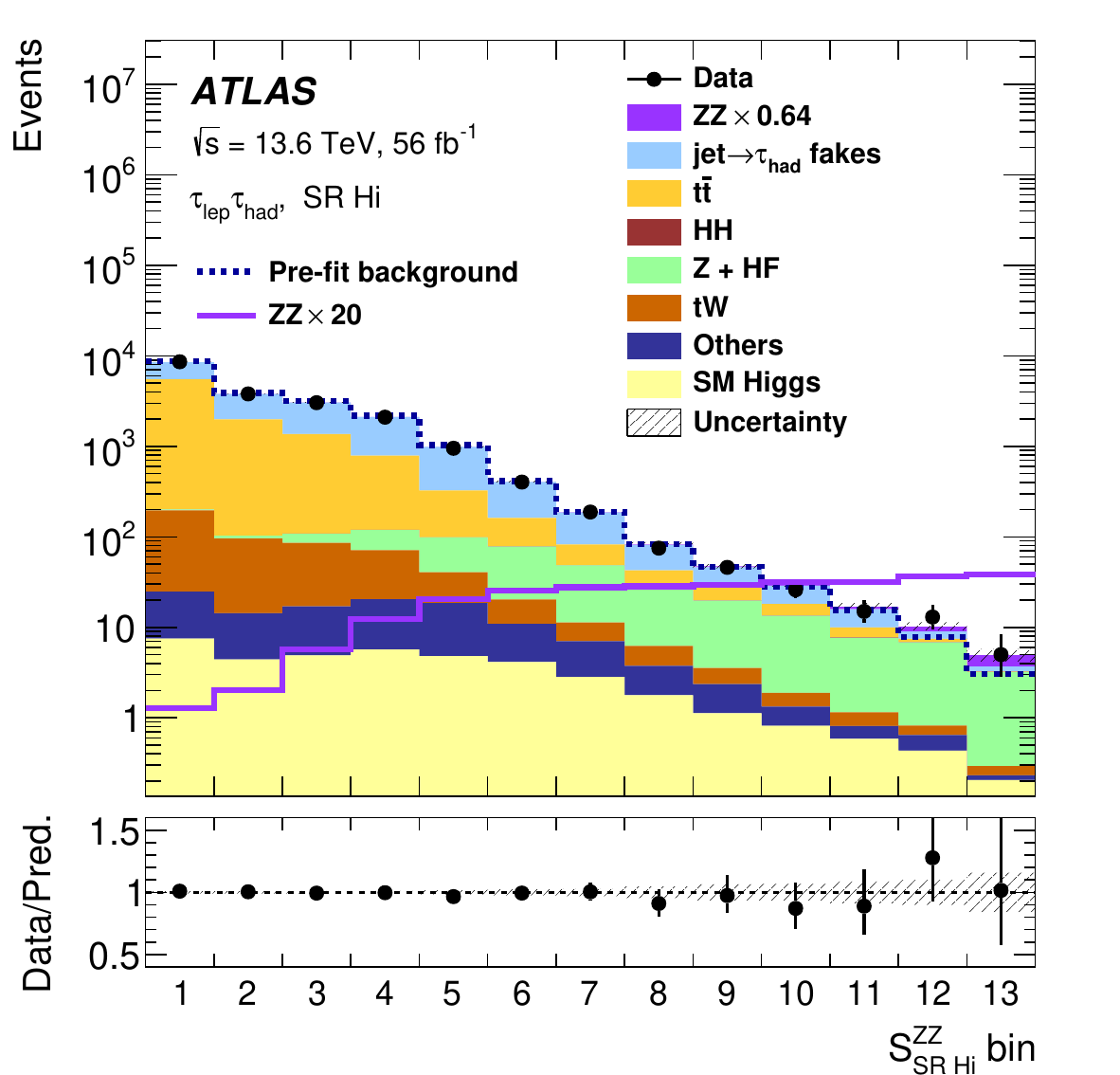}
\label{fig:BDT_3d}
}
\\   \vspace{-0.5em}
\caption[]{Post-fit classifier score distributions for the $ZZ$ measurement. The \SRHi regions are shown for \protect\subref{fig:BDT_3a} the Run~2 \hadhad channel, \protect\subref{fig:BDT_3b} the Run~3 \hadhad channel, \protect\subref{fig:BDT_3c} the Run~2 \lephad channel and \protect\subref{fig:BDT_3d} the Run~3 \lephad channel. The background contributions after the fit are shown as filled histograms. The reference signal is shown as a filled histogram on top of the fitted backgrounds normalised to the signal yield extracted from data, and unstacked as an unfilled histogram, scaled by a factor of 20 for better visualisation. The size of the combined statistical and systematic uncertainties for the sum of the fitted signal and background is indicated by the hatched band. The ratios of the data to the sum of the fitted signal and background are shown in the lower panels.}
\label{fig:postfitzzzh}
\end{figure}

\begin{figure}[ht!b]
\centering
\subfloat[]{
\includegraphics[width=\myfigwidthB]{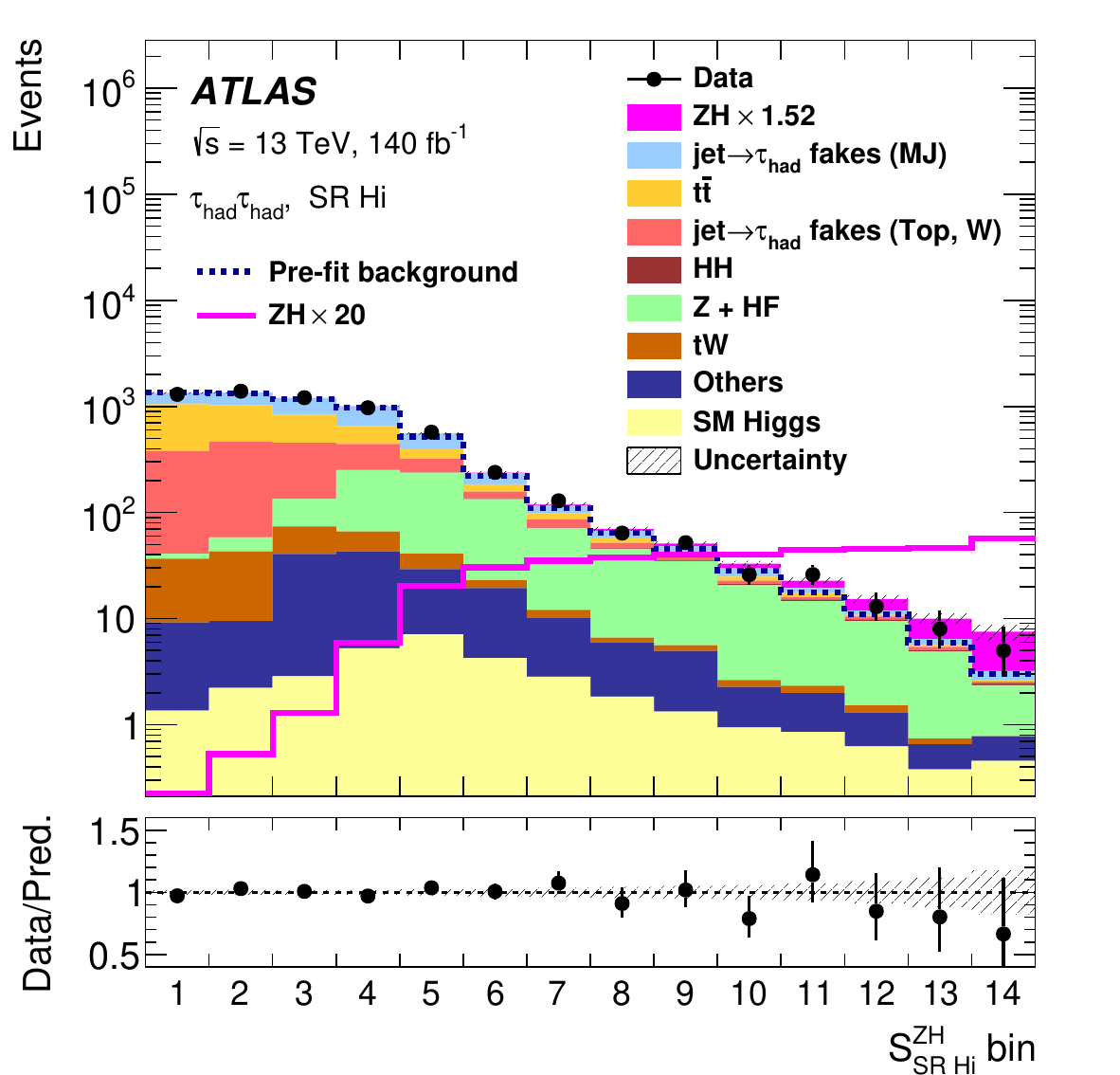}
\label{fig:BDT_4a}
}
\subfloat[]{
\includegraphics[width=\myfigwidthB]{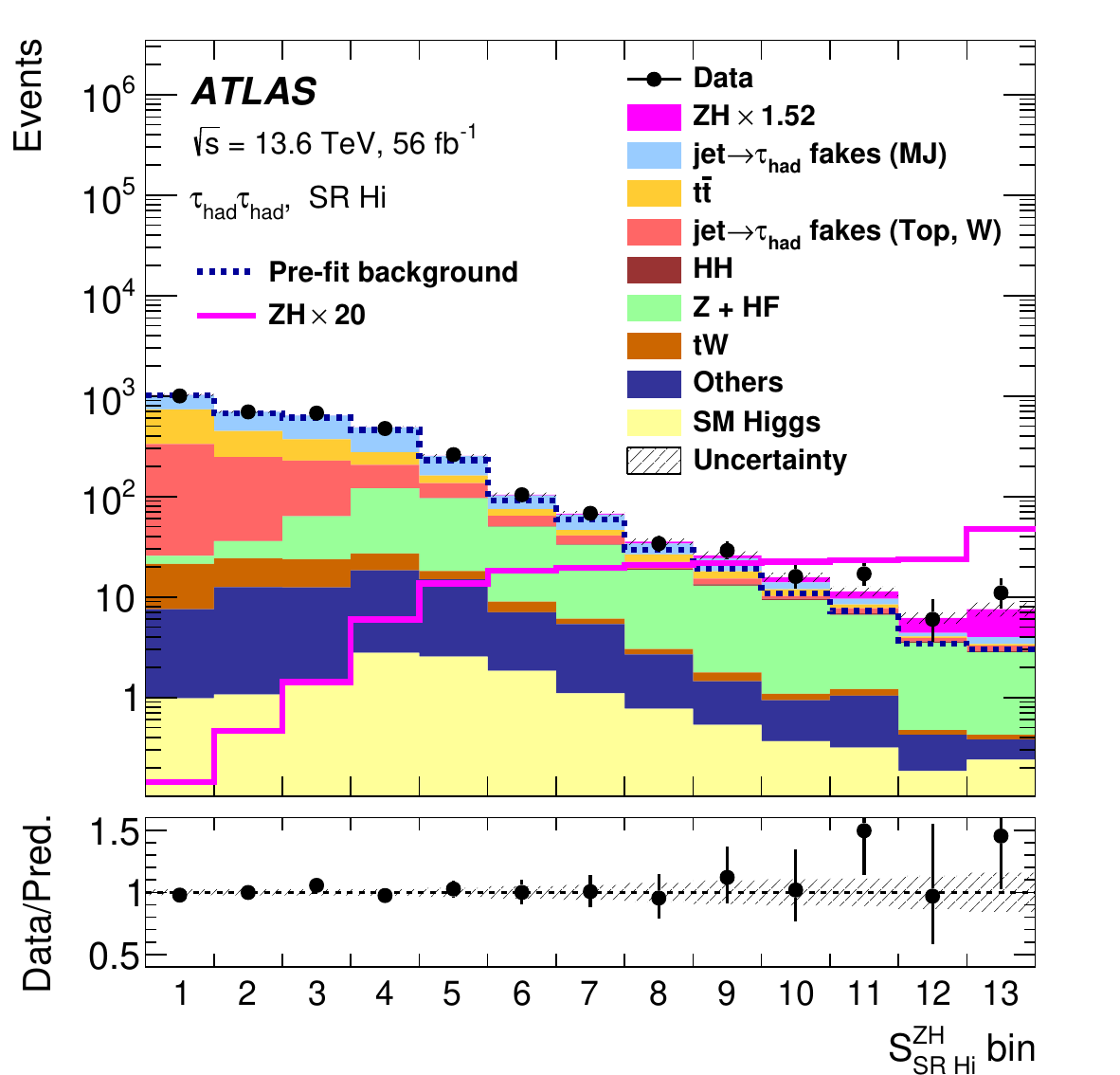}
\label{fig:BDT_4b}
}
\\    \vspace{-0.9em} %
\subfloat[]{
\includegraphics[width=\myfigwidthB]{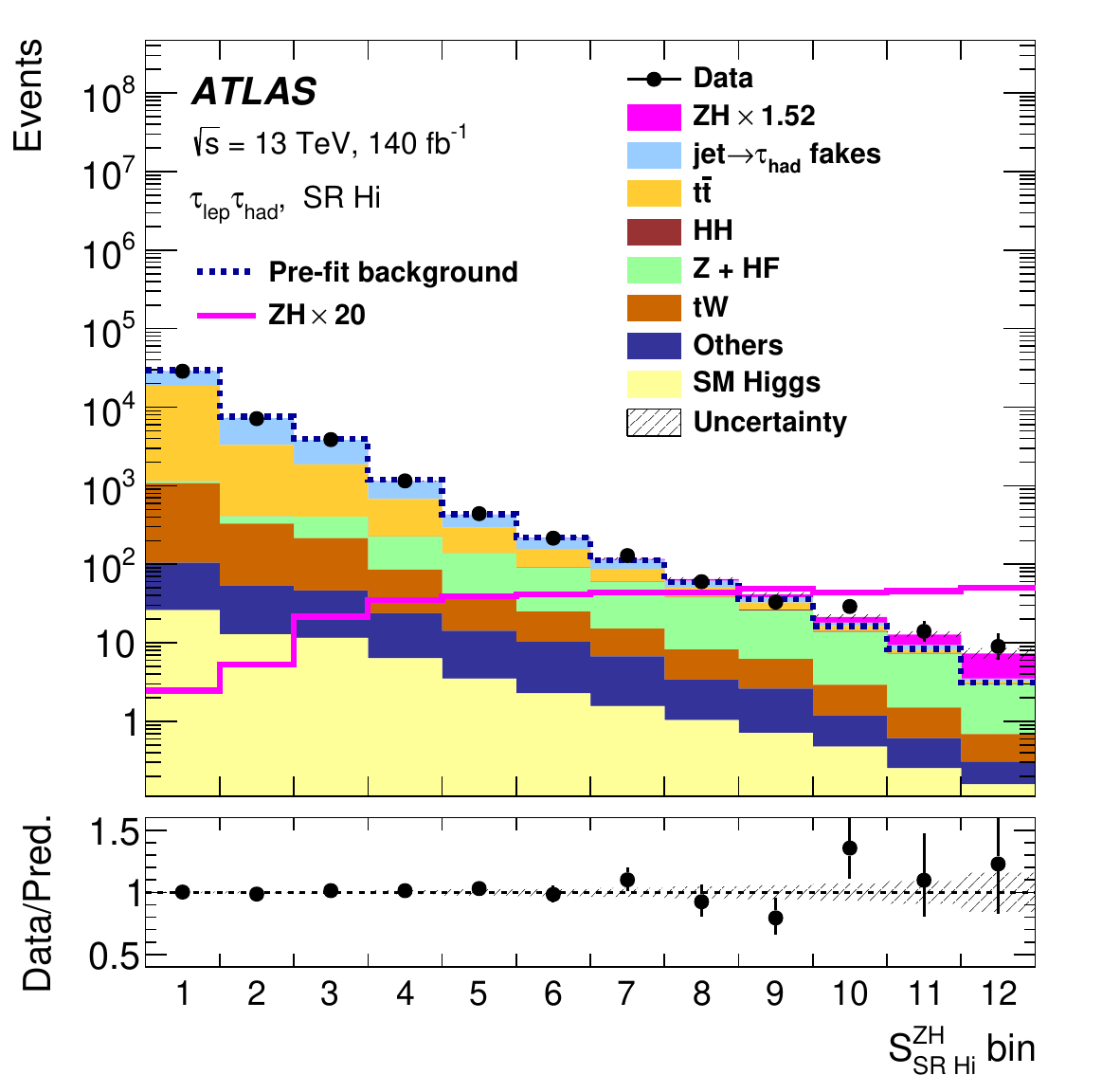}
\label{fig:BDT_4c}
}
\subfloat[]{
\includegraphics[width=\myfigwidthB]{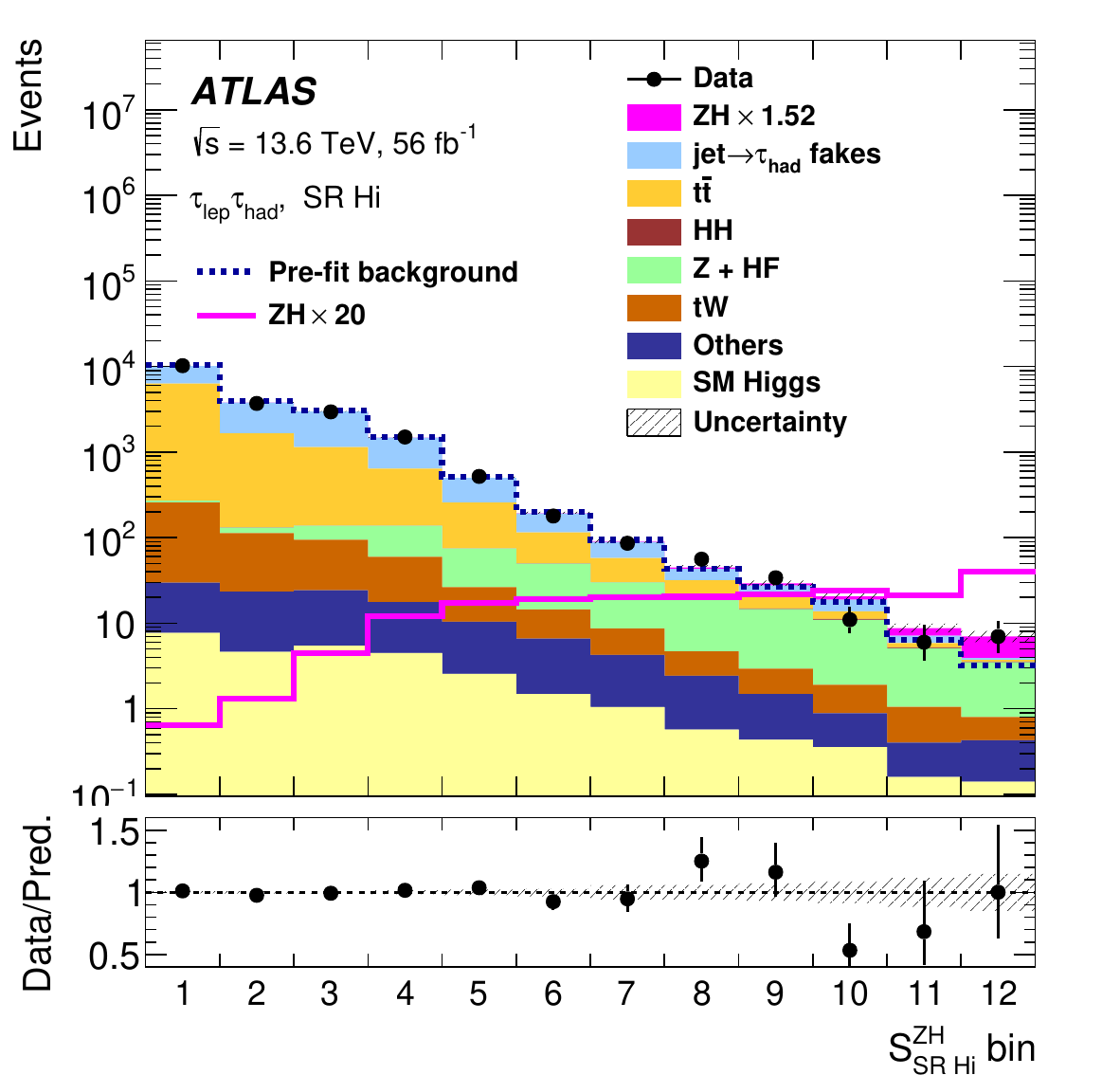}
\label{fig:BDT_4d}
}
\\   \vspace{-0.5em}
\caption[]{Post-fit classifier score distributions for the $ZH$ measurement. The \SRHi regions are shown for \protect\subref{fig:BDT_4a} the Run~2 \hadhad channel, \protect\subref{fig:BDT_4b} the Run~3 \hadhad channel, \protect\subref{fig:BDT_4c} the Run~2 \lephad channel and \protect\subref{fig:BDT_4d} the Run~3 \lephad channel. The background contributions after the fit are shown as filled histograms. The reference signal is shown as a filled histogram on top of the fitted backgrounds normalised to the signal yield extracted from data, and unstacked as an unfilled histogram, scaled by a factor of 20 for better visualisation. The size of the combined statistical and systematic uncertainties for the sum of the fitted signal and background is indicated by the hatched band. The ratios of the data to the sum of the fitted signal and background are shown in the lower panels.}
\label{fig:postfitzzzh2}
\end{figure}

\FloatBarrier


\section{Conclusion}

\label{sec:conclusion}

A search for non-resonant Higgs boson pair production in the \bbtt final state is presented, using 140~fb$^{-1}$ of $pp$ collision data recorded by the ATLAS detector at a centre-of-mass energy of 13~\TeV during Run~2 and 56~fb$^{-1}$ at 13.6~\TeV in Run~3. The analysis combines the \lephad and \hadhad decay channels and supersedes the previous ATLAS search in this final state based on the Run~2 dataset~\cite{HDBS-2019-27}. Relative to the previous publication, the expected sensitivity improves by 15\% to 65\%, depending on the parameter of interest. This improvement results from the addition of the Run~3 data, updated reconstruction and identification algorithms, and a re-optimised multivariate classification strategy.

With respect to the background-only hypothesis of no $HH$ production, a data excess is observed that corresponds to a significance of $2.6$ standard deviations, where $1.2$ standard deviations are expected from SM-like $HH$ production. The signal strength of Higgs boson pair production is found to be $\mu_{HH}=  \muHHRes^{+1.4}_{-1.0}$, compatible with the SM prediction of unity at a level of 1.65 standard deviations. In the absence of a statistically significant signal, limits are placed on enhanced $HH$ signal strengths. At the 95\% confidence level, an observed upper limit of $\mu_{HH} < \hhLimObs$ is placed, compared to an expected limit of $\mu_{HH} < \hhLimExp$ assuming the absence of $HH$ production, i.e. $\mu_{HH} = 0$. The corresponding observed (expected) 95\% confidence intervals for the self-coupling modifier are $\kl \in [\kLdoObsA, \kLupObsA] \cup [\kLdoObsB, \kLupObsB]$ ($\kl \in[\kLdoExp, \kLupExp]$), and that for the quartic-coupling modifier is $\kvv \in [\kVVdoObs, \kVVupObs ]$ ($\kvv \in[ \kVVdoExp, \kVVupExp ]$), respectively.

In addition, validation measurements are performed in the same final state using the same event selection.
The $ZH$ process is observed (expected) with a significance of 3.5 (2.4) standard deviations with respect to the background-only prediction. This marks the first time evidence for this process above 3 standard deviations has been obtained in the $\bbtt$ final state.
The $ZZ$ process is measured with an observed (expected) significance of 1.8 (2.7) standard deviations. Both results are consistent with the SM predictions for these processes, thus providing a strong validation of the analysis methodology.

\FloatBarrier


%
%
%
%
\section*{Acknowledgements}
%

%

%
%

%
%

We thank CERN for the very successful operation of the LHC and its injectors, as well as the support staff at
CERN and at our institutions worldwide without whom ATLAS could not be operated efficiently.

The crucial computing support from all WLCG partners is acknowledged gratefully, in particular from CERN, the ATLAS Tier-1 facilities at TRIUMF/SFU (Canada), NDGF (Denmark, Norway, Sweden), CC-IN2P3 (France), KIT/GridKA (Germany), INFN-CNAF (Italy), NL-T1 (Netherlands), PIC (Spain), RAL (UK) and BNL (USA), the Tier-2 facilities worldwide and large non-WLCG resource providers. Major contributors of computing resources are listed in Ref.~\cite{ATL-SOFT-PUB-2026-001}.

We gratefully acknowledge the support of ANPCyT, Argentina; YerPhI, Armenia; ARC, Australia; BMWFW and FWF, Austria; ANAS, Azerbaijan; CNPq and FAPESP, Brazil; NSERC, NRC and CFI, Canada; CERN; ANID, Chile; CAS, MOST, NSFC and FRFCU, China; Minciencias, Colombia; MEYS CR, Czech Republic; DNRF and DNSRC, Denmark; IN2P3-CNRS and CEA-DRF/IRFU, France; SRNSFG, Georgia; BMFTR, HGF and MPG, Germany; GSRI, Greece; RGC and Hong Kong SAR, China; ICHEP and Academy of Sciences and Humanities, Israel; INFN, Italy; MEXT and JSPS, Japan; CNRST, Morocco; NWO, Netherlands; RCN, Norway; MNiSW, Poland; FCT, Portugal; MNE/IFA, Romania; MSTDI, Serbia; MSSR, Slovakia; ARIS and MVZI, Slovenia; DSI/NRF, South Africa; MICIU/AEI, Spain; SRC and Wallenberg Foundation, Sweden; SERI, SNSF and Cantons of Bern and Geneva, Switzerland; NSTC, Taipei; TENMAK, T\"urkiye; STFC/UKRI, United Kingdom; DOE and NSF, United States of America.

Individual groups and members have received support from BCKDF, CANARIE, CRC and DRAC, Canada; CERN-CZ and PRIMUS, Czech Republic; COST, ERC, ERDF, Horizon 2020 and Marie Sk{\l}odowska-Curie Actions, European Union; Investissements d'Avenir Labex, Investissements d'Avenir Idex and ANR, France; DFG and AvH Foundation, Germany; Herakleitos, Thales and Aristeia programmes co-financed by EU-ESF and the Greek NSRF, Greece; BSF-NSF and MINERVA, Israel; NCN and NAWA, Poland; La Caixa Banking Foundation, CERCA and AGAUR programs from Generalitat de Catalunya and PROMETEO and GenT Programmes Generalitat Valenciana, Spain; G\"{o}ran Gustafssons Stiftelse, Sweden; The Royal Society and Leverhulme Trust, United Kingdom; Eric and Wendy Schmidt Fund for Strategic Innovation, United States of America.

In addition, individual members wish to acknowledge support from Chile: Agencia Nacional de Investigaci\'on y Desarrollo (ANID FONDECYT reg. 1230987, FONDECYT 1230812, FONDECYT 1240864, Fondecyt 3240661, Fondecyt Regular 1240721); China: Fundamental Research Funds for the Central Universities (010-63263105), Chinese Ministry of Science and Technology (MOST-2023YFA1605700, MOST-2023YFA1609300), National Natural Science Foundation of China (NSFC 12275265, NSFC-W2543005); Czech Republic: Czech Science Foundation (GACR - 24-11373S), Ministry of Education Youth and Sports (ERC-CZ-LL2327, FORTE CZ.02.01.01/00/22\_008/0004632); EU: H2020 European Research Council (ERC - 101002463); European Union: European Research Council (BARD No. 101116429, ERC - 101219398, ERC 101089007), European Regional Development Fund (HE COFUND GA No.101081355, ERDF), Marie Sklodowska-Curie Actions (GAP-101168829); France: Agence Nationale de la Recherche (ANR-22-EDIR-0002, ANR-24-CE31-0504-01); Germany: Deutsche Forschungsgemeinschaft (DFG - 469666862); China: Research Grants Council (GRF); Italy: Istituto Nazionale di Fisica Nucleare (LHC-MIUR - 28003/2025), Ministero dell'Università e della Ricerca (NextGenEU I53D23000820006 M4C2.1.1, SOE2024\_0000023); Japan: Japan Society for the Promotion of Science (JSPS KAKENHI  JP25H0063, JSPS KAKENHI JP22H04944, JSPS KAKENHI JP24K23939, JSPS KAKENHI JP24KK0251, JSPS KAKENHI JP25H00650, JSPS KAKENHI JP25H01291, JSPS KAKENHI JP25K01011, JSPS KAKENHI JP25K01023, JSPS KAKENHI JP25KK0047); Poland: Polish National Science Centre (NCN 2021/42/E/ST2/00350, NCN OPUS 2023/51/B/ST2/02507, NCN OPUS nr 2022/47/B/ST2/03059, NCN UMO-2019/34/E/ST2/00393, UMO-2022/47/O/ST2/00148, UMO-2023/49/B/ST2/04085, UMO-2023/51/B/ST2/00920, UMO-2024/53/N/ST2/00869); Spain: Agència de Gestió d'Ajuts Universitaris i de Recerca. (AGAUR - 2023 BP 00141), Ministry of Science and Innovation (RYC2019-028510-I, RYC2020-030254-I, RYC2021-031273-I, RYC2022-038164-I), Ministerio de Ciencia, Innovación y Universidades/Agencia Estatal de Investigaci\'on (EU NextGenerationEU (PRTR-C17.I1), PID2022-142604OB-C22); Sweden: Carl Trygger Foundation (Carl Trygger Foundation CTS 22:2312), Swedish Research Council (Swedish Research Council 2023-04654, VR 2021-03651, VR 2022-03845, VR 2022-04683, VR 2023-03403, VR 2024-05451, VR 2025-05940), Knut and Alice Wallenberg Foundation (KAW 2023.0366); United Kingdom: The Binks Trust, Royal Society (NIF-R1-231091); United States of America: U.S. Department of Energy (ECA DE-AC02-76SF00515), John Templeton Foundation (John Templeton Foundation 63206), Neubauer Family Foundation.

%
%


%
\clearpage
\printbibliography
\clearpage
 
\begin{flushleft}
\hypersetup{urlcolor=black}
{\Large The ATLAS Collaboration}

\bigskip

\AtlasOrcid[0000-0002-6665-4934]{G.~Aad}$^\textrm{\scriptsize 102}$,
\AtlasOrcid[0000-0001-7616-1554]{E.~Aakvaag}$^\textrm{\scriptsize 17}$,
\AtlasOrcid[0000-0002-5888-2734]{B.~Abbott}$^\textrm{\scriptsize 121}$,
\AtlasOrcid[0000-0002-0287-5869]{S.~Abdelhameed}$^\textrm{\scriptsize 83b}$,
\AtlasOrcid[0000-0002-1002-1652]{K.~Abeling}$^\textrm{\scriptsize 54}$,
\AtlasOrcid[0000-0001-5763-2760]{N.J.~Abicht}$^\textrm{\scriptsize 48}$,
\AtlasOrcid[0000-0002-8496-9294]{S.H.~Abidi}$^\textrm{\scriptsize 30}$,
\AtlasOrcid[0009-0003-6578-220X]{M.~Aboelela}$^\textrm{\scriptsize 44}$,
\AtlasOrcid[0000-0002-9987-2292]{A.~Aboulhorma}$^\textrm{\scriptsize 36e}$,
\AtlasOrcid[0000-0001-5329-6640]{H.~Abramowicz}$^\textrm{\scriptsize 154}$
% ...
\bigskip
\\

% ...
$^{17}$Department for Physics and Technology, University of Bergen, Bergen; Norway.\\
% ...
$^{30}$Physics Department, Brookhaven National Laboratory, Upton NY; United States of America.\\
% ...
$^{36}$$^{(a)}$Facult\'e des Sciences Ain Chock, Universit\'e Hassan II de Casablanca;$^{(b)}$Facult\'{e} des Sciences, Universit\'{e} Ibn-Tofail, K\'{e}nitra;$^{(c)}$Facult\'e des Sciences Semlalia, Universit\'e Cadi Ayyad, LPHEA-Marrakech;$^{(d)}$LPMR, Facult\'e des Sciences, Universit\'e Mohamed Premier, Oujda;$^{(e)}$Facult\'e des sciences, Universit\'e Mohammed V, Rabat;$^{(f)}$Institute of Applied Physics, Mohammed VI Polytechnic University, Ben Guerir; Morocco.\\
% ...
$^{44}$Physics Department, Southern Methodist University, Dallas TX; United States of America.\\
% ...
$^{48}$Fakult\"{a}t Physik, Technische Universit{\"a}t Dortmund, Dortmund; Germany.\\
% ...
$^{54}$II. Physikalisches Institut, Georg-August-Universit\"{a}t G\"ottingen, G\"ottingen; Germany.\\
% ...
$^{83}$$^{(a)}$Khalifa University of Science and Technology, Abu Dhabi;$^{(b)}$New York University Abu Dhabi, Abu Dhabi;$^{(c)}$United Arab Emirates University, Al Ain;$^{(d)}$University of Sharjah, Sharjah; United Arab Emirates.\\
% ...
$^{102}$CPPM, Aix-Marseille Universit\'e, CNRS/IN2P3, Marseille; France.\\
% ...
$^{121}$Homer L. Dodge Department of Physics and Astronomy, University of Oklahoma, Norman OK; United States of America.\\
% ...
$^{154}$Raymond and Beverly Sackler School of Physics and Astronomy, Tel Aviv University, Tel Aviv; Israel.\\
% ...
$^{a}$ Also at Affiliated with an institute formerly covered by a cooperation agreement with CERN.\\
$^{b}$ Also at An-Najah National University, Nablus; Palestine.\\
$^{c}$ Also at Borough of Manhattan Community College, City University of New York, New York NY; United States of America.\\
$^{d}$ Also at Center for Interdisciplinary Research and Innovation (CIRI-AUTH), Thessaloniki; Greece.\\
$^{e}$ Also at Centre of Physics of the Universities of Minho and Porto (CF-UM-UP); Portugal.\\
$^{f}$ Also at CERN, Geneva; Switzerland.\\
% ...

\end{flushleft}


%

%
%

%

%
%
%
%

\end{document}